\documentclass[]{aastex631}

\received{Jan 9, 2024}
\accepted{ApJ, Feb 29, 2024}

\begin{document}

\title{Reconnaissance with JWST of the J-region Asymptotic Giant Branch in Distance Ladder Galaxies: From Irregular Luminosity Functions to Approximation of the Hubble Constant}

\correspondingauthor{Siyang Li}
\email{sli185@jh.edu}

\author[0000-0002-8623-1082]{Siyang Li}
    \affil{Department of Physics and Astronomy, Johns Hopkins University, Baltimore, MD 21218, USA}

\author[0000-0002-6124-1196]{Adam G. Riess}
    \affil{Department of Physics and Astronomy, Johns Hopkins University, Baltimore, MD 21218, USA}
    \affil{Space Telescope Science Institute, Baltimore, MD, 21218, USA}    

\author{Stefano Casertano}
    \affil{Space Telescope Science Institute, Baltimore, MD, 21218, USA}

\author[0000-0002-5259-2314]{Gagandeep S. Anand}
    \affil{Space Telescope Science Institute, Baltimore, MD, 21218, USA}
    
\author[0000-0002-4934-5849]{Daniel M. Scolnic}
    \affil{Department of Physics, Duke University, Durham, NC 27708, USA}

\author[0000-0001-9420-6525]{Wenlong Yuan}
\affiliation{Department of Physics and Astronomy, Johns Hopkins University, Baltimore, MD 21218, USA}

\author[0000-0003-3889-7709]{Louise Breuval}
\affil{Department of Physics and Astronomy, Johns Hopkins University, Baltimore, MD 21218, USA}
    
\author[0000-0001-6169-8586]{Caroline D. Huang}
    \affil{Center for Astrophysics, Harvard \& Smithsonian, 60 Garden Street, Cambridge, MA 02138, USA} 
 
\begin{abstract}

We study stars in the J-regions of the asymptotic giant branch (JAGB) of near-infrared color magnitude diagrams in the maser host NGC 4258 and 4 hosts of 6 Type~Ia~supernovae (SN~Ia): NGC 1448, NGC 1559, NGC 5584, and NGC 5643. These clumps of stars are readily apparent near $1.0<F150W-F277W<1.5$ and $m_{F150W}$=22-25~mag with \textit{James Webb Space Telescope} NIRCam photometry. Various methods have been proposed to assign an apparent reference magnitude for this recently proposed standard candle, including the mode, median, sigma-clipped mean or a modeled luminosity function parameter. We test the consistency of these by measuring intra-host variations, finding differences of up to $\sim$0.2~mag that significantly exceed statistical uncertainties. Brightness differences appear intrinsic, and are further amplified by the non-uniform shape of the JAGB luminosity function, also apparent in the LMC and SMC. We follow a ``many methods' approach to consistently measure JAGB magnitudes and distances to the SN~Ia host sample calibrated by NGC 4258. We find broad agreement with distances measured from Cepheids, tip of the red giant branch (TRGB), and Miras. However, the SN host mean distance estimated via the JAGB method necessary to estimate $H_0$ differs by $\sim$0.19~mag amongst the above definitions, a result of different levels of luminosity function asymmetry. The methods yield a full range of $71-78$ km s$^{-1}$ Mpc$^{-1}$, i.e., a fiducial result of $H_0=74.7\pm2.1$(stat)$\pm$2.3(sys)($\pm$3.1 if combined in quadrature) km s$^{-1}$ Mpc$^{-1}$, with systematic errors limited by the differences in methods. Future work may seek to further standardize and refine this promising tool, making it more competitive with established distance indicators.
\end{abstract}

\keywords{}

\section{Introduction} \label{sec:Intro}

\begin{figure*}[ht!]
\epsscale{0.5}
\plotone{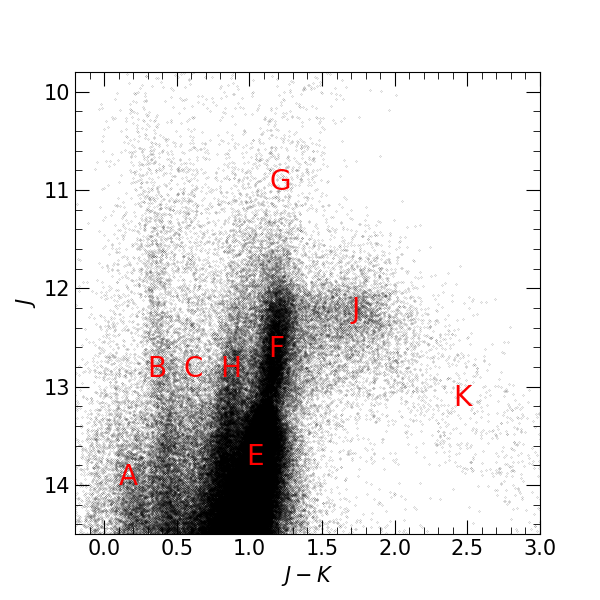}
\caption {$J$ vs. $J - K$ CMD using Large Magellanic Cloud (LMC) photometry from \cite{Macri_2015AJ....149..117M} and labelled with regions from \cite{Weinberg_2001ApJ...548..712W}.}
\label{fig:JAGB_LMC_CMD_Labelled}
\end{figure*}

The need for primary distance indicators that can reach tens of Mpc motivates efforts to identify alternative luminous standard candles, such as the recently proposed stars in the J-region Asymptotic Giant Branch (JAGB).  The JAGB can be identified as an enhanced density of stars or a ``clump'' in the near-infrared color magnitude diagram (CMD; often in $J$ vs. $J-K$ or a similar filter set) redward of the red giant branch (RGB) and brighter than the tip of the RGB. This region is labelled `J' in Fig. \ref{fig:JAGB_LMC_CMD_Labelled}, corresponding to the labels defined by \cite{Weinberg_2001ApJ...548..712W}, and is expected from stellar theory to be populated by thermally-pulsing carbon-rich (photospheric carbon-to-oxygen ratio $> 1$), dust-producing asymptotic giant branch stars with estimated amplitudes of $\sim$0.7~mag \citep{Madore_2020ApJ...899...66M} and J-band luminosities of $M_J\sim$ $-$6~mag. Various measures of the apparent magnitude of the JAGB region, such as the mode, mean, median or a fitted value of its luminosity function, have been suggested and used in the literature as a potential standard candle that is luminous enough to measure extragalactic distances at megaparsec scales \citep{Battinelli_2005AA...442..159B, Madore_2020ApJ...899...66M, Freedman_2020ApJ...899...67F, Ripoche_2020MNRAS.495.2858R,
Lee_MW_2021ApJ...923..157L, 
Lee_WLM_2021ApJ...907..112L,
Zgirski_2021ApJ...916...19Z, Parada_2021MNRAS.501..933P, Madore_2022ApJ...938..125M, Parada_2023MNRAS.522..195P, Lee_2023ApJ...956...15L, Lee_2024ApJ...961..132L}.

The theoretical foundation for expecting the mean of the JAGB to be a useful standard candle is rooted in the prediction that only a narrow range of stellar masses allows for a large quantity of carbon to be transported to the star's surface, which results in the formation of carbonaceous circumstellar dust. These stars are expected to be of intermediate age and metallicity, with ages ranging from 300 Myr to 1 Gyr and masses ranging from $\sim$ 2 to 5 $\mathrm{M}_\odot$ \citep{Marigo_2008A&A...482..883M, Karakas_2014MNRAS.445..347K, Madore_2020ApJ...899...66M}, although these ranges may change with metallicity of the stars \citep{Karakas_2014MNRAS.445..347K}. When a star's mass falls within the prescribed range, the third dredge-up events during post-main sequence mass transport lead to the movement of carbon to the surface, resulting in the formation of a carbon AGB star \citep{Groenewegen_2004agbs.book..105G}. Stars with too little mass ($\lesssim 1.2-1.3 \,\mathrm{M}_\odot$) \citep{Groenewegen_2004agbs.book..105G} do not have dredge up events that are efficient enough to bring enough carbon to the surface of the star to form a carbon star. Stars with too much mass will experience hot bottom burning, where carbon burns at the bottom of the convection zone before it can reach the surface.  Observationally, the C/O$>$1 feature of carbon stars is characteristically shown by carbon spectral bands such as CN, C$_2$, and CO, and lack of oxides \citep[see for instance][and references therein]{Gonneau_2017AA...601A.141G}.

\begin{figure*}[ht!]
\epsscale{0.5}
\plotone{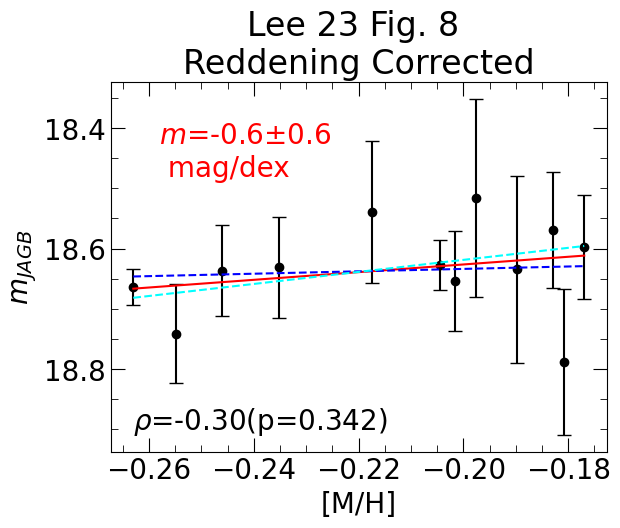}
\caption {Fig. 8 from \cite{Lee_2023ApJ...956...15L} for reddening corrected JAGBs measured with a mode-based method as a function of metallicity [M/H]. We fit the points by minimizing chi-squared and add the results in red. We also add two lines demonstrating example metallicity trends of $-$1 mag/dex (cyan) and $-$0.2 mag/dex \citep[blue; approximating the metallicity trend for Cepheids,][]{Breuval_2022ApJ...939...89B, Trentin_2023arXiv231003603T}. All three lines would be formally consistent with the fit shown in red. The Spearman coefficient is shown in the bottom left hand corner and matches the value found by \cite{Lee_2023ApJ...956...15L}.}
\label{fig:Lee23_JAGB_M/H}
\end{figure*}

Previous studies have indicated that the mass range required for an oxygen-rich asymptotic giant branch (AGB) star to transition into a carbon-rich AGB star can vary with the metallicity of the stellar population \citep{Karakas_2014MNRAS.445..347K, Ripoche_2020MNRAS.495.2858R, Parada_2021MNRAS.501..933P, Parada_2023MNRAS.522..195P}. \cite{Lee_2023ApJ...956...15L} showed that the JAGB does not show any dependency of its luminosity with metallicity over a very small explored range in [M/H] of 0.08 dex ($-0.18 < [M/H] < -0.26$). However, due to the small range in metallicity studied, we caution that their data is simultaneously consistent with no trend and with what could be a very large trend of $\sim-$1~mag/dex. To show this, we show in Fig.~\ref{fig:Lee23_JAGB_M/H} the data points from the reddening corrected JAGBs in Fig.~8 of \cite{Lee_2023ApJ...956...15L}; the nominal best-fit trend (red line) has a slope of $-0.6 \pm 0.6$~mag/dex.\footnote{We also show the Spearman's coefficient in the bottom left hand corner, which matches the value found in \cite{Lee_2023ApJ...956...15L}.}  The limited leverage due to the small range of metallicity results in a constraint that is consistent to within 1 $\sigma$ with both no metallicity dependence and what would be a very large metallicity dependence of $\sim$$-$1~mag/dex (cyan line in Fig. \ref{fig:Lee23_JAGB_M/H}). For reference, we also plot a line with slope of $-$0.2~mag/dex, corresponding to approximately the most widely found Cepheid metallicity dependence \citep[][and references therein]{Breuval_2022ApJ...939...89B, Trentin_2023arXiv231003603T}. All three of the lines shown in Fig.~\ref{fig:Lee23_JAGB_M/H} are consistent with the data presented in \cite{Lee_2023ApJ...956...15L}, indicating no strong constraint. For context, if the metallicity dependence was as large as $-$1~mag/dex, the effort to calibrate $H_0$ to 5\% precision without any metallicity correction would restrict the measurement to a field about 0.3 arcminutes wide at the distances of SN Ia host galaxies, based on typical host metallicity gradients of 0.05 dex/kpc \citep{Hoffmann_2016ApJ...830...10H}.

Thus it would be risky to assume the JAGB is not impacted by differences in host population metallicity (or age) at a level needed to support a few-percent-level determination of the Hubble constant. However, few studies have yet quantified the potential size of such intrinsic variations in the JAGB candle at the target scale \citep{Madore_2020ApJ...899...66M, Freedman_2020ApJ...899...67F, Lee_MW_2021ApJ...923..157L, Lee_WLM_2021ApJ...907..112L, Zgirski_2021ApJ...916...19Z, Lee_2022ApJ...933..201L,Lee_WLM_2021ApJ...907..112L, Lee_2023ApJ...956...15L}. \cite{Karakas_2014MNRAS.445..347K} used theoretical stellar evolutionary models of AGB stars to demonstrate that higher metallicities in stellar populations lead to less efficient dredge-up events.  

Another issue which challenges the use of JAGB as a new standard candle derives from the change in shape (or population mix) indicated in its luminosity function between different hosts.  \cite{Parada_2021MNRAS.501..933P} show significantly different levels of asymmetry in the JAGB LFs in different galaxies: the LMC and NGC 6822 exhibit highly skewed luminosity functions, in contrast to the symmetric luminosity functions in the SMC and IC 1613; these differences are not due to foreground or background contamination.  In the Local Group, the asymmetric LF in the LMC results produces a difference between its {\it mode} and {\it mean} of $\sim$ 0.16 mag, while this difference for the SMC is only $\sim 0.01$ mag.  The magnitude estimates based on the median or a tightly-framed mean are in good agreement with their relative geometric distances; the distances based on the mode are not \citep{Parada_2021MNRAS.501..933P}.  We will revisit this point in the Discussion.  Tight-framing of the mean (i.e., the selection of a small range to average), as employed by \cite{Madore_2020ApJ...899...66M} and \cite{Freedman_2020ApJ...899...67F}, can be used to avoid population contamination, and may offer good agreement with other distance indicators, but the lack of a defined prescription for these tight frames makes it difficult to evaluate and replicate this procedure. The non-trivial width of the J-region LF ($\sim$ 0.3 mag for a single epoch and $\sim$0.2~mag using time-averaged magnitudes) also suggests some diversity of the stellar populations in this region.  Despite these issues, the JAGB LF still offers one of the few possible standard candles luminous enough to rival the distance reach of Cepheids, TRGB, and Miras \citep{Lee_2024ApJ...961..132L}, motivating the present empirical study. 

 We find the JAGB clumps are readily apparent in the CMDs measured from JWST Cycle 1 program \citep[GO-1685, P.I.: A. Riess,][]{Riess_2021jwst.prop.1685R}, which observed four SN Ia hosts within 25 Mpc and NGC 4258 (obtained to measure Cepheids, Miras, and the TRGB). This further motivated us to investigate the feasibility of using the JAGB as a standard candle following the methods employed in the literature. We first investigate variations in the measured JAGB within hosts. We then measure distances and $H_0$ following a distance ladder and ``many methods" approach. In Sections \ref{sec:data} to \ref{sec:Crowding_Corrections} we describe the data processing used for this study. We introduce how we measure the JAGB in Section \ref{sec:Measurement_Basics}. In Section \ref{sec:Symmetry_Null_Tests_Outer_Fields}, we investigate the JAGB through null tests, by varying the region of the measurement within a host.   We characterize the variations in the JAGB measurements and then use them to determine distances to the SN Ia hosts in Section \ref{sec:JAGB_Distances}.  We discuss our results in Section \ref{sec:Discussion}. 

\section{Measurement} \label{sec:Measurement}

\subsection{Data Selection} \label{sec:data}

We retrieved $JWST$ \emph{F090W} (0.9 $\mu$m), \emph{F150W} (1.5 $\mu$m), and \emph{F277W} (2.8 $\mu$m) images of NGC 4258, NGC 1448, NGC 1559, NGC 5584, and NGC 5643 from Cycle 1 program GO-1685 \citep[P.I.: A. Riess,][]{Riess_2021jwst.prop.1685R} from the Mikulski Archive for Space Telescopes\footnote{\href{https://archive.stsci.edu/}{https://archive.stsci.edu/}}. 
Recently, STScI implemented a major update to their image pipeline with new flat fields and zeropoints in the \texttt{jwst\_1125.pmap} and \texttt{jwst\_1126.pmap} context files. After the release of \texttt{jwst\_1126.pmap}, there have been no major updates that would significantly affect NIRCam photometry. 
For this study, we use images processed with context files released later than \texttt{jwst\_1126.pmap}. The versions of these context files can be found in Table \ref{tab:Observations_Table}.

The primary purpose of these $JWST$ observations was to observe a large sample of Cepheids in the disks and TRGB in the halos of the anchor galaxy, NGC 4258, as well as multiple SN Ia host galaxies to construct a distance ladder to measure $H_0$.  Consequently, these observations contain two visits, visit 1 and visit 2.  Because the JAGB is measured in the outer disk (regions younger than old halos, hence with a population of AGB stars), this program necessarily samples regions between the sites of Cepheids and the TRGB. 

We provide a summary of the observations analyzed in this study in Table \ref{tab:Observations_Table} and show their footprints in Fig. \ref{fig:Footprints}. For all JAGB measurements presented below, we correct for foreground extinction by retrieving $E(B-V)$ values using the \texttt{dustmaps} package \citep{Green_2018JOSS....3..695M} in Python and galaxy coordinates from the NASA/IPAC Extragalactic Distance Database (NED)\footnote{\href{https://ned.ipac.caltech.edu/}{https://ned.ipac.caltech.edu/}. The NASA/IPAC Extragalactic Database (NED) is funded by the National Aeronautics and Space Administration and operated by the California Institute of Technology.} and applying the extinction coefficients listed in Table \ref{tab:Foreground_Extinction} assuming $R_V$=3.1 and $A_{\lambda}/A_V$ = 0.6021 and 0.2527 for \emph{F150W}, and \emph{F277W}, respectively. We measure all JAGBs in the main body of this paper in \emph{F150W}, noting that the $F150W$ filter is not a straight equivalent of the Two Micron All Sky Survey (2MASS) J-band that had been originally used for JAGB measurements such as in \cite{Madore_2020ApJ...899...66M}. We discuss using the $F150W$ band for JAGB reference magnitude magnitudes and color relations in Appendix \ref{Apdx: J-region_Tilts}.

\begin{deluxetable*}{ccccccc}
\label{tab:Observations_Table}
\tablehead{\colhead{Galaxy} & \colhead{Observation} & \colhead{Observation Date} & \colhead{Filters} & \colhead{Exposure Time [s]} & \colhead{Context File Number}}
\startdata
NGC 1448 & 13  & 2023-08-02 & \emph{F090W}/\emph{F277W} & 418.7 x 4 & 1146 \\
NGC 1448 & 13  & 2023-08-02 & \emph{F150W}/\emph{F277W} & 526.1 x 4 & " \\
NGC 1448 & 14  & 2023-08-18 & \emph{F090W}/\emph{F277W} & 418.7 x 4 & " \\
NGC 1448 & 14  & 2023-08-18 & \emph{F150W}/\emph{F277W} & 526.1 x 4 & " \\ 
NGC 1559 & 1  & 2023-06-30 & \emph{F090W}/\emph{F277W} & 418.7 x 4 & " \\
NGC 1559 & 1  & 2023-06-30 & \emph{F150W}/\emph{F277W} & 526.1 x 4 & "  \\
NGC 1559 & 2  & 2023-07-15 & \emph{F090W}/\emph{F277W} & 418.7 x 4 & "  \\
NGC 1559 & 2  & 2023-07-15 & \emph{F150W}/\emph{F277W} & 526.1 x 4 & "  \\
NGC 4258 & 5  & 2023-05-02 & \emph{F090W}/\emph{F277W} & 257.7 x 4 & 1147 \\
NGC 4258 & 5  &	2023-05-02 & \emph{F150W}/\emph{F277W} & 365.1 x 4 & " \\
NGC 4258 & 6  & 2023-05-17 & \emph{F090W}/\emph{F277W} & 257.7 x 4 & " \\
NGC 4258 & 6  & 2023-05-17 & \emph{F150W}/\emph{F277W} & 365.1 x 4 & " \\
NGC 5584 & 9  & 2023-01-30 & \emph{F090W}/\emph{F277W} & 418.7 x 4 & 1132 \\
NGC 5584 & 9  & 2023-01-30 & \emph{F150W}/\emph{F277W} & 526.1 x 4 & "  \\
NGC 5584 & 10 & 2023-02-21 & \emph{F090W}/\emph{F277W} & 418.7 x 4 & "  \\
NGC 5584 & 10 & 2023-02-21 & \emph{F150W}/\emph{F277W} & 526.1 x 4 & "  \\
NGC 5643 & 11 & 2023-07-07 & \emph{F090W}/\emph{F277W} & 311.4 x 4 & 1146 \\
NGC 5643 & 11 & 2023-07-07 & \emph{F150W}/\emph{F277W} & 418.7 x 4 & " \\
NGC 5643 & 12 &	2023-07-22 & \emph{F090W}/\emph{F277W} & 311.4 x 4 & " \\
NGC 5643 & 12 & 2023-07-22 & \emph{F150W}/\emph{F277W} & 418.7 x 4 & " \\
\enddata
\caption{Summary table for the $JWST$ observations from GO-1685 used in this study. Columns from left to right are: galaxy name, observation number, observation date, filters, and exposure time, and context file number.}
\end{deluxetable*}

\begin{figure*}[ht!]
\epsscale{1}
\plotone{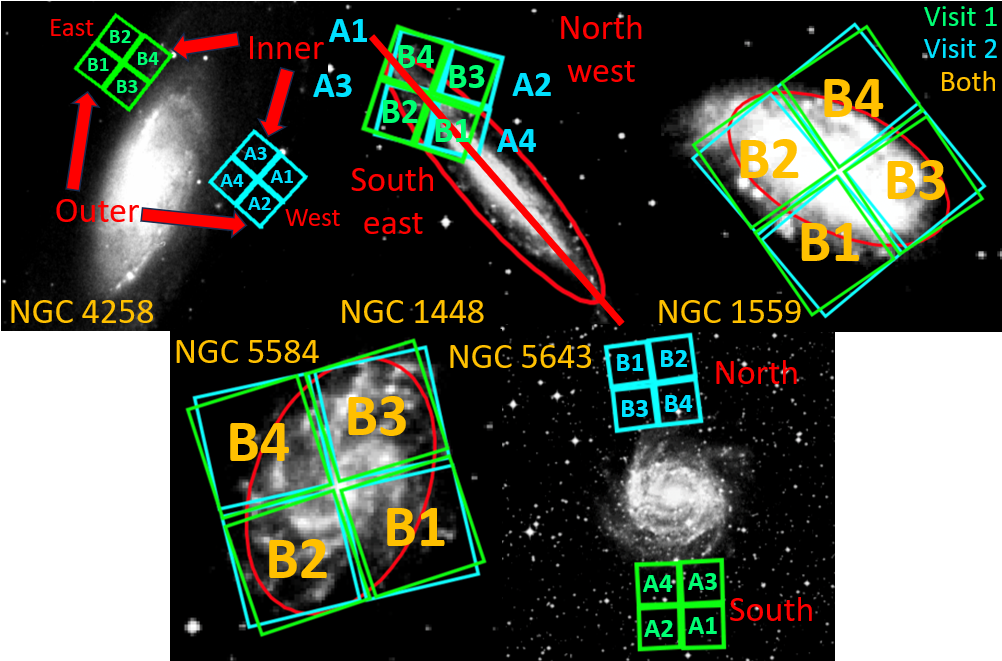}
\caption {Footprints for the NGC 4258, NGC 1448, NGC 1559, NGC 5584, and NGC 5643 fields analyzed in this study. Fields corresponding to the first and second visits are outlined and labelled in green and cyan, respectively. Yellow chip labels correspond to both visits 1 and 2. JAGBs are measured using stars that fall outside the red ellipses having parameters listed in Table \ref{tab:Disk_Field_Parameters}. We also show with red text names of the fields we use to perform symmetry null tests in Section \ref{sec:Symmetry_Null_Tests_Outer_Fields}. The red line shown in the image of NGC 1448 is the line used to separate the Northwest and Southeast fields for Section \ref{sec:Symmetry_Null_Tests_N1448_N5643}. Galaxy images are from the ESO Digitized Sky Survey.}
\label{fig:Footprints}
\end{figure*}

\begin{deluxetable*}{cccccc}
\tablecaption{}
\label{tab:Foreground_Extinction}
\tablehead{\colhead{Galaxy} & \colhead{RA} & \colhead{Dec} & \colhead{E(B-V)}   & \colhead{$A_{F150W}$} & \colhead{$A_{F277W}$}} 
\startdata
NGC 4258 & 184.73958 & 47.30397 & 0.016  & 0.010 & 0.004 \\
NGC 1448 & 56.13300 & $-$44.64483 & 0.014 & 0.009 & 0.004 \\
NGC 1559 & 64.39904 & $-$62.78367 & 0.030 & 0.018 & 0.008 \\
NGC 5584 & 215.59904 & $-$0.38767 & 0.039 & 0.024 & 0.010 \\
NGC 5643 & 218.16975 & $-$44.17442 & 0.169  & 0.102 & 0.043
\enddata
\caption{Summary table for the foreground extinction corrections applied for all JAGB and TRGB measurements in this paper. E(B$-$V) values were retrieved using the \texttt{dustmaps} package in Python \citep{Green_2018JOSS....3..695M, Green_2019ApJ...887...93G}, and galaxy coordinates were retrieved from NED.}
\end{deluxetable*}

\subsection{DOLPHOT Photometry} \label{sec:photometry}

We measure PSF photometry using the $JWST$ NIRCam version of DOLPHOT \citep{Dolphin_2000PASP..112.1383D, Dolphin_2016ascl.soft08013D} in the Vega-Vega system. For this analysis, we used the beta version of the NIRCam module of DOLPHOT released on April 6, 2023. The most recent update as of writing is from December 2, 2023; this includes two changes since the April 6, 2023 version that we do not expect to significantly affect our results. The first change was the inclusion of the `Sirius-Vega' zero-point. In this paper, we choose to remain with the original `Vega-Vega' zero-point. The second change incorporates the ``$-$etctime" option, which allows the user to adjust the exposure times provided in the image headers. The choice of exposure times do not affect the photometry, although it may affect the photometric errors estimated with Poisson statistics. The photometric errors for our observations are in general too small to have a significant impact on the JAGB luminosity function, which has a typical width of 0.2~mag or more. See also \cite{Anand_2024arXiv240104776A} for a discussion of these features. In addition, any future updates will likely minimally affect the results presented here because our measurements are all relative (i.e., between SN Ia hosts and NGC 4258 or within a host).

We use stage 3 \emph{F150W} *i2d.fits frames as reference frames, which are created from combining multiple visit images, and perform photometry on stage 2 *cal.fits frames, which are the single frames corresponding to each dither. We use the DOLPHOT parameters recommended in the NIRCam DOLPHOT manual\footnote{\href{http://americano.dolphinsim.com/dolphot/dolphotNIRCam.pdf}{http://americano.dolphinsim.com/dolphot/dolphotNIRCam.pdf}} and apply the same quality cuts to the \emph{F090W} and \emph{F150W} sample as used by \cite{Riess_2023ApJ...956L..18R}, which were modified from \cite{Warfield_2023RNAAS...7...23W}: 1) Crowding $\leq$ 0.5, 2) Sharpness$^2 \leq$ 0.01, 3) Object Type $\leq$ 2, 4) S/N $\geq$ 3, 5) Error Flag $\leq$ 2.  As described in \cite{Riess_2023ApJ...956L..18R}, there is currently an issue in the $JWST$ NIRCam DOLPHOT source detections when long wavelength images (\emph{F277W}) are combined with with short wavelength images (\emph{F090W}, \emph{F150W}) when performing photometry. To circumvent this issue, we perform photometry using the ``warmstart" option in DOLPHOT and perform a first round of photometry on the short wavelength images (\emph{F090W} and \emph{F150W}), then use the resulting source list to find stars in a second round of photometry that includes all three filters. For our final measurements, we use the \emph{F150W} photometry from the first run and the \emph{F277W} from the second warmstart run (see \citealt{Riess_2023ApJ...956L..18R} for a detailed explanation of this setup).

\subsection{Crowding Corrections} \label{sec:Crowding_Corrections}

Because of the moderate stellar density present in the outer disks where we measured the JAGB, we estimated the impact of crowding on photometry by measuring ``crowding corrections''.
We derive and apply crowding corrections for all stars located inside the outer disk fields listed in Table \ref{tab:Disk_Fields} using DOLPHOT artificial star measurements. To determine these crowding corrections, we divide each image in these disk fields into a 10x10 grid. Next, we place 200 artificial stars in each grid cell at three magnitudes spaced 1~mag apart, with the center magnitude at the approximate magnitude of the JAGB determined after iteration (for instance, for NGC 1448 stars are placed at 23.35, 24.35, and 25.35~mag in \emph{F150W}); the central magnitudes for \emph{F150W} and \emph{F277W} are listed in Table \ref{tab:Disk_Fields}. This results in a total of 600 stars per grid cell. We then calculate the mean difference between the input and output magnitudes and fit this bias as a function of input magnitude using an unweighted linear least squares fit. For each star, we use this fit to solve for the `true' magnitude before bias, that is,
\begin{equation}
    m_{true} = (m_{obs} + b) / (1 - m)
\end{equation}
\noindent where $m_{true}$ is the magnitude of the JAGB stars after crowding correction, $m_{obs}$ is the observed magnitude of the same star, $m$ is the slope of the fit between the crowding bias and magnitude, and $b$ is the intercept of that same fit. We use the Rockfish cluster at Advanced Research Computing at Hopkins (ARCH) together with GNU parallel \citep{Tange_2011a} to significantly reduce the time needed to run these artificial star measurements. In addition to determining the crowding bias, we also use the artificial star test results for NGC 5584, the furthest of the galaxies analyzed here, to test for possible incompleteness of the photometry. At the faintest test point of 1~mag fainter than 24.9~mag (25.9~mag) in \emph{F150W} for NGC 5584 we recover approximately 97\% of artificial stars in the region used to measure the JAGB. We provide the photometry used to measure the JAGB after crowding correction and crossmatching in the Github repository (upon publication), \url{https://github.com/siyangliastro/JWST_JAGB_GO1685}, and Zenodo repository, \dataset[DOI]{10.5281/zenodo.10666761}{} 10.5281/zenodo.10666761, and the CMDs using this photometry in Fig. \ref{fig:Host_Galaxy_CMDs}.

\begin{deluxetable*}{cccccccc}
\tablecaption{}
\label{tab:Disk_Fields}
\tablehead{\colhead{Galaxy} & \colhead{Visit} &\colhead{Module} & \colhead{Input \emph{F150W}} & \colhead{Input \emph{F277W}}  & \colhead{Mean Crowd F150W} & \colhead{Mean Crowd F277W}}
\startdata
NGC 4258 & 1, 2 & B, A  & 22.5 & 21.3 & 0.04 & 0.04\\
NGC 1448 & 1, 2 & B, A & 24.35 & 23.05& 0.02 & 0.01\\
NGC 1559 & 1, 2 & B  & 24.4 & 23.2& 0.09 & 0.06\\
NGC 5584 & 1, 2 & B  & 24.9 & 23.7& 0.06 & 0.04\\
NGC 5643 & 1, 2 & A, B  & 23.55 & 22.35  & 0.04 & 0.04\\
\enddata
\caption{Summary table for the outer disk fields and center magnitudes in \emph{F150W} and \emph{F277W} used for the artificial star tests to derive crowding corrections determined with DOLPHOT artificial star tests for J-region stars, and mean crowding correction for the JAGB stars used in this study. The footprints for these fields can be seen in Fig.~\ref{fig:Footprints}.}
\end{deluxetable*}

\medskip

\subsection{Selection of a JAGB Reference Magnitude} \label{sec:Measurement_Basics}

Perhaps the most challenging issue, based on the study of past efforts, is to define the JAGB standard candle, i.e., the procedure by which we assign a {\it reference magnitude} and its uncertainty to represent the apparent clumping of stars in the JAGB regions of the NIR color-magnitude diagram. We start with the example of the CMDs in NGC 4258, an important reference galaxy for distance measurements, to illustrate possible procedures.

\begin{figure*}[ht!]
  \centering
  \begin{tabular}{cc}
    \includegraphics[width=0.45\linewidth]{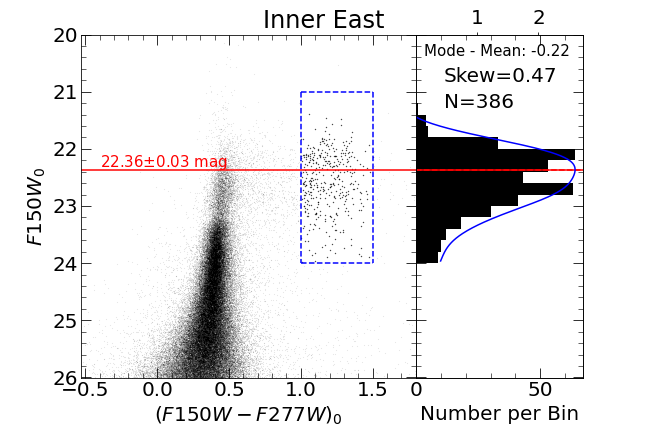} &
    \includegraphics[width=0.45\linewidth]{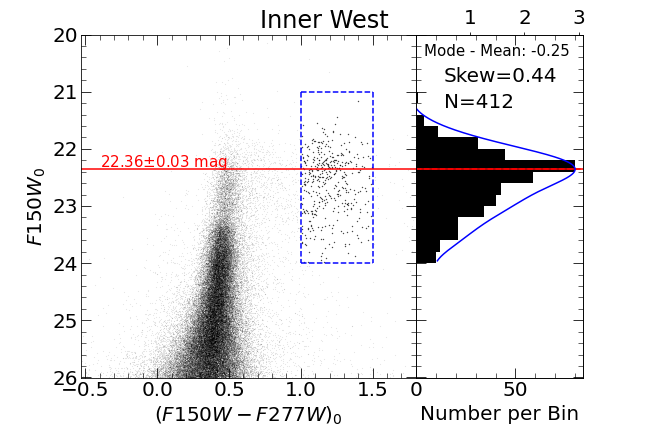} \\
    \includegraphics[width=0.45\linewidth]{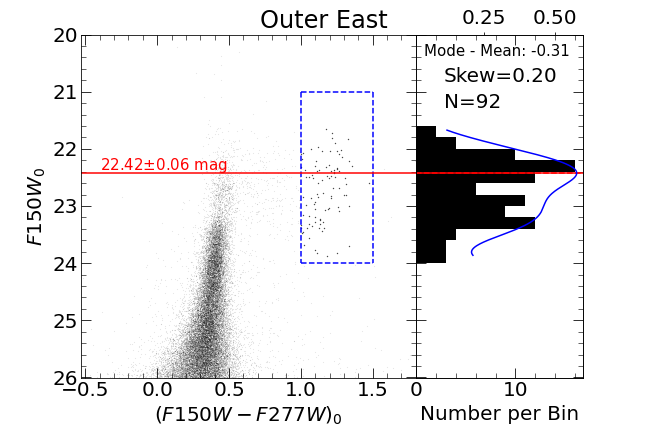} &
    \includegraphics[width=0.45\linewidth]{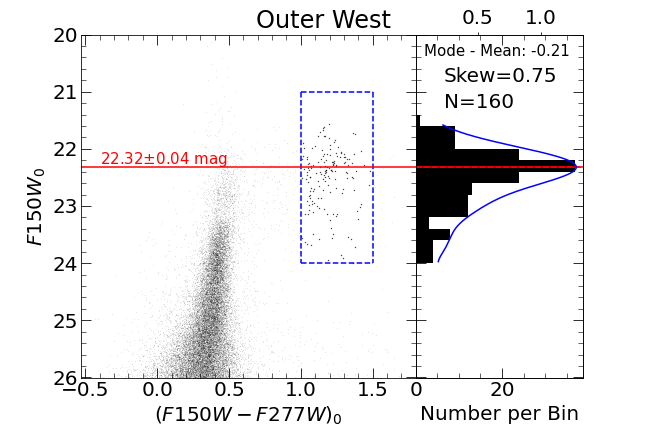} \\
    \includegraphics[width=0.45\linewidth]{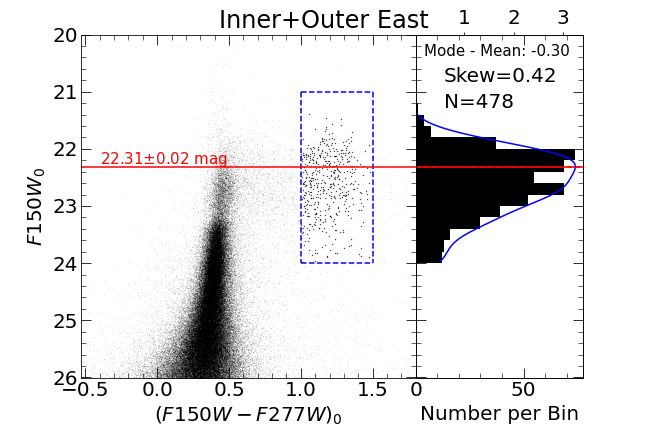} &
    \includegraphics[width=0.45\linewidth]{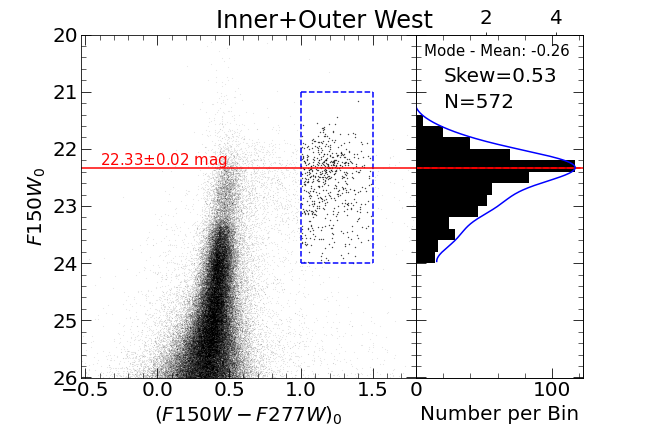} \\
  \end{tabular}
\caption{CMDs and J-region luminosity functions in \emph{F150W} vs. $F150W - F277W$ for the East Inner (B3, B4) and West Inner (A3, A4), East Outer (B1, B2) and West Outer (A1, A2), and East and West combined Inner and Outer fields in NGC 4258. The blue dashed box surrounds the J-region used for the JAGB measurement, and the points inside the blue dashed box are enlarged for emphasis. J-region luminosity functions on the right panels of each subplot are binned with bin widths of 0.2~mag (black). To provide an example JAGB measurement, we use the mode based method from \cite{Lee_2022ApJ...933..201L, Lee_2023ApJ...956...15L, Lee_2024ApJ...961..132L}, smooth the luminosity function binned in 0.01~mag widths and apply GLOESS with a smoothing parameter of $s$=0.25~mag. The blue line corresponds to the smoothed luminosity function, and the red line and corresponding red label marks the location of the JAGB measured using this method. The tick marks below and above the luminosity function correspond to the 0.2~mag binned and 0.01~mag binned then smoothed luminosity functions, respectively.}
\label{fig:Intro_Example_CMDs}
\end{figure*}

In the top row of Fig. \ref{fig:Intro_Example_CMDs}, we show CMDs for NGC 4258 containing the J-region in \emph{F150W} vs. $F150W - F277W$ (see Fig. \ref{fig:Footprints}) on the West (chips A3+A4) and East (chips B3+B4) sides of the galaxy. We will refer to these two regions as the `Inner' regions, as they appear to be on the inner disk and they are inside the ``Outer'' chips, A1, A2, B1, B2, at or near the halo.  JAGB studies in the literature typically apply color and magnitude cuts to isolate the enhanced density of stars in the J-region (see, for instance, \citealt{Zgirski_2021ApJ...916...19Z}); we approximate past efforts to put a ``box'' around the JAGB clump by defining a color cut of $F150W-F277W$ of 1 to 1.5~mag, and for NGC 4258 we use magnitude cuts of \emph{F150W} = 21 to 24~mag (spanning about $\pm$ 5 standard deviations, i.e., a broad frame).  This is the region marked as a blue dashed box in Fig. \ref{fig:Intro_Example_CMDs}. We enlarge the points in the blue dashed box for emphasis. A binned luminosity function corresponding to stars inside the box can be seen in the right hand side of each subplot. 

In the literature, the reference magnitude has been measured from this luminosity function in a variety of ways, such as from its mode, median, mean, or from a model, in some cases with additional, unstated specifications which thus make it difficult for us employ here.   For example \cite{Madore_2020ApJ...899...66M} and \cite{Freedman_2020ApJ...899...67F} use a mean but do not specify its range in magnitude or color {\it a priori};  rather the region of the CMD to average is chosen on a case-by-case basis from inspection of each CMD (based on the variety of selection boxes presented in their CMDs).  As a result, the width in magnitude varies from 0.8 to 2.0 mag, and the red $J-K$ color boundary from $\sim$ 1.8 to 2.2.
   
To illustrate a specific prescription, we can use the mode-based method from \cite{Lee_2024ApJ...961..132L}, which bins the luminosity function in 0.01~mag widths, applies Gaussian-windowed, locally weighted scatterplot smoothing \citep[GLOESS;][]{Loader_RePEc:zbw:caseps:200412, Persson_2004AJ....128.2239P} to the binned luminosity function with a smoothing scale of 0.25~mag, and uses the \textit{mode} of the smoothed luminosity function to define the reference magnitude of the JAGB. The uncertainty is set to the standard deviation of the LF divided by the square root of the number of stars in contains. We show the smoothed luminosity function for these measurements in blue and the corresponding mode-based JAGB measurement with the red line and label in Fig. \ref{fig:Intro_Example_CMDs}. 

Unfortunately, the luminosity functions we observed for the JAGB in NGC 4258 are quite {\it asymmetric} (skew $\sim$ 0.4 for the Inner fields) which makes the selection of the measure of the reference magnitude {\it non-trivial} because its value will differ substantially depending on the choice of method, and this choice will not cancel when we compare to other LFs if their shapes are not the same.  The asymmetry seen here is not a fluke or the result of incompleteness or extinction; in fact, a varying degree of one-sided asymmetry appears to be {\it intrinsic} to JAGB luminosity functions.  As previously noted, asymmetric examples for nearby galaxies include the LMC and NGC 6822 (skew $\sim$ $-$0.4 to $-$0.5), while symmetric examples include the SMC and IC 1613 \citep[skew $\sim 0.0$,][]{Parada_2021MNRAS.501..933P}.
   
Depending on the level of LF asymmetry, we note that even the measurement of its mode will vary depending on the smoothing scale used as illustrated in the Appendix \ref{Apdx: Effects of Smoothing on JAGB Luminosity Functions}, giving us an additional choice to consider. It is also far from obvious which measure of the LF will yield the {\it better standard candle}; this depends on whether the breadth of the LF contains multiple populations, the most common of which may be the standard candle, or whether it results from the spreading of a population whose mean is the true candle. Literature studies generally choose one method but none have provided a systematic study demonstrating the superiority of one or another\footnote{Although as previously noted, the mean is preferred over the mode for comparing JAGB and the geometric distances for the LMC and SMC which we review in the Discussion}. Therefore we will explore several different measures and retain a ``many methods" approach to comparing these.

We use the following four measure types of the JAGB LF: median, 3-sigma clipped mean, mode, and Gaussian$+$Quadratic model fit as used (or modified) from recent JAGB literature \citep{Madore_2020ApJ...899...66M, Freedman_2020ApJ...899...67F, Ripoche_2020MNRAS.495.2858R, Zgirski_2021ApJ...916...19Z, Parada_2021MNRAS.501..933P, Lee_MW_2021ApJ...923..157L,  Lee_2022ApJ...933..201L, Lee_WLM_2021ApJ...907..112L, Parada_2023MNRAS.522..195P, Lee_2023ApJ...956...15L}. We use 3$\sigma$ clipping to the mean to reduce sensitivity to outliers and box boundaries, features already inherent with the median and mode.  In addition, we use the simple median instead of the modified median combined with skew from \cite{Ripoche_2020MNRAS.495.2858R, Parada_2021MNRAS.501..933P, Parada_2023MNRAS.522..195P}. We will also vary the color range used to select the J-region stars in five different ways.  In all we will use 25 combinations (5 measures, 5 color ranges) as listed in Table \ref{tab:Variants} to provide a comprehensive view.  

Ideally we seek a standard candle which is least sensitive to variations in underlying, intrinsic properties we cannot (yet) easily measure, such as age and metallicity; therefore  we proceed to assess how the 25 possible combinations of JAGB measures vary in different fields of the same host.

\section{JAGB Reconnaissance: Tests within Hosts} \label{sec:Symmetry_Null_Tests_Outer_Fields}

\begin{figure*}[ht!]
  \centering
  \begin{tabular}{ccc}
    \includegraphics[width=0.3\linewidth]{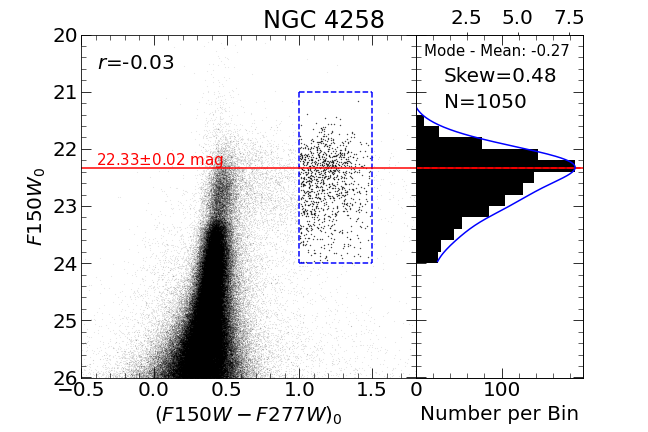} &
    \includegraphics[width=0.3\linewidth]{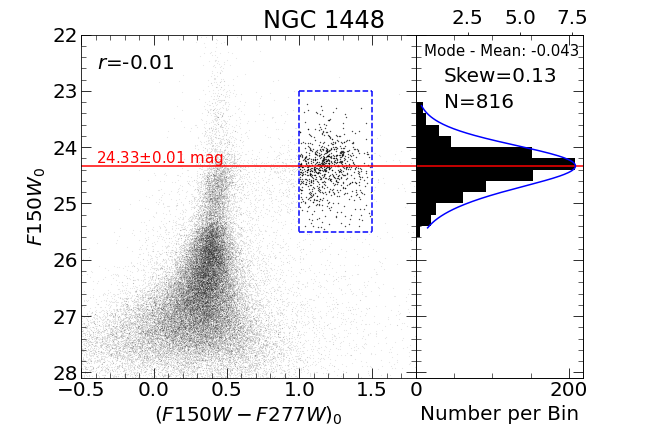} &
    \includegraphics[width=0.3\linewidth]{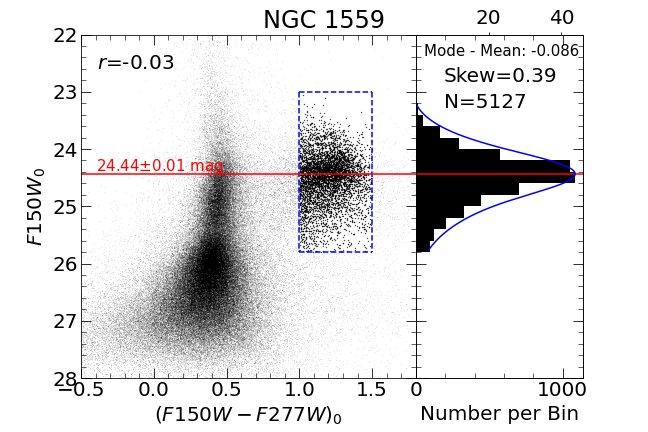} \\
    \includegraphics[width=0.3\linewidth]{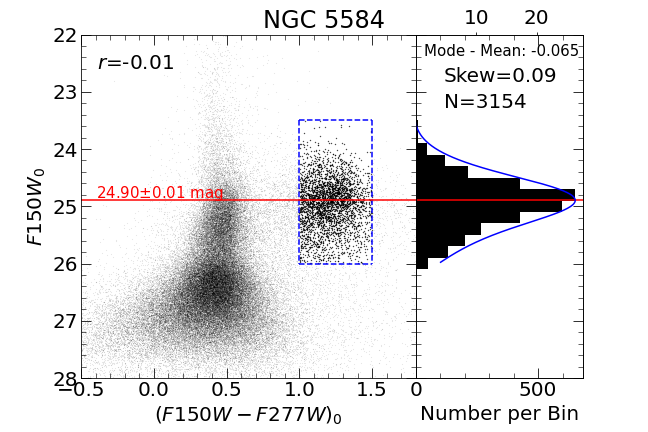} &
    \includegraphics[width=0.3\linewidth]{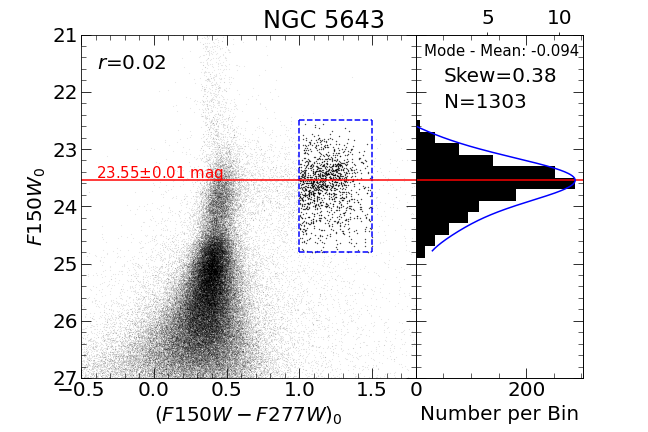} \\
  \end{tabular}
\caption{CMDs and J-region luminosity functions in \emph{F150W} vs. $F150W - F277W$ for NGC 4258, NGC 1448, NGC 1559, NGC 5584, and NGC 5643. The blue, dashed box surrounds the region defined as the J-region and used for the JAGB measurement. Points inside the dashed blue box are enlarged for emphasis. J-region luminosity functions on the right panels of each subplot are binned with bin widths of 0.2~mag (black). To provide an example JAGB measurement, we smooth the luminosity function binned in 0.01~mag widths and apply GLOESS with a smoothing parameter of $s$=0.25~mag. The blue line corresponds to the smoothed luminosity function, and the red line and corresponding red label marks the location of the JAGB measured using this method. The tick marks below and above the luminosity function correspond to the 0.2~mag binned and smoothed luminosity functions, respectively. We show the skew of the J-region luminosity functions as calculated using the Fisher-Pearson coefficient using \texttt{skew} and the Pearson correlation coefficient using \texttt{pearsonr}, both from the Python sub-package \texttt{scipy.stats}, in the top right and left sides of each plot, respectively.}
\label{fig:Host_Galaxy_CMDs}
\end{figure*}

We first define a reference JAGB luminosity function in the {\it outer} disk of each host before proceeding to divide the measures within a host for testing.   For each host, we select stars with the same color cuts described in Section \ref{sec:Measurement} (1 to 1.5~mag) and outside ellipses which contain the main disks of the host (see Figure 3) with parameters and magnitude cuts listed in Table \ref{tab:Disk_Field_Parameters}. We discuss these choices of ellipse parameters and the effect of the measured JAGB with radius in Appendix \ref{Apdx: Stability of the JAGB with Radius}. We plot the CMDs for all host galaxies in Fig. \ref{fig:Host_Galaxy_CMDs} together with the luminosity function in the J-region and its mode-based magnitude; we also list the differences between mode and mean, the skewness, and the number of stars in that region. We can also test for the possibility of a tilt to the J-region in the $F150W$ band using the Pearson correlation coefficient for the stars enclosed in each dashed box. Pearson coefficients of $\pm$ 1 would indicate a linear relationship, while a coefficient of 0 would indicate a lack thereof. In the top left corner of each subplot in Fig. \ref{fig:Host_Galaxy_CMDs}, we show using the Python \texttt{scipy.stats.pearsonr} routine the Pearson correlation coefficient. We find that all five galaxies have Pearson correlation coefficients close to zero. (see discussion in Appendix \ref{Apdx: J-region_Tilts}). 

We can test the stability of the JAGB by comparing fields around each host that are relatively equidistant from their main disks.  The observation orientations for each host (see Figure 3) provide for such a test by comparing the fields 180 degrees apart for NGC 4258, NGC 1448 and NGC 5643.  For NGC 5584 and NGC 1559 we will compare fields around the circumference of each disk.

\begin{deluxetable*}{cccccc}
\tablehead{\colhead{Variant} & \colhead{} & \colhead{}  & \colhead{Range Considered} & \colhead{} & \colhead{}}
\startdata
\multicolumn{1}{c|}{Method}  & \multicolumn{1}{c|}{Median}  & \multicolumn{1}{c|}{3$\sigma$ clipped mean} & \multicolumn{1}{c|}{Mode} & \multicolumn{1}{c|}{Gaussian+Quadratic Model Fit} &\\
\tableline
\multicolumn{1}{c|}{Smoothing}  & \multicolumn{1}{c|}{}  & \multicolumn{1}{c|}{} & \multicolumn{1}{c|}{0.25, 0.35, 0.4} & \multicolumn{1}{c|}{} & \\
\tableline
Color Range & Narrow  & Baseline & Wide & Blue 5\% &  Red 5\%  \\
(All 5 used for & Both sides 5\% narrower && Both sides 5\% wider & Blue side 5\% redder &Red side 5\% bluer \\
each method) & $1.05-1.425$~mag & $1-1.5$~mag & $0.95-1.575$~mag & $1.05-1.5$~mag & $1-1.425$~mag \\
\enddata
\label{tab:Variants}
\caption{Summary table for the measurement and selection variants explored in this analysis. We note that we do not apply smoothing variants when the measurement method corresponds to the median, 3$\sigma$ clipped mean, or Gaussian+Quadratic model fit. The baseline color range corresponds to 1~mag $< F150W - F277W <$ 1.5~mag.}
\end{deluxetable*}

\subsection{NGC 4258} \label{sec:Symmetry_Null_Tests_N4258}

We first compare measured JAGBs in the East vs. West fields on opposite sides of NGC 4258 (see footprints in Fig. \ref{fig:Footprints}). We construct three cases for this comparison using different combinations of chips: Outer (A1, A2 vs. B1, B2), Inner (A3, A4 vs. B3, B4), and Inner+Outer (Full A module vs. Full B module). 

\begin{figure*}[ht!]
  \centering
  \begin{tabular}{ccc}
    \includegraphics[width=0.27\linewidth]
    {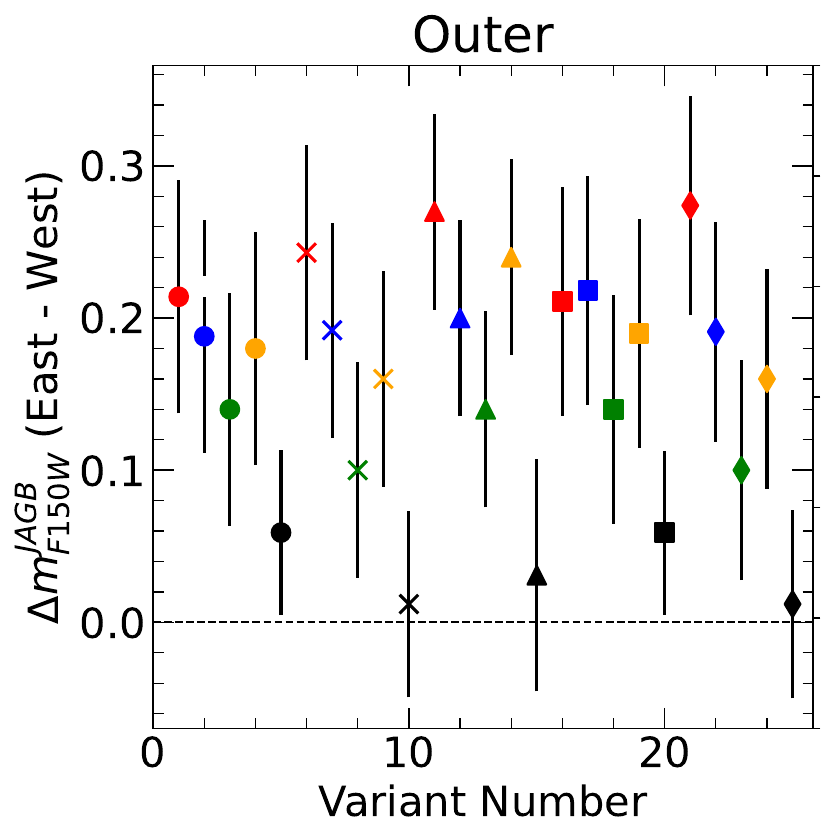} &
    \includegraphics[width=0.27\linewidth]{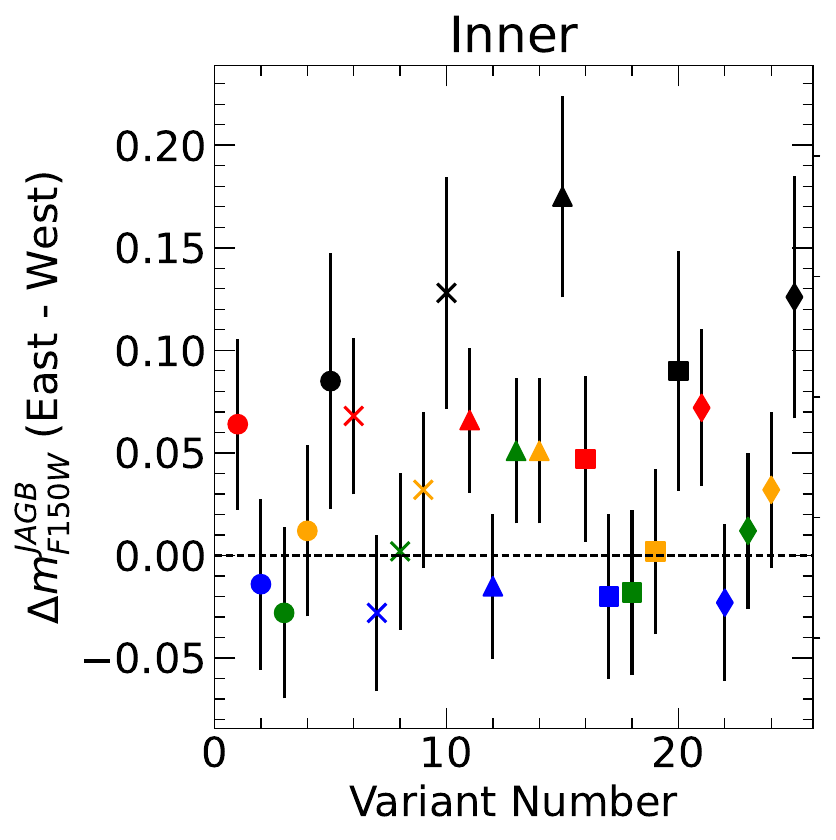} &
    \includegraphics[width=0.41\linewidth]{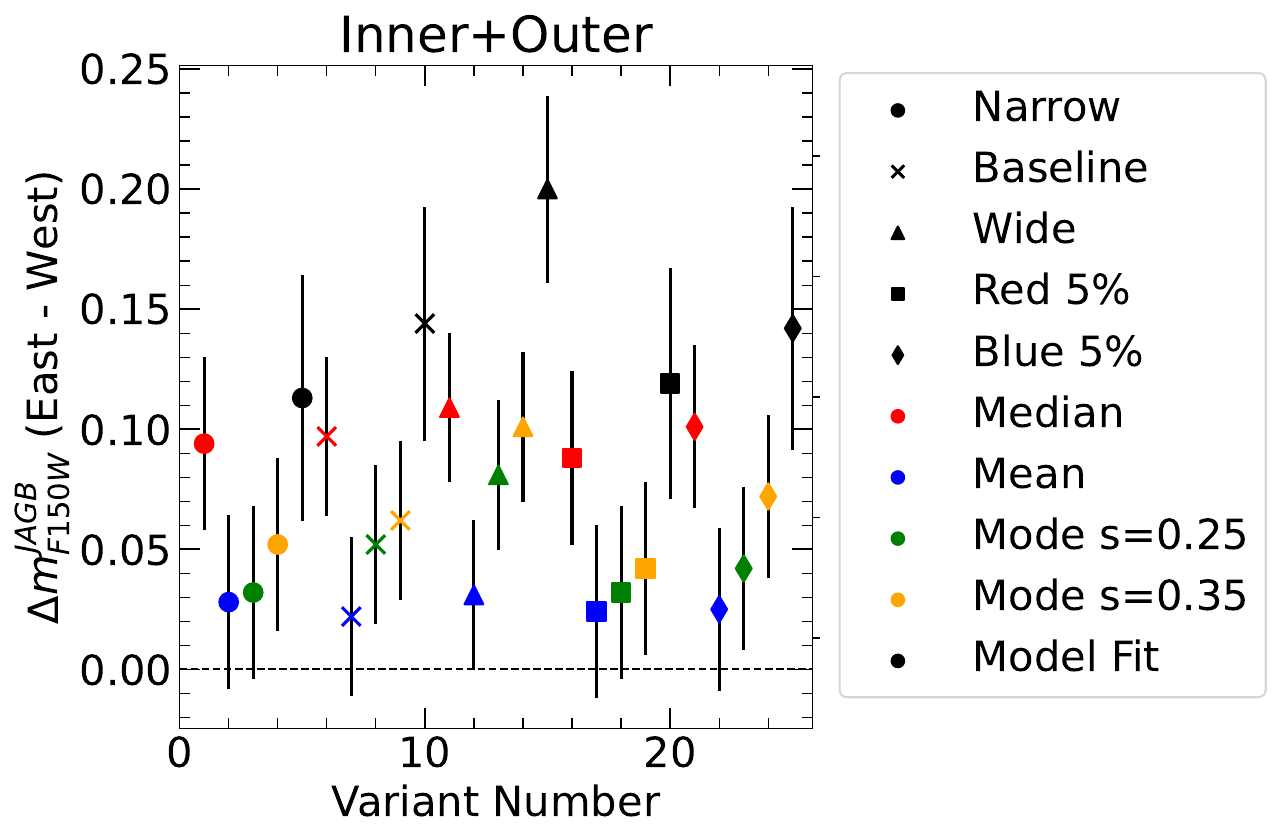} \\
  \end{tabular}
\caption{Differences between the measured JAGBs (East $-$ West) plotted against their measurement variants for the NGC 4258 Outer (A1, A2 vs. B1, B2), Inner (A3, A4 vs. B3, B3), and Inner+Outer (Full A modules vs. Full B module) cases. Different shapes correspond to different color ranges, while different colors correspond to different measurement methods. We add a dashed black line at $\Delta m^{JAGB}_{F150W}$ = 0 for reference.}
\label{fig:Null_Test_Outer_Fields_Results}
\end{figure*}

The CMDs for these cases can be seen in Fig. \ref{fig:Intro_Example_CMDs}. From these figures, we can immediately see an asymmetry (skew $\sim$ 0.4, similar to the LMC) in the J-region luminosity functions at a level which produces a difference of $\sim$ 0.3 mag   between the mode and mean, as shown in the upper right hand corners of each subplot.  We also plot the differences between the measured JAGBs in the East and West fields for all three cases, together with measurement and selection variants, in Fig. \ref{fig:Null_Test_Outer_Fields_Results}. For the Outer fields {\it model} fits only, the sparsity of stars makes it difficult to achieve convergence and so we bin the luminosity function with widths of 0.2~mag instead of widths of 0.1~mag. From Fig. \ref{fig:Null_Test_Outer_Fields_Results}, we find that the Outer fields show variations between the sides of NGC 4258 on the order of $\sim$0.1-0.2~mag for all methods; these are larger than their statistical uncertainties, indicating that the differences are intrinsic. The Inner fields show better agreement, although agreement to within their statistical uncertainties depends on measurement and selection choices. It was noted by \cite{Lee_2024ApJ...961..132L} that the JAGB magnitude may converge at greater radii from the host, but this may run counter to the sense here of the better agreement for the Inner fields.  In the interest of greater field-to-field averaging, we will use the combined Inner+Outer fields from both sides to calibrate other host JAGB measurements (see Section \ref{sec:JAGB_Distances} for more discussion). 

\subsection{NGC 1448 \& 5643} \label{sec:Symmetry_Null_Tests_N1448_N5643}

\begin{figure*}[ht!]
  \centering
  \begin{tabular}{ccc}
    \includegraphics[width=0.45\linewidth]{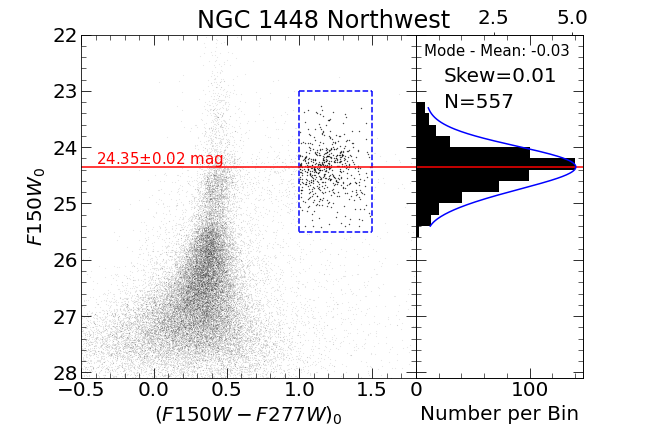} &
    \includegraphics[width=0.45\linewidth]{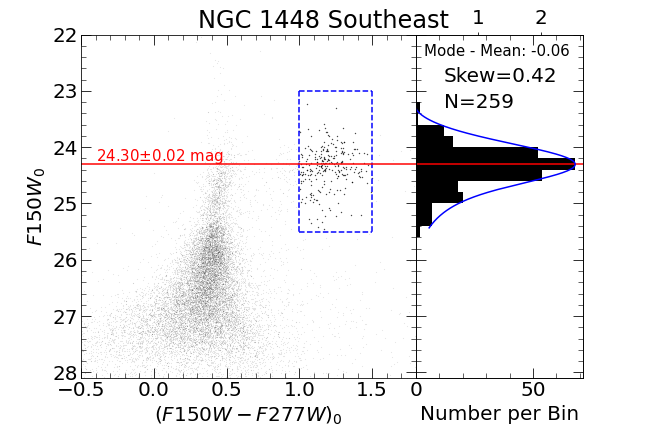} \\
    \includegraphics[width=0.45\linewidth]{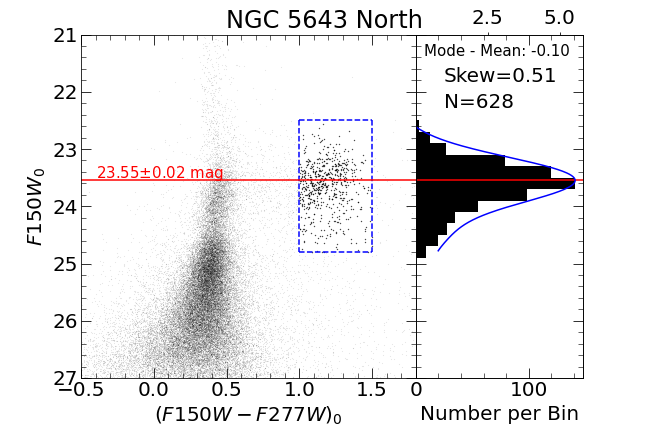} &
    \includegraphics[width=0.45\linewidth]{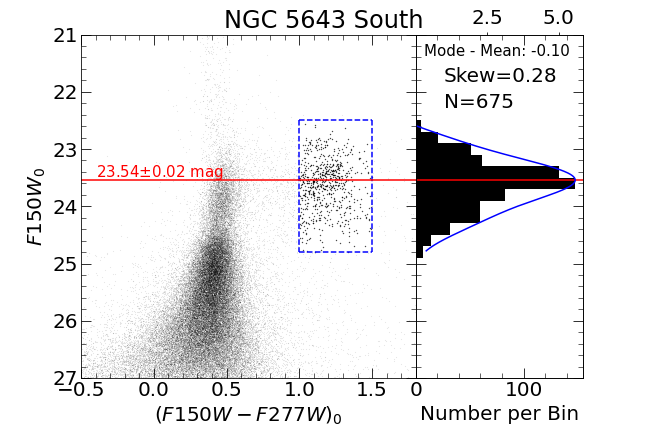} \\
  \end{tabular}
\caption{\emph{F150W} vs. $F150W - F277W$ CMDs and \emph{F150W} J-region luminosity functions for the fields used for the  NGC 1448 and 5643 symmetry null tests. The blue, dashed box surrounds the region defined as the J-region and used for the JAGB measurements. Points inside the blue dashed box are enlarged for emphasis. J-region luminosity functions on the right panels of each subplot are binned with bin widths of 0.2~mag (black). To provide an example JAGB measurement, we smooth the luminosity function binned in 0.01 mag widths and apply GLOESS with a smoothing parameter of $s$=0.25 mag. The blue curve corresponds to the smoothed luminosity function, and the red line and corresponding red label marks the location of the JAGB measured the mode of the smoothed luminosity function. The tick marks below and above the luminosity function correspond to the 0.2 mag binned and 0.01 mag binned then smoothed luminosity functions, respectively}
\label{fig:Null_Test_CMDs_Outer_Fields_N1448_5643}
\end{figure*}

Next, we repeat the symmetry null test described in Section \ref{sec:Symmetry_Null_Tests_N4258} for NGC 1448 and NGC 5643. For NGC 1448, we combine photometry from visit 1, module B and Visit 2, module A, and average the magnitudes for overlapping stars.  Following the recommendation of using only stars in the outer disk for JAGB measurements from \cite{Lee_2022ApJ...933..201L, Lee_2023ApJ...956...15L, Lee_2024ApJ...961..132L}, we select stars in the outer disk that lie outside the ``disk'' ellipse with parameters listed in Table \ref{tab:Disk_Field_Parameters}; we then further split the outer disk region into the Northwest and Southeast regions, as labeled in Fig.~\ref{fig:Footprints}. For NGC 5643, we compare the North and South modules (visit 2 module B vs.~visit 1 module A) as labelled in Fig. \ref{fig:Footprints}.  We plot the CMDs in Fig. \ref{fig:Null_Test_CMDs_Outer_Fields_N1448_5643}; the differences in measured JAGBs between the Northwest and Southeast fields for NGC 1448 and the North and South fields for NGC 5643 are shown in Fig.~\ref{fig:Null_Test_CMDs_Outer_Fields_N1448_5643}. We find that the region differences in the apparent JAGB reference magnitudes between the two fields within NGC 1448 and NGC 4258 are centered at approximately $\sim \Delta m^{JAGB}_{F150W} = 0.04$~mag and $ \sim \Delta m^{JAGB}_{F150W} = 0.05$~mag, respectively, while the JAGB reference magnitudes for NGC 5643 mostly agree to within one sigma and generally $<$ 0.05 mag, suggesting that the consistency of the JAGB is better in some galaxies compared to others.

\begin{figure*}[ht!]
  \centering
  \begin{tabular}{ccc}
    \includegraphics[width=0.4\linewidth]
    {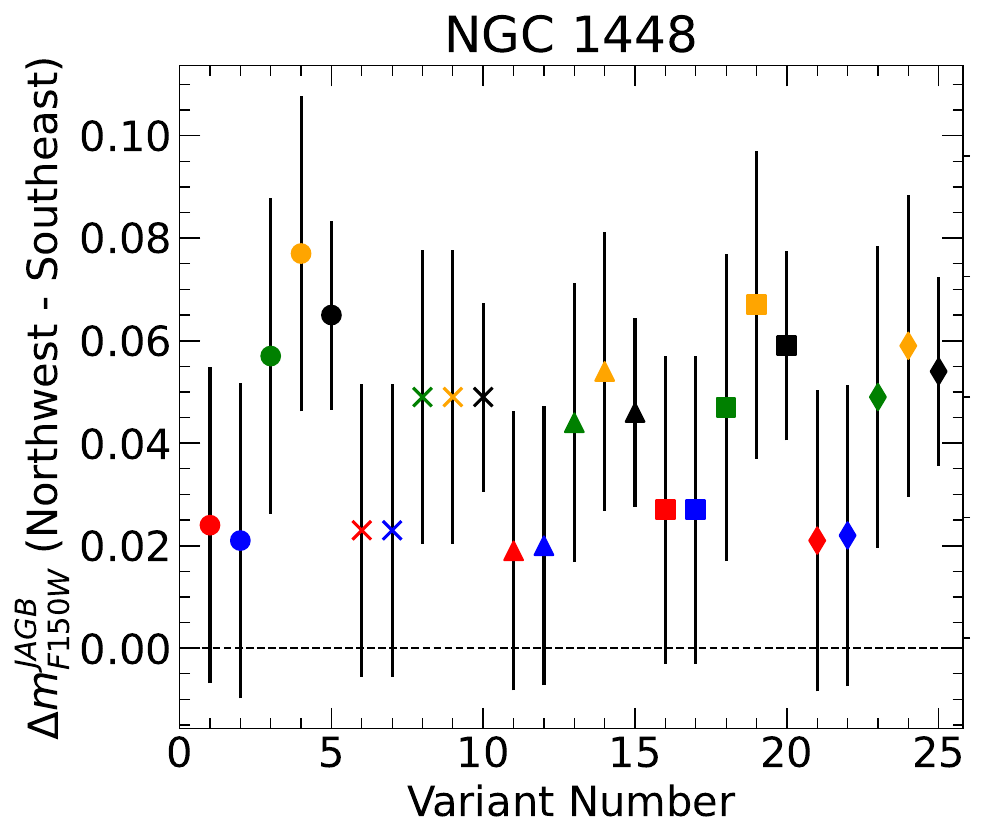} &
    \includegraphics[width=0.58\linewidth]{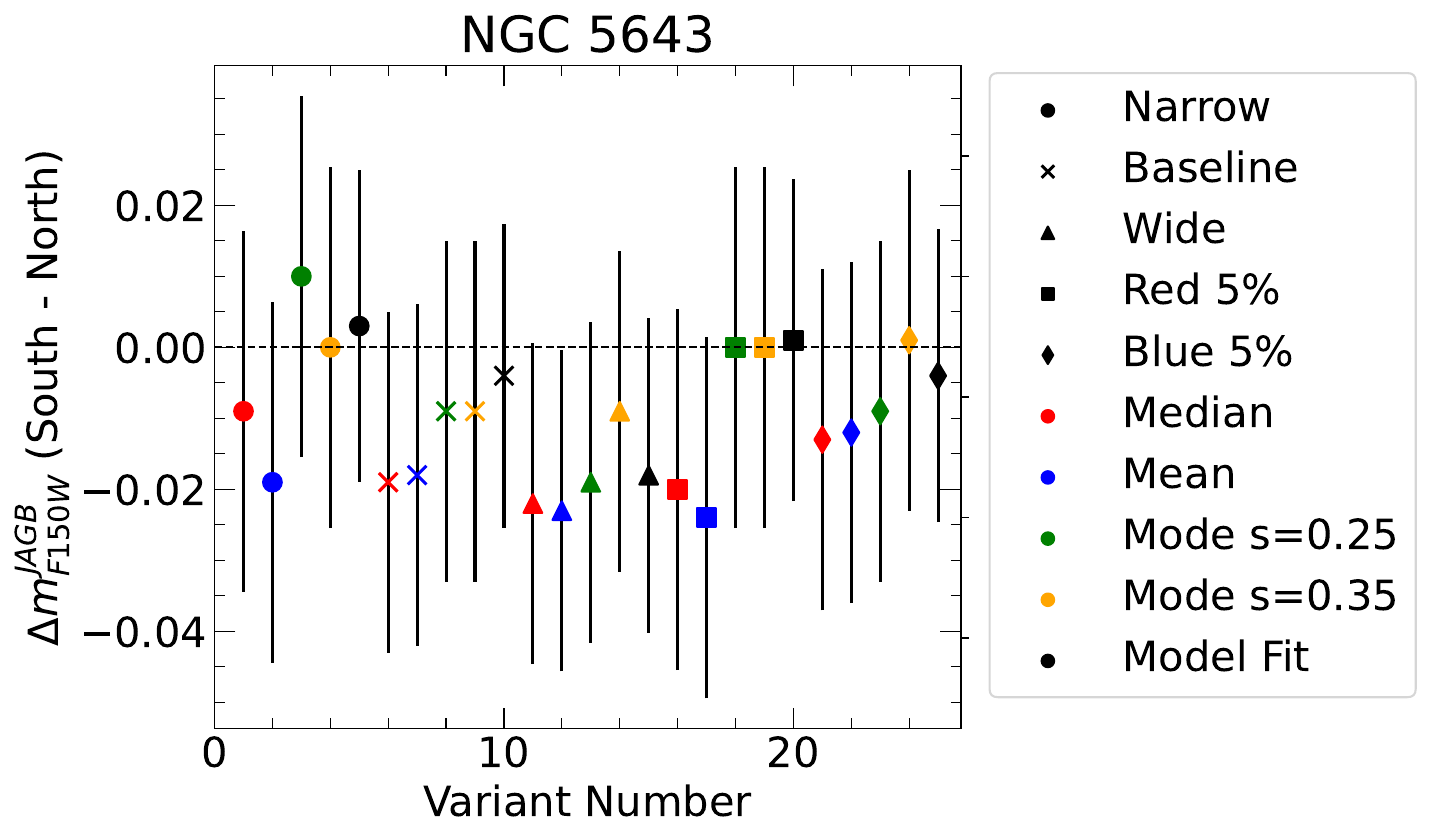}  \end{tabular}
\caption{Differences between the measured JAGBs plotted in the Northwest and Southeast fields for NGC 1448 and the South and North fields for NGC 5643. The differences are plotted against the measurement variants listed in Table \ref{tab:Variants}.}
\label{fig:Null_Test_Outer_Fields_Results_N1448_5643}
\end{figure*}

\subsection{Angular Symmetry Tests} \label{sec:Angular_Symmetry_Tests}

Because the measured JAGBs in Section \ref{sec:Symmetry_Null_Tests_Outer_Fields} did not all pass the test of consistency between 180 degree opposite fields, we decided to further investigate their variation with angle around NGC 1559 and NGC 5584, whose outer disks are well contained within the NIRCam frame. We do not use NGC 1448 or NGC 5643 because they are larger in size such that the GO-1685 observations do not include the perimeters of their outer disks.

\begin{figure*}[ht!]
  \centering
  \begin{tabular}{cccc}
    \includegraphics[width=0.5\linewidth]{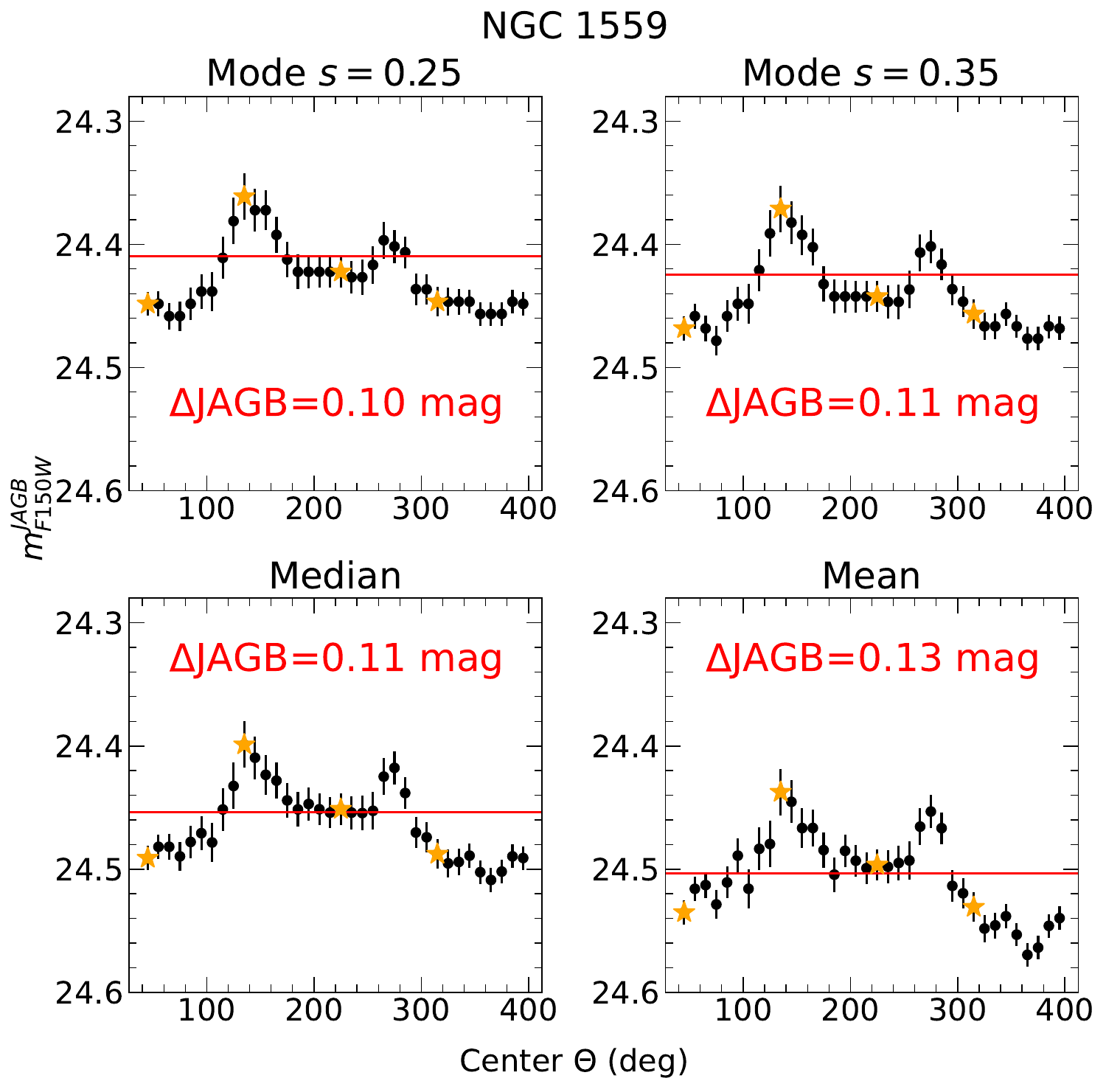} &
    \includegraphics[width=0.5\linewidth]{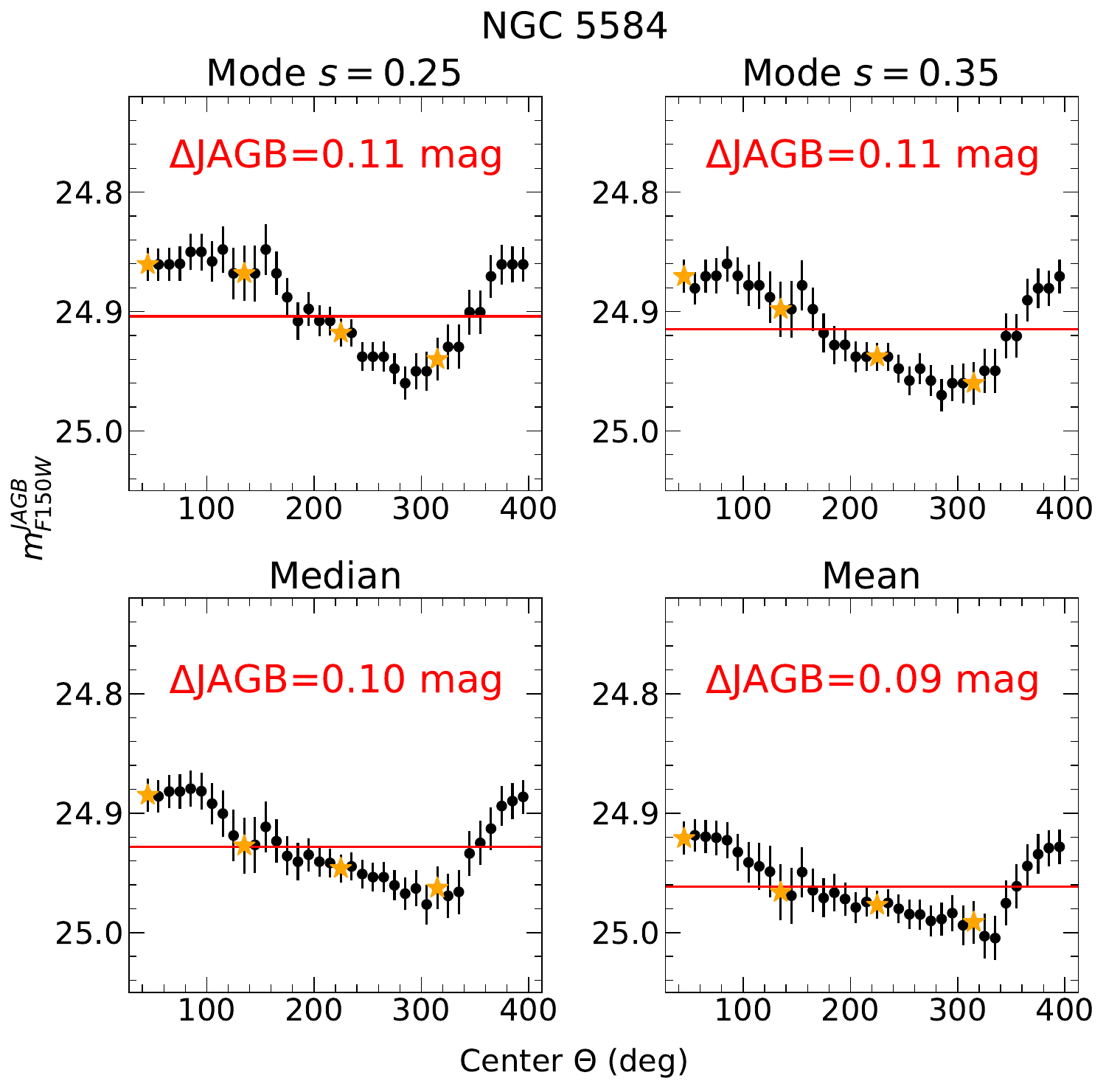} &
  \end{tabular}
\caption{Measured JAGBs in \emph{F150W} as a function of the center angle of 90$^\circ$ windows moved around the galaxy in 10$^\circ$ increments for NGC 1559 and NGC 5584. Only stars in the outer disk were used for these tests. Yellow stars mark windows that do not overlap.  We repeat using four measurement methods (mode with 0.25 and 0.35~mag smoothings, median, and sigma-clipped mean) for each galaxy. Uncertainties are calculated using the error on the mean.}
\label{fig:Angle_Test_Outer_Disk}
\end{figure*}

We first combine the two epochs of observations for NGC 1559 and NGC 5584 and average stars that are observed twice.  We then define a window 90$^\circ$ wide beginning with a position angle of 0$^\circ$, where 0$^\circ$ corresponds to the unit coordinate of (1, 0). We then slide each window by 10$^\circ$ and repeat the JAGB measurement until lower bound of the window edges reaches 350$^\circ$. We repeat this test for each galaxy using four different representative measurement methods (mode with 0.25 and 0.35~mag smoothings, median, and sigma-clipped mean). We do not show the model fit as the mode with $s=0.35$~mag behaves very similarly. In addition, we choose these methods to best approximate the methodological choices from the literature \citep[for instance, from][]{Madore_2020ApJ...899...66M, Ripoche_2020MNRAS.495.2858R, Lee_2024ApJ...961..132L}. We also impose a minimum threshold of 100 stars to minimize fluctuations due to low number statistics and plot the results in Fig. \ref{fig:Angle_Test_Outer_Disk}. Yellow stars mark windows that do not overlap and are uncorrelated. Statistical uncertainties are calculated using the error on the mean. We observe differences of  $\sim$0.1~mag that are not consistent with statistical uncertainties and persist across measurement methods, again suggesting some intrinsic differences in the measured JAGB.  

From the tests described above, we find evidence of spatial variations in the JAGB reference magnitude. These variations depend on the method used to fix the JAGB reference magnitude and may be reduced by observing a larger portion of the outer disks. We discuss these variations in a broader context in Section \ref{sec:Discussion}.

\begin{deluxetable*}{cccccc}
\label{tab:Disk_Field_Parameters}
\tablehead{\colhead{Galaxy} & \colhead{Semi-Major Axis (SMA)} & \colhead{Minor/Major Axis Ratio}& \colhead{PA} & \colhead{$F150W_0$ Cuts}}
\startdata
NGC 1448 &  200$''$ & 0.224 & 41$^{\circ}$ & 23 to 25.5~mag \\
NGC 1559 &  75$''$ & 0.523 & 65$^{\circ}$ & 23 to 25.8~mag \\
NGC 5584 &  75$''$ & 0.64 & 158$^{\circ}$ & 23.5 to 26~mag  \\
NGC 5643  &  75$''$ & 0.871 & 89$^{\circ}$ & 22.5 to 24.8~mag  \\
\enddata
\caption{Summary table for the selections applied to the outer disk JAGB analyses. Ellipses are centered on the galaxy coordinates listed in Table \ref{tab:Foreground_Extinction}.}
\end{deluxetable*} 

\section{Distance Ladder and $H_0$} \label{sec:JAGB_Distances}

\begin{figure*}[ht!]
  \centering
  \begin{tabular}{cc}
    \begin{minipage}[c]{0.35\linewidth}
      \centering
      \includegraphics[width=\linewidth]{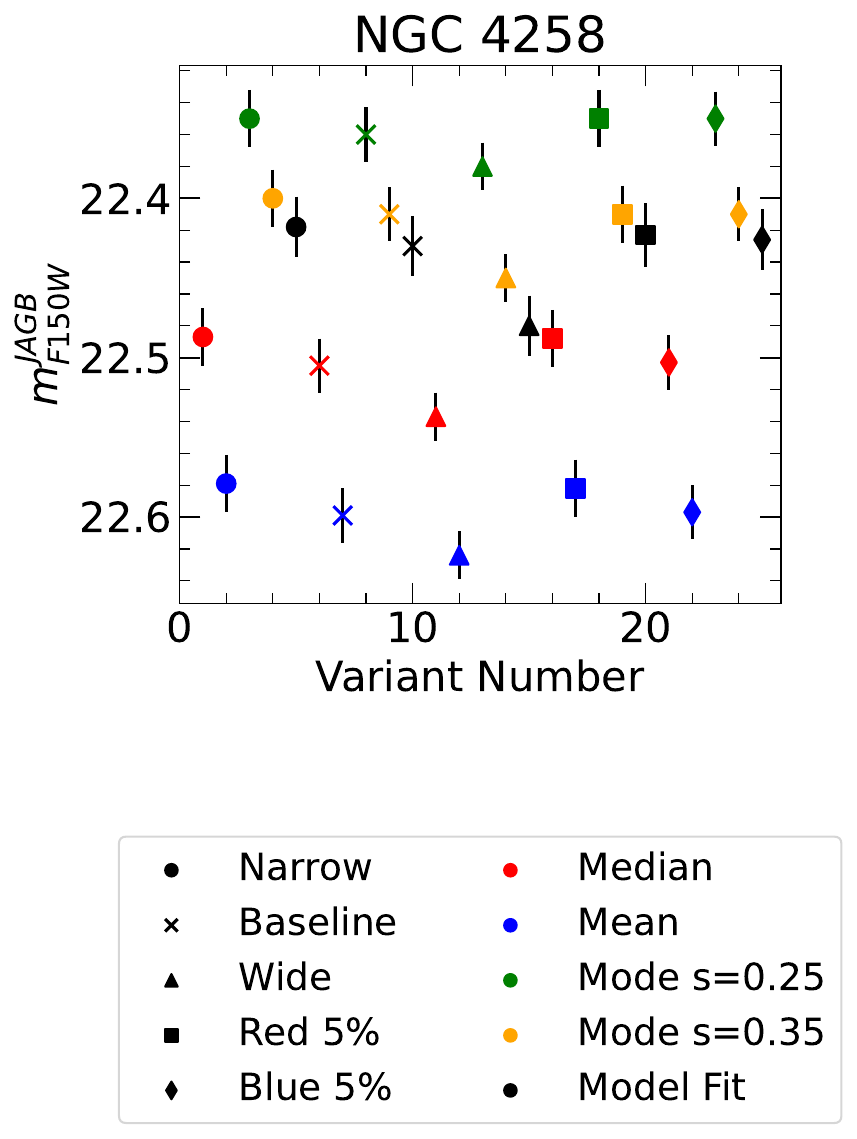} 
    \end{minipage}
    &
  \begin{tabular}{ccc}
    \includegraphics[width=0.3\linewidth]{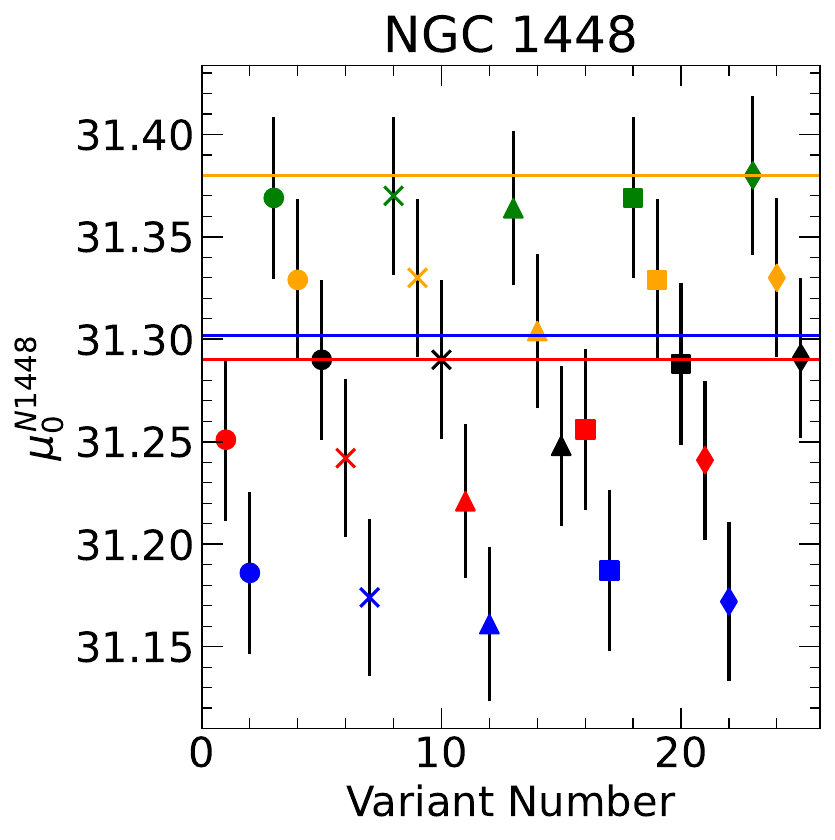} &
    \includegraphics[width=0.29\linewidth]{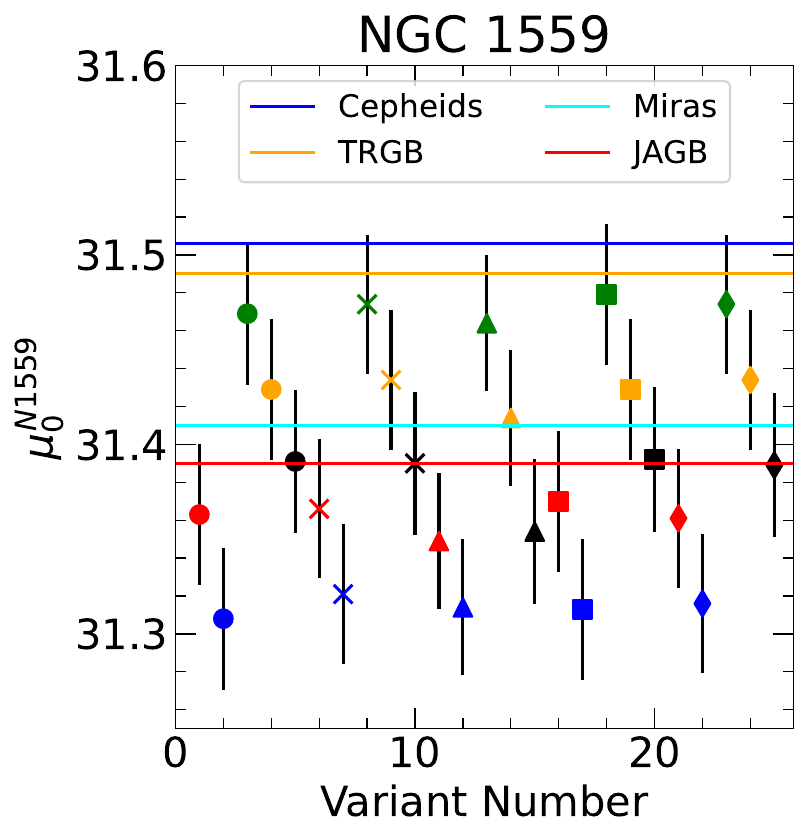} \\
    \includegraphics[width=0.3\linewidth]{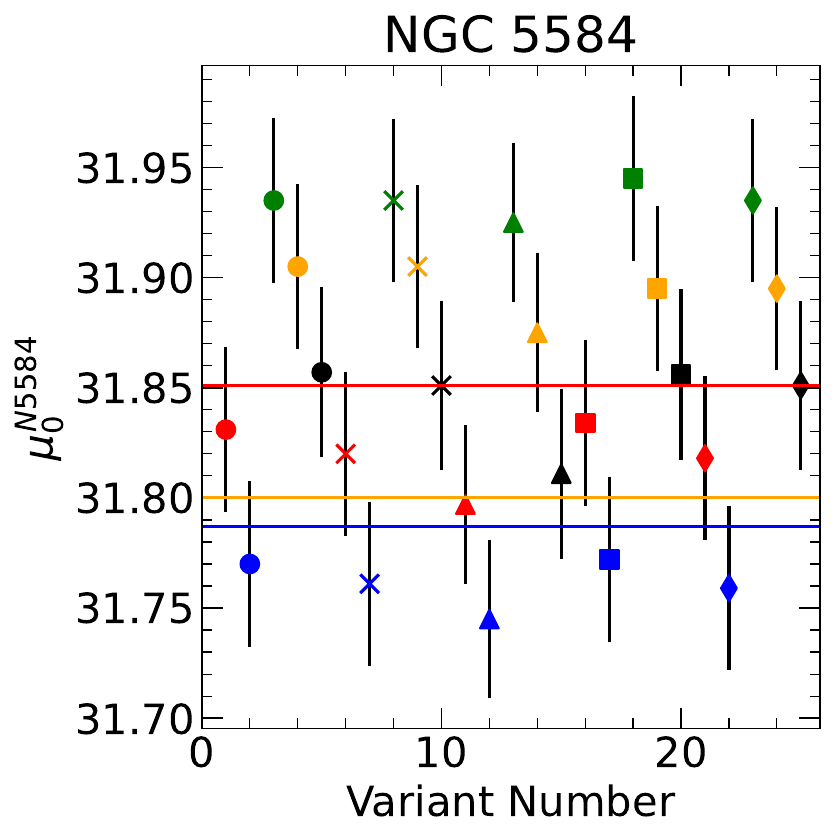} &
    \includegraphics[width=0.3\linewidth]{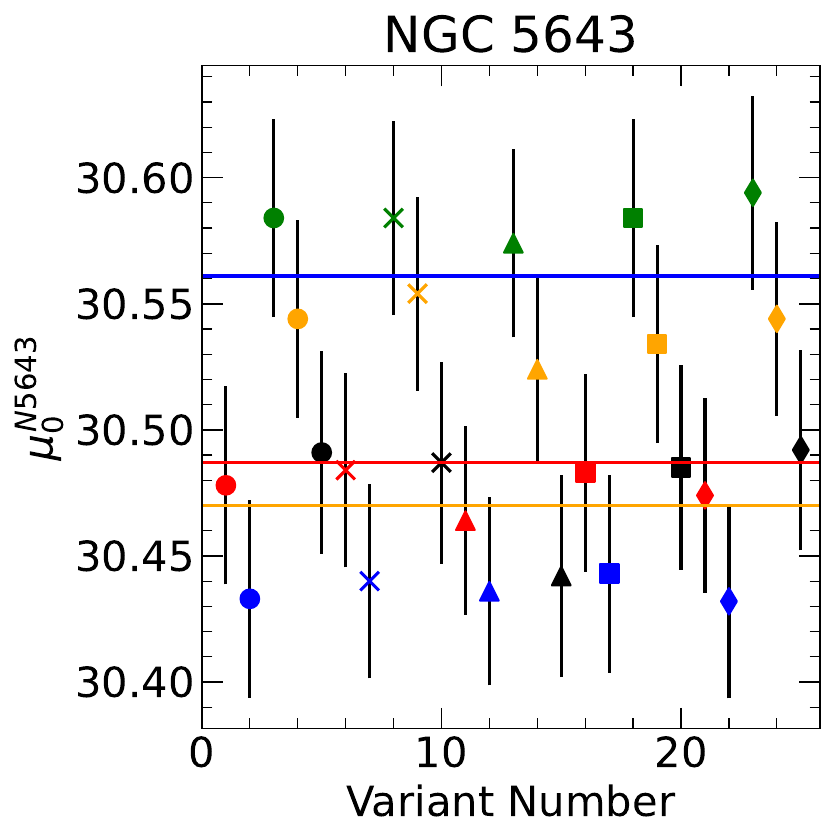} \\
  \end{tabular}
  \end{tabular}

\caption{JAGB reference magnitudes for NGC 4258 and JAGB distances to NGC 1448, NGC 1559, NGC 5584, and NGC 5643 using the measured JAGBs and their variants described in Sections \ref{sec:Symmetry_Null_Tests_Outer_Fields}. We anchor these distances to the NGC 4258 Inner+Outer fields to best approximate the outer disk. We show the Cepheid distances of $\mu_{N1448}$ = 31.302~mag, $\mu_{N1559}$ = 31.506~mag, $\mu_{N5584}$ = 31.787~mag, and $\mu_{N5643}$ = 30.561~mag, all of which are from \cite{Riess_2022ApJ...934L...7R}  and adjusted by adding 0.015~mag for a NGC 4258 anchor only distance, with the solid blue lines as reference. We show the median of the JAGB distance variants for each galaxy with the red solid line. For NGC 1559, we show the Mira distance from \cite{Huang_2020ApJ...889....5H} in cyan and TRGB distances from EDD and \cite{Anand_2024arXiv240104776A} in orange.}
\label{fig:JAGB_Distances}
\end{figure*}

In this section, we measure distances to NGC 1448, NGC 1559, NGC 5584, and NGC 5643 using the JAGB methods defined in the previous sections. For each measurement method, we compare the SN host JAGB magnitudes to those from the combined NGC 4258 Inner+Outer fields, which best approximates the outer disk, and use the geometric maser distance of $\mu_{N4258}$ = 29.397 $\pm$ 0.032~mag from \cite{Reid_2019ApJ...886L..27R} to calibrate the SN host distances.  The JAGB magnitudes for NGC 4258 across all variants are listed in Table \ref{tab:Anchor_Variants} and plotted in the leftmost panel of Fig.~\ref{fig:JAGB_Distances}.  The other four panels of Fig.~\ref{fig:JAGB_Distances} give the corresponding distance modulus to each of the four galaxies for all variants, with the median shown as a red horizontal line.  

For reference, we also give the distance moduli obtained from Cepheids (blue), TRGB (orange), and Miras (cyan) where available; the Cepheid distances are obtained from the solution in \citep{Riess_2022ApJ...934L...7R}. To eliminate calibration differences we adjust these distances upwards by 0.02 mag to reflect the distance scale based on NGC 4258 as the sole anchor.  The resulting Cepheid distance moduli are $\mu_{N1448}$ = 31.31~mag, $\mu_{N1559}$ = 31.51~mag, $\mu_{N5584}$ = 31.79~mag, and $\mu_{N5643}$ = 30.56~mag, with the solid blue lines.  The Mira distance to NGC 1559 is $\mu_{N1559} =31.41~$mag \citep{Huang_2020ApJ...889....5H}.  The TRGB distances of $\mu_{N1448}$ = 31.38~mag and $\mu_{N5643}$ = 30.47~mag are from the Extragalactic Distance Database (EDD) \citep{Tully_2009AJ....138..323T, Anand_2021AJ....162...80A}, while $\mu_{N1559}$ = 31.49~mag and $\mu_{N5584}$ = 31.80~mag are from JWST measurements from \cite{Anand_2024arXiv240104776A}. 

We find broad agreement between JAGB distances and those based on the Cepheid, TRGB, and Miras, with the variation in the distances spanning approximately 0.2~mag. As expected, the mode is closer to the mean and median measures for the case of larger LF smoothing, likely due to the GLOESS smoothing removing asymmetry in the luminosity function. The model fit also tends to agree better with a higher smoothed mode, likely due to the Gaussian component of the model fit. 

\begin{deluxetable*}{cccccc}
\label{tab:Anchor_Variants}
\tablehead{\colhead{$m^{JAGB}_{F150W}$} & \colhead{$\sigma$} & \colhead{Method}& \colhead{Color Selection} & \colhead{Smoothing}}
\startdata
22.487 & 0.018   & Median & Narrow & nan \\
22.579 & 0.018  & 3$\sigma$ Clipped Mean & Narrow & nan \\
\textbf{22.35} & 0.018  & Mode & Narrow & 0.25 \\
22.4 & 0.018  & Mode & Narrow & 0.35 \\
22.418 & 0.019  & Model Fit & Narrow  & nan \\
22.505 & 0.017  & Median & Baseline  & nan \\
22.599 & 0.017  & 3$\sigma$ Clipped Mean & Baseline  & nan \\
22.36 & 0.017  & Mode & Baseline & 0.25 \\
22.41 & 0.017  & Mode & Baseline  & 0.35 \\
22.43 & 0.019   & Model Fit & Baseline & nan \\
22.537 & 0.015  & Median & Wide & nan \\
\textbf{22.624} & 0.015  & 3$\sigma$ Clipped Mean & Wide & nan \\
22.38 & 0.015  & Mode & Wide &  0.25 \\
22.45 & 0.015  & Mode & Wide &  0.35 \\
22.48 & 0.019  & Model Fit & Wide  & nan \\
22.488 & 0.018  & Median & Blue 5$\%$  & nan \\
22.582 & 0.018  & 3$\sigma$ Clipped Mean & Blue 5$\%$  & nan \\
22.35 & 0.018  & Mode & Blue 5$\%$  & 0.25 \\
22.41 & 0.018  & Mode & Blue 5$\%$  & 0.35 \\
22.423 & 0.02  & Model Fit & Blue 5$\%$ & nan \\
22.503 & 0.017  & Median & Red 5$\%$  & nan \\
22.597 & 0.017  & 3$\sigma$ Clipped Mean & Red 5$\%$ & nan \\
22.35 & 0.017  & Mode & Red 5$\%$ &  0.25 \\
22.41 & 0.017  & Mode & Red 5$\%$ &  0.35 \\
22.426 & 0.019  & Model Fit & Red 5$\%$   & nan \\
\enddata
\caption{JAGB reference magnitudes for the combined East and West Inner+Outer NGC 4258 fields across all measurement and selection variants. These measurements are used to anchor the distances presented in Section \ref{sec:JAGB_Distances}. We bold the brightest and faintest JAGB reference magnitudes, which differ by $\sim$ 0.27 mag, in the left most column for reference.}
\end{deluxetable*}

\begin{figure*}[ht!]
  \centering
  \begin{tabular}{ccc}
    \includegraphics[width=0.7\linewidth]{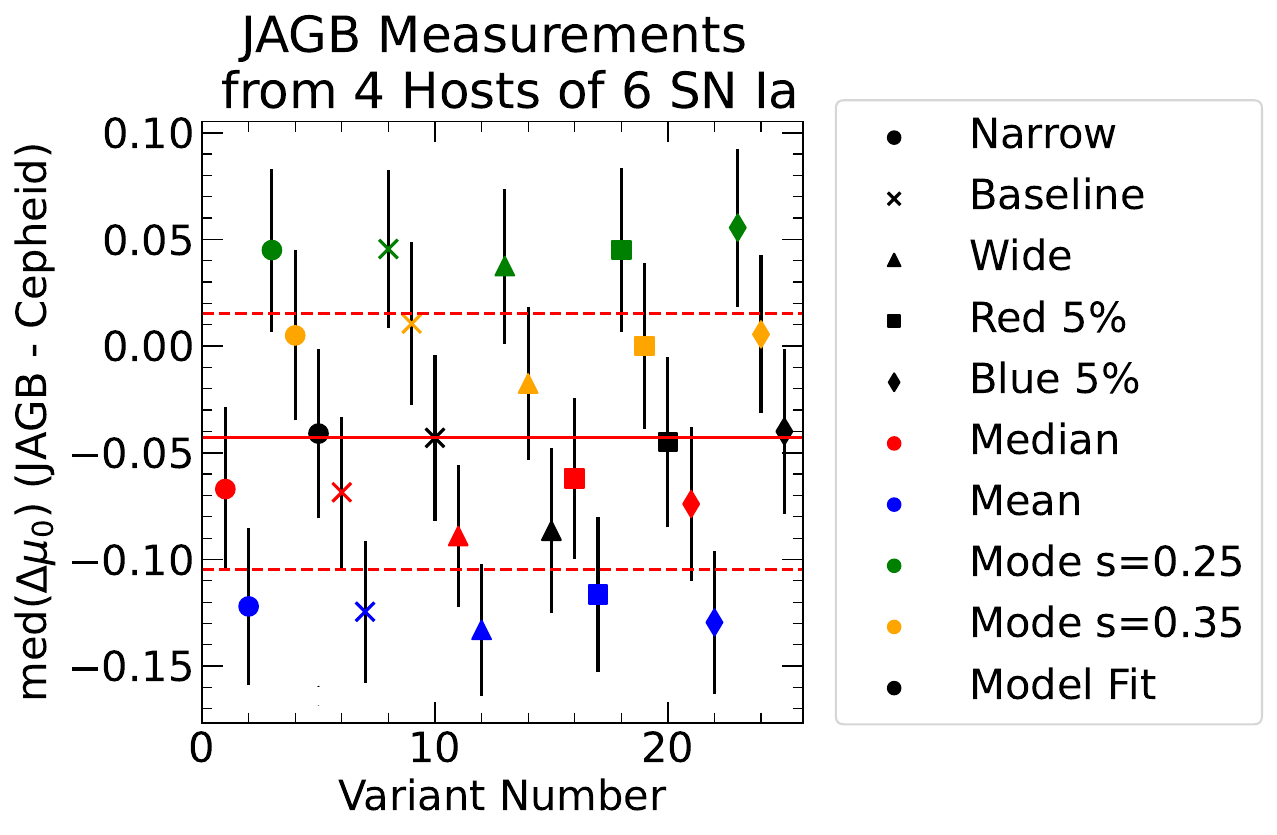}
  \end{tabular}
\caption{JAGB distances relative to the Cepheid distances from \cite{Riess_2022ApJ...934L...7R} for the median \emph{F150W} measurement variants over NGC 1448, NGC 1559, NGC 5584, and NGC 5643. We show the median of all variants with the red horizontal line and the standard deviations with the dashed red lines.}
\label{fig:JAGB_Distances_Average_to_Cepheids}
\end{figure*}

Next, we take the median over the four hosts and compare the median JAGB distances and Cepheid-based distances as shown in Fig. \ref{fig:JAGB_Distances_Average_to_Cepheids} for each measurement variant. This sample allows us to determine the value of $H_0$.

Unfortunately, as seen in Fig. \ref{fig:JAGB_Distances_Average_to_Cepheids}, this sample difference does depend quite significantly on which measurement method we choose, with a full range of $\sim$ 0.19 mag.  In the next section we identify the source of this methodological uncertainty as originating from the varying degree of asymmetry of the JAGB LFs.  All of the luminosity functions exhibit to varying degrees a positive skew and negative difference between the mode and mean-based JAGB measurements, as shown in Fig. \ref{fig:Host_Galaxy_CMDs}.  The fact that the calibrator, NGC 4258, has the largest skew (but not unusual in degree as it is comparable to that of the LMC and NGC 6822 from \cite{Parada_2021MNRAS.501..933P}), with a mean minus mode of 0.27 mag (versus the mean SN host with mean minus mode of 0.1 mag), produces a systematic bias that depends on the measure employed.

Despite the method-dependent bias, the JAGB observations in the anchor NGC 4258 and four SN~Ia host galaxies still provide interesting constraints on the Hubble constant.  We may adopt a fiducial result and uncertainty from the median of the differences between the JAGB and Cepheid distances as shown in Fig. \ref{fig:JAGB_Distances_Average_to_Cepheids} to derive $H_0$ using a similar approach in \cite{Scolnic_2023ApJ...954L..31S}. We calculate the weighted mean SN~Ia luminosity, $M^0_B$, calibrated to the median JAGB distances shown in Fig. \ref{fig:JAGB_Distances} and using the 6 SN~Ia (SN~2001el, SN 2021pit, SN~2005df, SN~2007af, SN~2013aa, SN~2017cbv) with standardized brightnesses, $m^0_B$, listed in Table 6 from \cite{Riess_2022ApJ...934L...7R} to find $M^0_B = -19.21 \pm 0.05$~mag. We then use this value to solve for $H_0$ using Equation 8 in \cite{Scolnic_2023ApJ...954L..31S} and adopt the intercept of the Hubble diagram, $a_b = 0.71448 \pm 0.0012$ from \cite{Riess_2022ApJ...934L...7R}. We find $H_0=74.7$ $\pm$ $2.1$ (stat) $\pm$ 2.3 (sys)  ($\pm$ 3.1 combined) km s$^{-1}$ Mpc$^{-1}$ with a range from 71 $-$ 78 km s$^{-1}$ Mpc$^{-1}$ from variations in the choice of measurement method. We list the error budget in Table \ref{tab:H0_Error_Budget}.   In the Discussion we consider whether there is compelling evidence to select any specific method over the spread of methods as to indicate a better result.

\begin{deluxetable*}{ccc}
\label{tab:H0_Error_Budget}
\tablehead{\colhead{Error} & \colhead{Value (stat) [mag]} & \colhead{Value (sys) [mag]}}
\startdata
NGC 4258 Maser Distance &  & 0.03 \\
Methodological Variations &  & 0.06 \\
JAGB Median four host Measurement & 0.04 &\\
$M^0_B$ & 0.05 & \\ 
Rest of DL: SN Ia, local and Hubble Flow & 0.01 & \\
\tableline
Total & 0.06 & 0.07 \\
\enddata
\caption{Error budget in magnitudes for the JAGB-based $H_0$ presented in this paper.}
\end{deluxetable*}

\medskip

\section{Discussion} \label{sec:Discussion}

We investigated the JAGB using $JWST$ observations of NGC 4258, NGC 1448, NGC 1559, NGC 5584, and NGC 5643 from GO-1685 \citep[P.I.: A. Riess,][]{Riess_2021jwst.prop.1685R}, testing the degree of internal consistency of the measurements from different subsets of the JAGB samples.  A series of symmetry and angular uniformity tests yield clear evidence of variations in JAGB luminosities larger than their measurement uncertainties and confirms past evidence \citep{Parada_2021MNRAS.501..933P} of differences in the degree of asymmetry of their luminosity functions. We further verified that the LF asymmetry is that of the host stars in the JAGB selection box and not due to contamination from background galaxies which are well-resolved by JWST and identified by DOLPHOT's star/galaxy discriminator (with compact, background galaxies also generally too blue to appear in the selection box).  We also note that foreground stars are also far too sparse to contaminate the host CMDs, as a similar field size of the Hubble Deep (or Ultra Deep) Field at similar galactic latitude as these hosts yields only O($10^1$) stars in the relevant magnitude range of $H=23-28$, and fewer in the selected JAGB color range.

Our findings are supported by the results of \cite{Parada_2021MNRAS.501..933P}, who had also found both variations in the JAGB luminosity and differences in the shape of J-region luminosity functions via skew, with the LMC and NGC 6822 being more asymmetric than the SMC or IC 1613.  Likewise, \cite{Lee_2022ApJ...933..201L} identified variations in the asymmetry and luminosity of the JAGB LF in M33 which varied with radial distance from the host with the LF being most symmetric in a ``sweet spot'' (region 3, between regions 1, 2 and 4).  In Appendix \ref{Apdx: Stability of the JAGB with Radius}, we present further study of the radial differentiation of the JAGB.

\begin{figure*}[ht!]
  \centering
  \begin{tabular}{ccc}
    \includegraphics[width=0.8\linewidth]{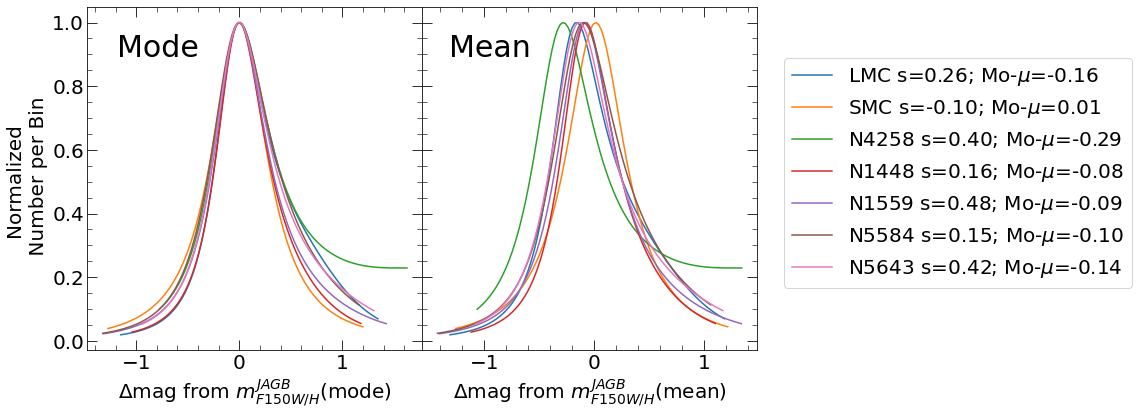}
  \end{tabular}
\caption{J-region luminosity functions for the LMC, SMC, NGC 4258, NGC 1448, NGC 1559, NGC 5584, and NGC 5643. The LMC and SMC luminosity functions are in the H-band, and the NGC 4258, NGC 1448, NGC 1559, NGC 5584, and NGC 5643 luminosity functions are in \emph{F150W}. The luminosity functions shown here are the fitted Lorentzians using the form from \cite{Parada_2021MNRAS.501..933P}. In the left hand subplot, we offset the LF with their modes. In the right hand subplot, we offset the LF using their sigma-clipped means to demonstrate how the mode and mean measurements can differ when the LF is skewed. }
\label{fig:Overlayed_LF}
\end{figure*}

In Fig. \ref{fig:Overlayed_LF} we compare the skewness of the JAGB luminosity functions examined in this paper by fitting them with the  Lorentzian model described in \cite{Parada_2021MNRAS.501..933P}.  We also include the LF for the LMC and SMC with H-band photometry retrieved from \cite{Macri_2015AJ....149..117M} and \cite{Ripoche_2020MNRAS.495.2858R}, respectively.  The legend provides the Fisher-Pearson skew and the difference between the JAGB measured using the mode of the fit and the 3$\sigma$ clipped mean.  We offset the LFs in magnitude based on their modes (left panel) or their sigma-clipped mean (right panel); the difference clearly illustrates how mode- and mean-based measurements differ for skewed LFs.  We note that the $F110W$ band JAGB can also be skewed by similar amounts, as shown in \cite{Lee_2023ApJ...956...15L}; the same holds for the J-band LF of the LMC \citep{Parada_2021MNRAS.501..933P}.  This shows that the skew effect is not unique to a particular NIR band. In this comparison, we use the Lorentzian fit to the LFs, because GLOESS smoothing tends to naturally reduce the LF skew; this can be seen in Appendix \ref{Apdx: Effects of Smoothing on JAGB Luminosity Functions}.  As a result, a larger smoothing window used before calculating the mode produces a result closer to the mean and median measurements (see for instance, Fig. \ref{fig:JAGB_Distances_Average_to_Cepheids}). The source of variation in inferred $H_0$ primarily comes from the difference in the skews between NGC 4258 and host galaxies. In NGC 4258, the difference between the mode (using GLOESS) and mean is $-$0.27~mag, while in the host galaxies, it is $\sim$0.1~mag resulting in a $\sim$0.2~mag difference across the distance ladder.

\subsection{Asymmetry in the Clouds}

It is instructive to further scrutinize the JAGB luminosity functions in the Magellanic Clouds for further guidance, as these exhibit a disparity in asymmetry similar to our target galaxies, while having the virtue that their relative distances are well-established geometrically and they have been well studied in the literature using various measurement methods.  In Table \ref{tab:LMC_SMC_JAGBs} we provide JAGB reference magnitudes in the LMC and SMC taken from the literature, together with our measurements using the same suite of methods used in our analysis (i.e., mode with $s=0.25, 35$, sigma-clipped mean and the median) on the same datasets as prior authors. For the LMC, which has an asymmetric JAGB LF (\cite{Parada_2021MNRAS.501..933P} reports a skew of $-0.43 \pm 0.06$), past studies of the J vs $J-K$ CMD from several authors \citep{Madore_2020ApJ...899...66M,Freedman_2020ApJ...899...67F,Ripoche_2020MNRAS.495.2858R,Parada_2021MNRAS.501..933P,Zgirski_2021ApJ...916...19Z} give an apparent reference magnitude of $J=12.23-12.38$, a range of 0.15 mag; the difference between these results is much larger than the typical quoted statistical uncertainty of $\sim$ 0.004 mag, which is derived from the standard deviation of the distribution (0.2--0.3 mag) divided by the square root of the number of stars (typically several thousands).  With our measurements, we similarly find that the asymmetry produces a large, method-dependent range of $>$ 0.1 mag.  The choice of smoothing for the mode-based method also slightly shifts the JAGB reference magnitude. Another example of this effect can be seen in Appendix \ref{Apdx: Effects of Smoothing on JAGB Luminosity Functions}. The SMC, which has a symmetric luminosity function, has a far more stable JAGB reference magnitude that also does not change with different smoothing values. We find that the mean and median for the J-region luminosity function using the \cite{Macri_2015AJ....149..117M} data differ by 0.06~mag.

\begin{figure*}[ht!]
  \centering
  \begin{tabular}{ccc}
    \includegraphics[width=0.38\linewidth]{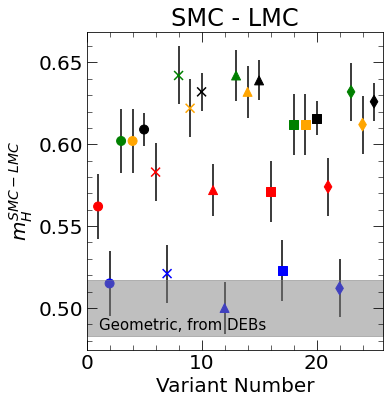} &
    \includegraphics[width=0.6\linewidth]{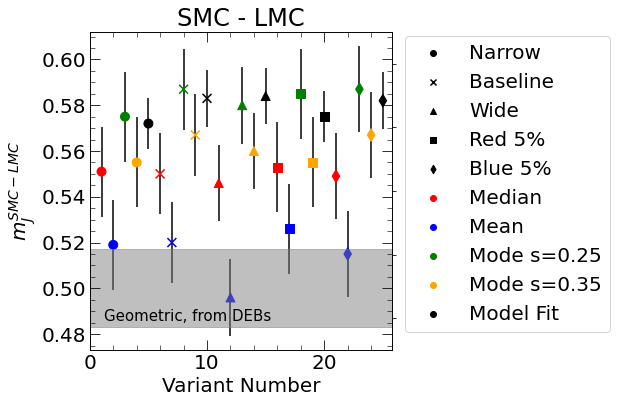} \\
  \end{tabular}
\caption{Differences between SMC and LMC JAGB reference magnitudes in the J and H bands using the variants described in Section \ref{sec:Measurement}. LMC and SMC data are both from \citep{Ripoche_2020MNRAS.495.2858R}. We correct for reddening using the \cite{Skowron_2021ApJS..252...23S}} dust maps. We show the difference between the geometric distances to the SMC and LMC from \cite{Graczyk_2020ApJ...904...13G, Pietrzynski_2019Natur.567..200P} of 0.500 $\pm 0.01 7$~mag as the gray horizontal shaded region. 
\label{fig:SMC_minus_LMC}
\end{figure*}

We can check which method provides a more standardized reference magnitude by comparing the differences between the geometric distances and JAGB reference magnitudes between the SMC and LMC from \cite{Graczyk_2020ApJ...904...13G} and \cite{Pietrzynski_2019Natur.567..200P} using the LMC and SMC samples from \cite{Ripoche_2020MNRAS.495.2858R}. We restrict the SMC sample to stars within 1 degree within the SMC center to cover the same region as the detached eclipsing binaries while avoiding the complexity of the depth of the SMC outside its core region.  We correct for reddening using dust maps from \cite{Skowron_2021ApJS..252...23S} and measure the JAGB reference magnitudes using the same variants described above and plot the differences between the SMC and LMC reference magnitudes in Fig. \ref{fig:SMC_minus_LMC}. We shade in gray the area corresponding to the difference between the SMC and LMC geometric distances of 0.500~$\pm$ 0.017~mag. We find that in general, the clipped mean method agrees most closely with the geometric differences; this method also tends to produce the least variation in the null and angular tests shown in Section \ref{sec:Symmetry_Null_Tests_Outer_Fields}.  Methods which try to measure the peak of the LF, such as the mode (especially with little LF smoothing) or a model fit to the peak, are in significantly larger disagreement with the geometric difference.

Non-uniformity in the color-dependence of the luminosity function can also introduce some variation in the standard candle.  In all JWST host galaxies we found that the reference magnitude varies somewhat with color, not linearly, but tending to be brighter near the middle of the color range (see Appendix \ref{Apdx: J-region_Tilts}).   If the color dependence is uniform across hosts, it will have no impact on distance measurements determined by comparing different hosts as distance measurements are constructed by comparing magnitudes relative to the anchor.  In practice, the relation between the reference magnitude and color appears less important than the shape of the luminosity function based on our comparisons of different color ranges.

\begin{deluxetable*}{cccccccccc}
\label{tab:LMC_SMC_JAGBs}
\tablehead{\colhead{Galaxy} & \colhead{Study} & \colhead{Sample} & \colhead{Method} & \colhead{$m^{JAGB}_{J}$} & \colhead{$A_J^*$} & \colhead{$A_J$ Ref.} & \colhead{Anchor Dist} & \colhead{Anchor Ref.} & \colhead{$M^{JAGB}_{J}$} \\
\colhead{(1)} & \colhead{(2)} & \colhead{(3)} & \colhead{(4)} & \colhead{(5)} & \colhead{(6)} & \colhead{(7)} & \colhead{(8)} & \colhead{(9)}& \colhead{(10)}}
\startdata
LMC & MF20 & M15 & Selected range mean & 12.31 & 0.05 & NED & 18.477 & P19 & -6.22 \\
LMC & Here & M15 & Median & 12.37 & 0.08 & S21 & 18.477 & P19 & -6.21 \\
LMC & Z21 & K09 & Model Fit & 12.38 & 0.11 & Go20 & 18.477 & P19 & -6.21 \\
LMC & R20 & R20 & ML Median & 12.31 & 0.11 & Go20 & 18.477 & P19 & -6.28 \\
LMC & P21 & R20& Mode (Lorentzian model) & 12.23 & 0.11 & Go20 & 18.477 & P19 & -6.36 \\
LMC & Here & R20 & Median & 12.31 & 0.08 & S21 & 18.477 & P19 & -6.25 \\
LMC & Here & R20& Mode $s$ = 0.25 & 12.29 & 0.08 & S21 & 18.477 & P19 & -6.27 \\
LMC & Here & R20& Mode $s$ = 0.35 & 12.31 & 0.08 & S21 & 18.477 & P19 & -6.26 \\
LMC & Here & R20& Model Fit & 12.29 & 0.08 & S21 & 18.477 & P19 & -6.27 \\
\hline
SMC & MF20 & S06 & Selected range mean & 12.81 & 0.03 & NED & 18.965 & Gr14 & -6.18 \\
SMC & Z21 & K09 & Model Fit & 12.85 & 0.07 & Go20 & 18.977 & Gr20 & -6.20 \\
SMC & R20 & R20& ML Median & 12.87 & 0.07 & Go20 & 18.96 & S16 & -6.16 \\
SMC & P21 & R20& Mode (Lorentzian model) & 12.90 & 0.07 & Go20 & 18.96 & S16 & -6.15 \\
SMC & Here & R20 & Median & 12.83 & 0.04 & S21 & 18.977 & Gr20 & -6.19 \\
SMC & Here & R20 & Mode $s$ = 0.25 & 12.85 & 0.04 & S21 & 18.977 & Gr20 & -6.17 \\
SMC & Here & R20 & Mode $s$ = 0.35 & 12.85 & 0.04 & S21 & 18.977 & Gr20 & -6.17 \\
SMC & Here & R20 & Model Fit & 12.85 & 0.04 & S21 & 18.977 & Gr20 & -6.17 \\
\enddata
\caption{Summary table for JAGB reference magnitude measurements in the LMC and SMC from \cite{Madore_2020ApJ...899...66M} (MF20), \cite{Ripoche_2020MNRAS.495.2858R} (R20), \cite{Parada_2021MNRAS.501..933P} (P21), \cite{Zgirski_2021ApJ...916...19Z} (Z21), and here. These measurements are made using the \cite{Macri_2015AJ....149..117M} (M15), 2MASS from IRSA \cite{Skrutskie_2006AJ....131.1163S} (S06), Infrared Survey Facility (IRSF) Magellanic Clouds Point Source Catalog from \cite{Kato_2007PASJ...59..615K} (K09), and R20 samples. 
The method of MF20 is to select a narrow range by visual inspection of the CMD and average only that region (see also \citep{Freedman_2020ApJ...899...67F}.
Reddening corrections refer to those from NED, \cite{Skowron_2021ApJS..252...23S} (S21), and \cite{Gorski_2020ApJ...889..179G} (Go20). Anchor distances are from \cite{Pietrzynski_2019Natur.567..200P} (P19), \cite{Graczyk_2014ApJ...780...59G} (Gr14), \cite{Graczyk_2020ApJ...904...13G} (Gr20), and \cite{Skowcroft_2016ApJ...816...49S} (S16). Columns left to right: (1) Galaxy (LMC or SMC), (2) study, (3) dataset, (4) method used to fix the JAGB reference magnitude (simple mean, median, Gaussian+Quadratic model fit (see \cite{Zgirski_2021ApJ...916...19Z}, maximum likelihood (ML) median (see \cite{Ripoche_2020MNRAS.495.2858R}, and mode using a Lorentizan fit), and mode with $s=0.25, 0.35$, (5) J-band JAGB reference magnitude, (6) J-band extinction, (7) J-band extinction reference, (8) anchor distance, (9) anchor distance reference, and (10) J-band JAGB luminosity. For the extinctions, $A_J$, using dust maps from \cite{Skowron_2021ApJS..252...23S}, we list the mean $A_J$ for the sample. JAGB measurements labelled `Here' use a color range of 1.4 to 2~mag.} We note that \cite{Parada_2021MNRAS.501..933P} do not use the mode from their Lorentzian fit as a final JAGB reference magnitude.  $^*$-for MF20 extinction is from MW only, other studies include both MW and Cloud extinction.
\end{deluxetable*}

\subsection{The JAGB in Perspective}

We hypothesize that these variations in the shapes of the JAGB luminosity function may result from intrinsic, astrophysical differences in the distribution of carbon star demographics. 
\cite{Parada_2021MNRAS.501..933P} suggested that some of the field-to-field variations in the measured JAGB luminosity seen in the LMC, listed in their Table 5 and described in their Section~5.1, may be due to variations in the star formation history in different regions.  The differences are unlikely to result from interstellar reddening, due to the use of reddening maps to remove absorption.  We would also not expect interstellar reddening to play a role in the host studied here, based on estimates of the variation in interstellar reddening with radial separation presented by \cite{Menard_2010MNRAS.405.1025M} (see equation 29). For example, at semi-minor axes separations of 39$''$ and 48$''$ from the centers of NGC 1559 and NGC 5584 corresponding to the ellipses in Table \ref{tab:Disk_Field_Parameters}, we expect NIR internal extinctions of 6~mmag and 5~mmag, respectively. \cite{Parada_2021MNRAS.501..933P} also suggest that metallicity can drive the skewness of the J-region luminosity function and the JAGB luminosity, and choose either to use the LMC or SMC as anchors based on the skewness of the J-region luminosity functions in their target galaxies. We also note that it may also be possible that the J-region contains a mixed population of stars, such that not all groups of stars have the same mean magnitude. Further work is needed to investigate the origin of the variations in the measured JAGB and to demonstrate that these differences in treatment of different targets do not create undue variations and uncertainties in the estimated relative distances.  Because we do not know whether the often asymmetric shape of the LF is due to the spread of a single population or the combination of multiple symmetric populations, it is not clear {\it a priori} which reference flux should provide the best standard candle, something that is crucial to take into consideration when using the JAGB as a standard candle.

The investigations of variations in the JAGB presented above highlight several issues to consider when using the JAGB to measure distances. It is evident that the shape of the JAGB luminosity function, as it is currently selected, is non-uniform and often asymmetric.  This produces systematic uncertainties in distance measurements from the data presented here (and in the Clouds) resulting from the different and credible options of defining a reference flux.    We do not think it is plausible to select just one method or define a somewhat different measurement prescription or magnitude range for each host and ignore the uncertainty arising from the freedom in these choices. 

\begin{figure*}[ht!]
  \centering
  \begin{tabular}{ccc}
    \includegraphics[width=0.6\linewidth]{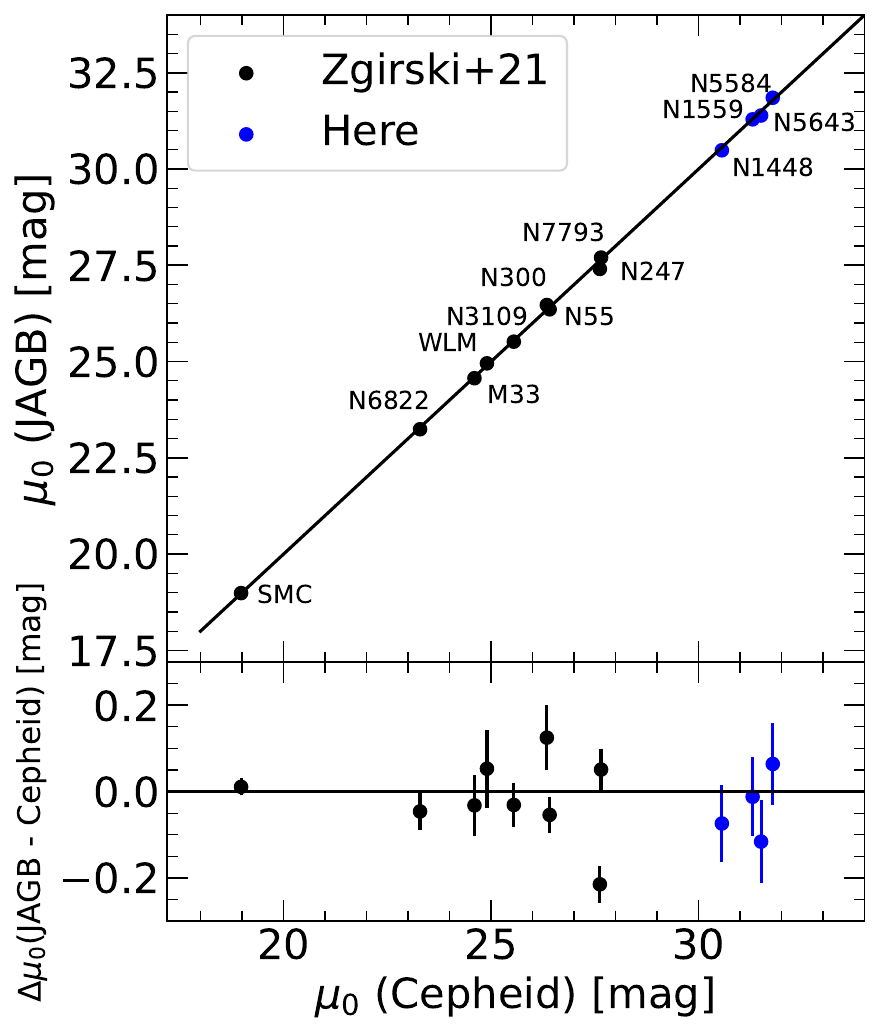}
  \end{tabular}
\caption{JAGB distances plotted against Cepheid distances from this paper, \cite{Riess_2022ApJ...934L...7R}, and those measured and compiled in \cite{Zgirski_2021ApJ...916...19Z}. The SMC is compared against a detached eclipsing binary distance and is shown for reference. We plot the residuals in the bottom subplot.}
\label{fig:JAGB_vs_Ceph_Distances}
\end{figure*}

For an overall perspective, we plot in Fig.~\ref{fig:JAGB_vs_Ceph_Distances} the JAGB distances measured here against the Cepheid distances from \cite{Riess_2022ApJ...934L...7R}, adjusted for a NGC 4258 only anchor, together with the JAGB and Cepheid distances from \cite{Zgirski_2021ApJ...916...19Z}.  For our distance measurements we use the median of the distances shown in Fig.~\ref{fig:JAGB_Distances}, and include the contribution of methodological variation in the estimated errors.  We find that the JAGB and Cepheid distances show reasonable agreement within their error bars. 

The JWST data presented above demonstrate that until JAGB luminosity function variations are standardized or understood, they should be included in the uncertainties when the JAGB is used to measure extragalactic distances and the Hubble constant.   At present, this issue lay in the critical path to reach the few percent level precision in measuring the Hubble constant.  

Such issues are not unexpected for the development and refinement of a new standard candle.  Recent work has already made inroads into improving the choice of radius at which to measure JAGB \citep{Lee_2023ApJ...956...15L} (see discussion comparing radii in Appendix~\ref{Apdx: Stability of the JAGB with Radius}) as well as how best to match the shape of the luminosity function to measure distances \citep{Parada_2023MNRAS.522..195P}. Future studies and JAGB observations are likely to continue improving this promising standard candle.

\begin{acknowledgments}

We thank Yukei Murakami and Abigail Lee for their helpful conversations. SL is supported by the National Science Foundation Graduate Research Fellowship Program under grant number DGE2139757. DS is supported by Department of Energy grant DE-SC0010007, the David and Lucile Packard Foundation, the Templeton Foundation and Sloan Foundation. 

This research is based on observations made with the NASA/ESA James Webb Space Telescope obtained from the Space Telescope Science Institute, which is operated by the Association of Universities for Research in Astronomy, Inc., under NASA contract NAS 5-03127. These observations are associated with program GO-1685.

Some of the data presented in this article were obtained from the Mikulski Archive for Space Telescopes (MAST) at the Space Telescope Science Institute. The specific observations analyzed can be accessed via \dataset[DOI]{https://doi.org/10.17909/sd84-nr38} 10.17909/sd84-nr38.

This work was carried out at the Advanced Research Computing at Hopkins (ARCH) core facility (rockfish.jhu.edu), which is supported by the National Science Foundation (NSF) grant number OAC 1920103.

This research has made use of the NASA/IPAC Extragalactic Database (NED), which is funded by the National Aeronautics and Space Administration and operated by the California Institute of Technology. This work has also made use of the data extraction tool WebPlotDigitizer from https://github.com/ankitrohatgi/WebPlotDigitizer.

\end{acknowledgments}

\vspace{5mm}
\facilities{JWST (NIRCam)}

\software{astropy \citep{Astropy_2013A&A...558A..33A, Astropy_2018AJ....156..123A, Astropy_2022ApJ...935..167A}, SAOImageDS9 \citep{SAO_Image_2003ASPC..295..489J}, DOLPHOT \citep{Dolphin_2000PASP..112.1383D, Dolphin_2016ascl.soft08013D},  GNU Parallel \citep{Tange_2011a}, dustmaps \citep{Green_2018JOSS....3..695M}}

\appendix

\section{Asymmetry and J-region Tilts} \label{Apdx: J-region_Tilts}

In this section, we investigate the shape and structure of the J-region in the CMD and its affect on its use as a standard candle. Ideally the shape would be simple (i.e., a fixed point) or if not, at least consistent between hosts.

\begin{figure*}[ht!]
  \centering
  \begin{tabular}{cccc}
    \includegraphics[width=0.32\linewidth]{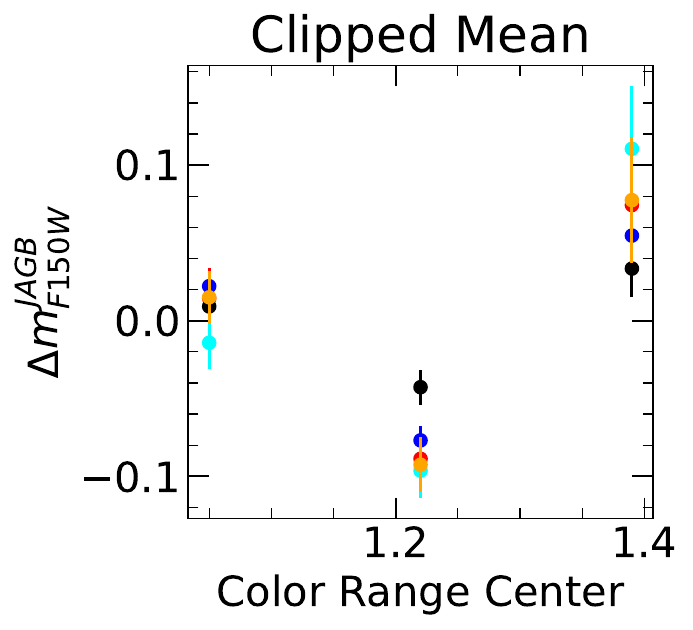} &
    \includegraphics[width=0.45\linewidth]{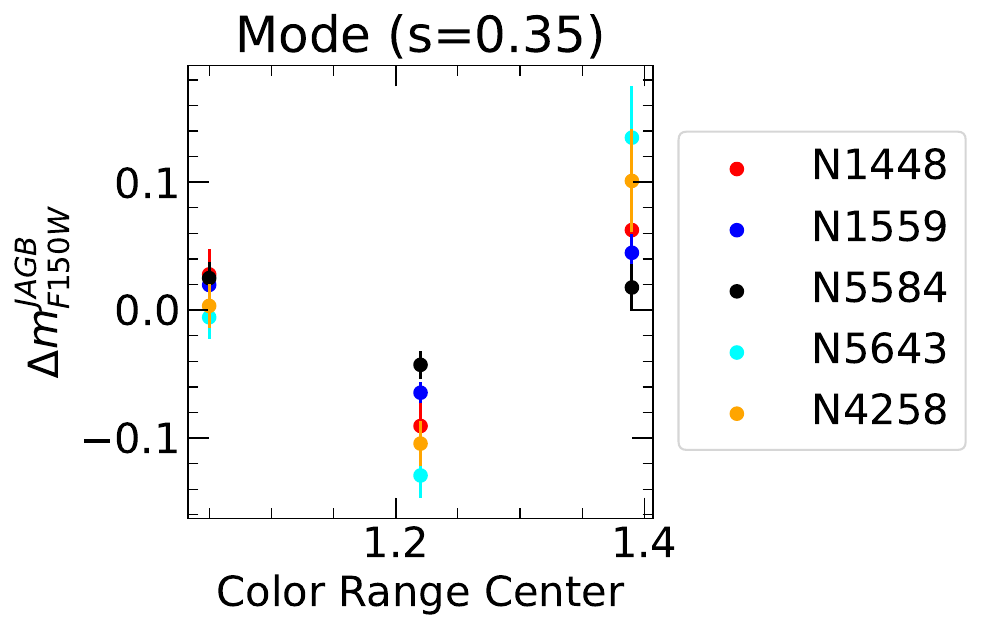} \\
    \includegraphics[width=0.36\linewidth]{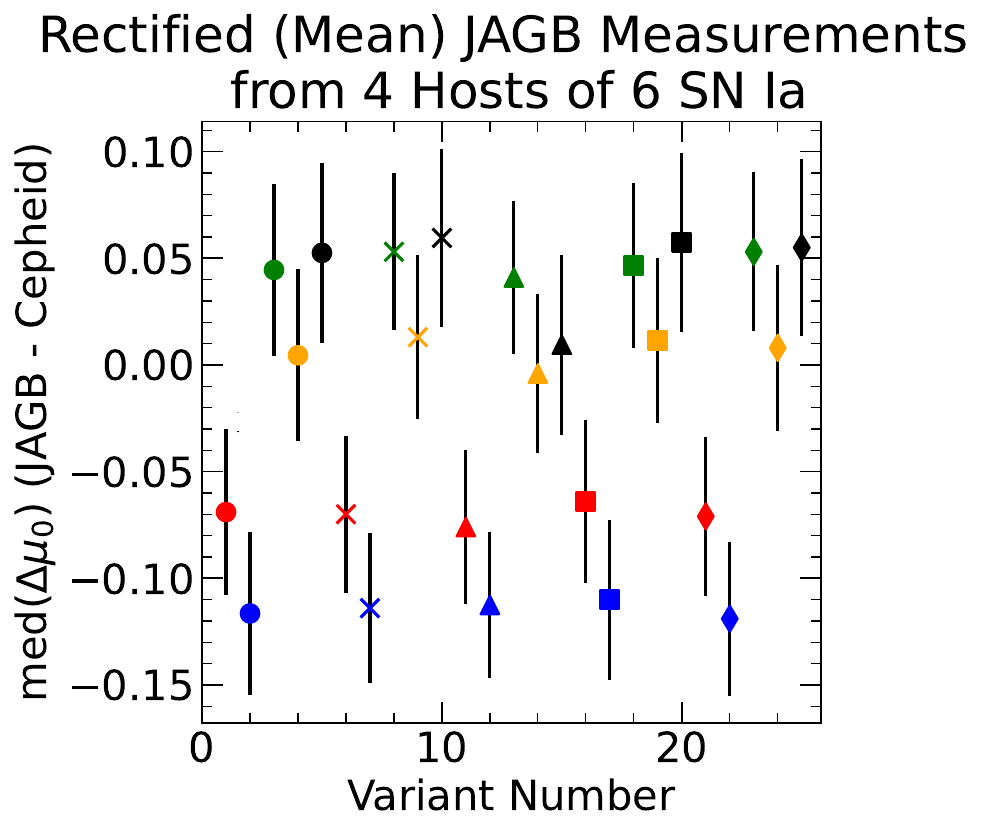} &
    \includegraphics[width=0.45\linewidth]{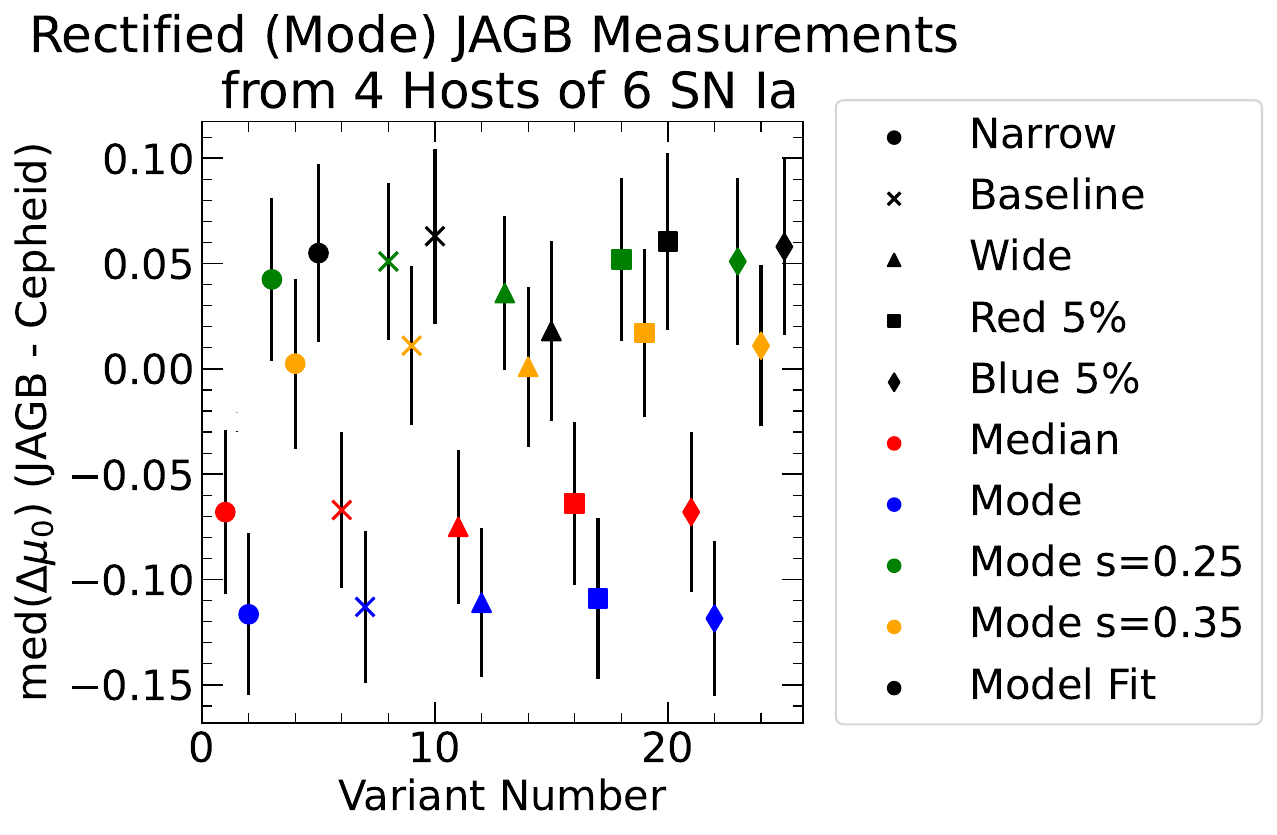} \\    
  \end{tabular}
\caption{The top row shows JAGB reference magnitudes, relative to their means, as a function of color for the four host galaxies measured in this study.  The y-axis shows the difference in magnitude between the mean JAGB for each galaxy. The bottom row shows the median difference between the JAGB and Cepheid distances across variants, similar to in Fig. \ref{fig:JAGB_Distances_Average_to_Cepheids}, but with quadratic color rectifications derived using the clipped mean and mode ($s=0.35)$.}
\label{fig:JAGB_vs_Color}
\end{figure*}

\begin{figure*}[ht!]
  \centering
  \begin{tabular}{ccc}
    \includegraphics[width=0.3\linewidth]{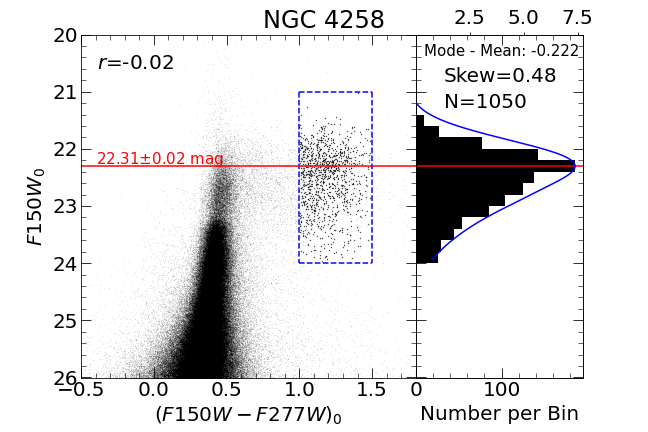} &
    \includegraphics[width=0.3\linewidth]{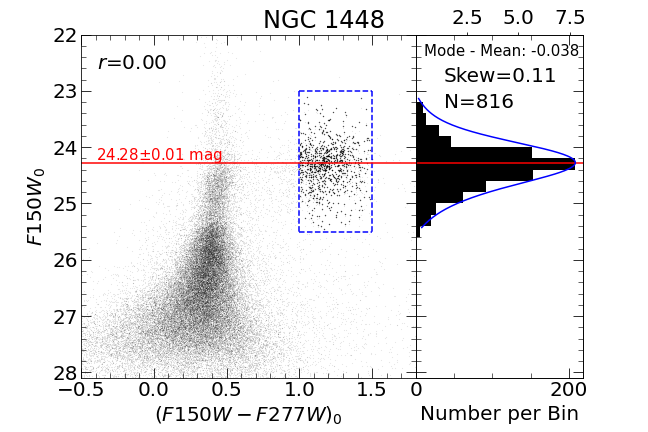} &
    \includegraphics[width=0.3\linewidth]{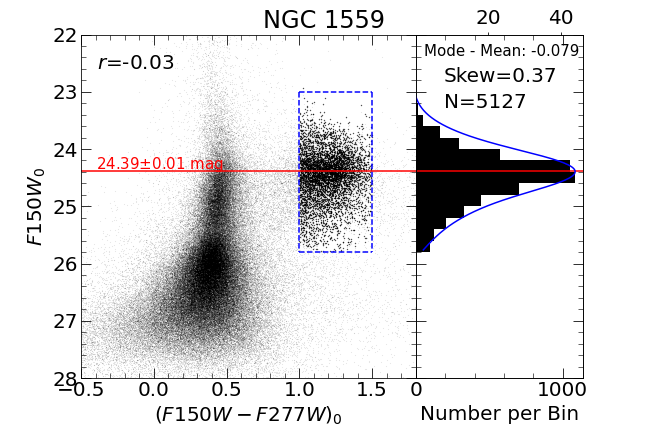} \\
    \includegraphics[width=0.3\linewidth]{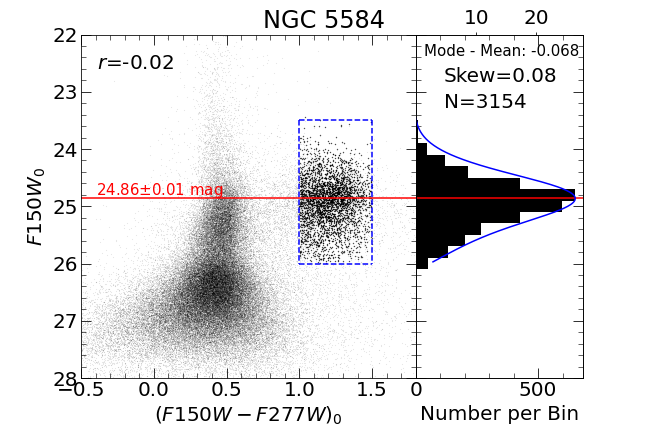} &
    \includegraphics[width=0.3\linewidth]{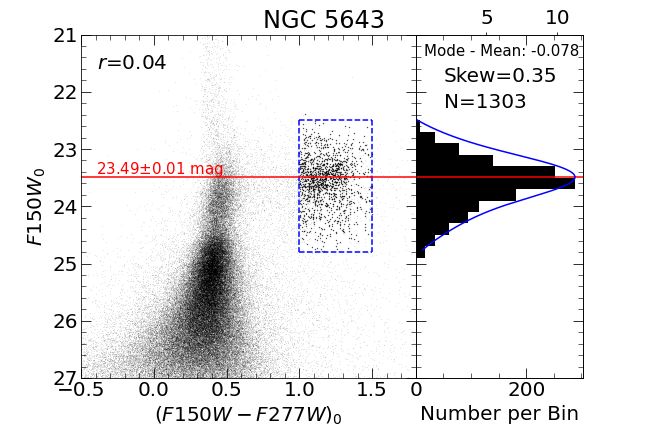} \\
  \end{tabular}
\caption{All CMDs and J-region luminosity functions in \emph{F150W} vs. $F150W - F277W$ for NGC 4258, NGC 1448, NGC 1559, NGC 5584, and NGC 5643 after using the mode ($s=0.35$) based color rectification described in Appendix \ref{Apdx: J-region_Tilts}. The blue, dashed box surrounds the region defined as the J-region and used for the JAGB measurement. Points inside the dashed blue box are enlarged for emphasis. J-region luminosity functions on the right panels of each subplot are binned with bin widths of 0.2~mag (black). To provide an example JAGB measurement, we smooth the luminosity function binned in 0.01~mag widths and apply GLOESS with a smoothing parameter of $s$=0.25~mag. The blue line corresponds to the smoothed luminosity function, and the red line and corresponding red label marks the location of the JAGB measured using this method. The tick marks below and above the luminosity function correspond to the 0.2~mag binned and smoothed luminosity functions, respectively. We show the skew of the J-region luminosity functions as calculated using the Fisher-Pearson coefficient using \texttt{skew} and the Pearson correlation coefficient using \texttt{pearsonr}, both from the Python sub-package \texttt{scipy.stats}, in the top right and left hand corners of each plot, respectively.}
\label{fig:Host_Galaxy_CMDs_Rectification}
\end{figure*}

We first quantify the color dependence of the JAGB reference magnitude in our data, showing in the upper row of Fig. \ref{fig:JAGB_vs_Color} the JAGB reference magnitudes for three equal color bins in the range of 1 $< F150W - F277W <$ 1.5~mag relative to their means.
The JAGB reference magnitudes for the bins use the clipped mean and mode ($s=0.35$) as illustrative examples. For the five hosts we see a semi-quadratic relation between magnitude and color, with the center color bin being brighter than the mean by $0.05-0.1$ mag. This finding is consistent with the lack of any significant Pearson correlation between brightness and color (i.e., not a linear trend) shown in Fig. \ref{fig:Host_Galaxy_CMDs}. We perform a rectification of the JAGB magnitudes based on color and a quadratic fit to these points (calculated separately for the clipped mean or mode) and plot the CMDs containing the rectified J-regions in Fig. \ref{fig:Host_Galaxy_CMDs_Rectification}. The Pearson correlation coefficients remain near zero after rectification.

We recalculate the median difference between the JAGB and Cepheid distances after color rectification in Fig. \ref{fig:JAGB_vs_Color}.  Compared to without rectification, see Figure \ref{fig:JAGB_Distances_Average_to_Cepheids}, we find little change.   This is not surprising since the relation between color and luminosity is fairly consistent across hosts so the quadratic structure largely cancels when comparing measures over the same color range. We also find that removing this quadratic color structure does not change the asymmetry (most strongly present in the LF of NGC 4258), as evidenced by seeing the same $\sim$ 0.15 mag difference between the mode and mean.

\begin{figure*}[ht!]
  \centering
  \begin{tabular}{cccc}
    \includegraphics[width=0.3\linewidth]{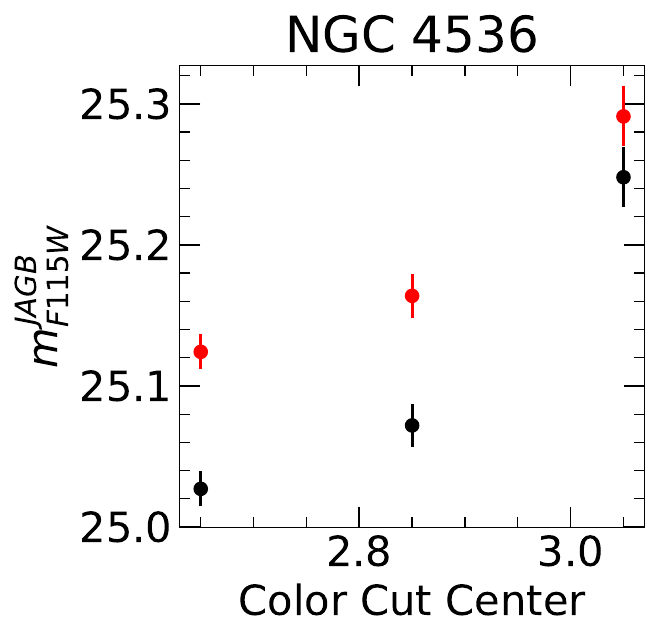} &
    \includegraphics[width=0.52\linewidth]{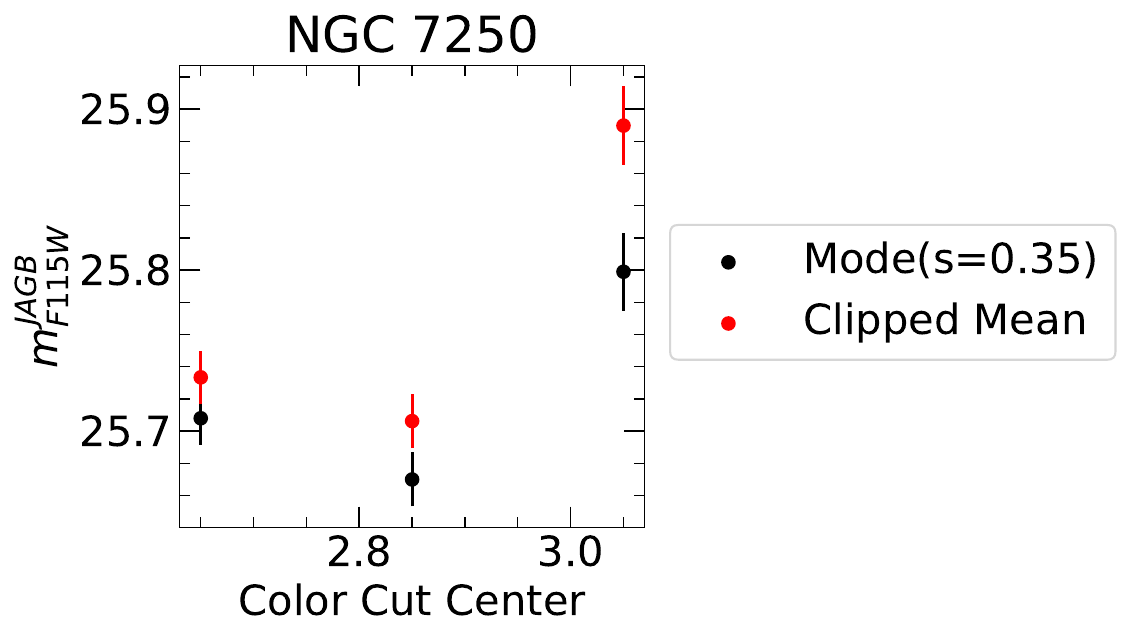} \\
  \end{tabular}
\caption{JAGB reference magnitudes as a function of color for NGC 4536 and NGC 7250 using the clipped mean and mode ($s=0.35$~mag) methods.}
\label{fig:JAGB_vs_Color_F115W}
\end{figure*}

For reference we repeat the same analysis for observations of the JAGB in the $F115W$ (1.15 microns) band, bluer but closer to the conventional $J$-band (1.25 microns) than $F150W$ (1.50 microns).  To do this we retrieved \emph{JWST} GO-1995 observations of NGC 4536 and NGC 7250 \citep[PI: W. Freedman][]{Freedman_2021jwst.prop.1995F} from MAST and perform photometry using DOLPHOT \citep{Dolphin_2000PASP..112.1383D, Dolphin_2016ascl.soft08013D} previously studied by \cite{Lee_2024ApJ...961..132L}.  We use the same pipeline described in Section \ref{sec:photometry} and quality cuts and disk region selection from \cite{Lee_2024ApJ...961..132L} as applicable and measure the JAGB reference magnitude as a function of three color bins across $F115W-F444W$=$2.6-3.2$ ~mag as studied by \cite{Lee_2024ApJ...961..132L}.  We show these measurements in Fig. \ref{fig:JAGB_vs_Color_F115W}. We find that the JAGB reference magnitudes in $F115W$ are not consistent with a flat line similar, to what we see with $F150W$. The relative changes in the JAGB reference magnitudes with color in $F115W$ vs. $F115W - F444W$ are perhaps less consistent across galaxies than what we found with $F150W$ vs. $F150W - F277W$ in Fig. \ref{fig:JAGB_vs_Color}.

The non-constant magnitude of the JAGB with color
demonstrated above for half a dozen galaxies in $F115W$ and $F150W$ leads us to revisit the expectation of a constant magnitude in the ground-based $J$-band (1.25 microns) claimed in for example \cite{Madore_2020ApJ...899...66M}.  \cite{Weinberg_2001ApJ...548..712W} observed that the J-region in the K-band followed the relation:

\begin{equation} \label{eq:K_band_tilt}
    K = D_o-0.99(J-K)
\end{equation}

\noindent where $D_0$ is a zero-point, and $J$ and $K$ refer to their respective bands. \cite{Madore_2020ApJ...899...66M} concluded from this relation that adopting a unit slope and rearranging Equation \ref{eq:K_band_tilt} yields J = $D_0$, a constant.  However, in detail this may not be broadly true for other hosts or in general.

\begin{figure*}[ht!]
  \centering
  \begin{tabular}{cccc}
    \includegraphics[width=0.5\linewidth]{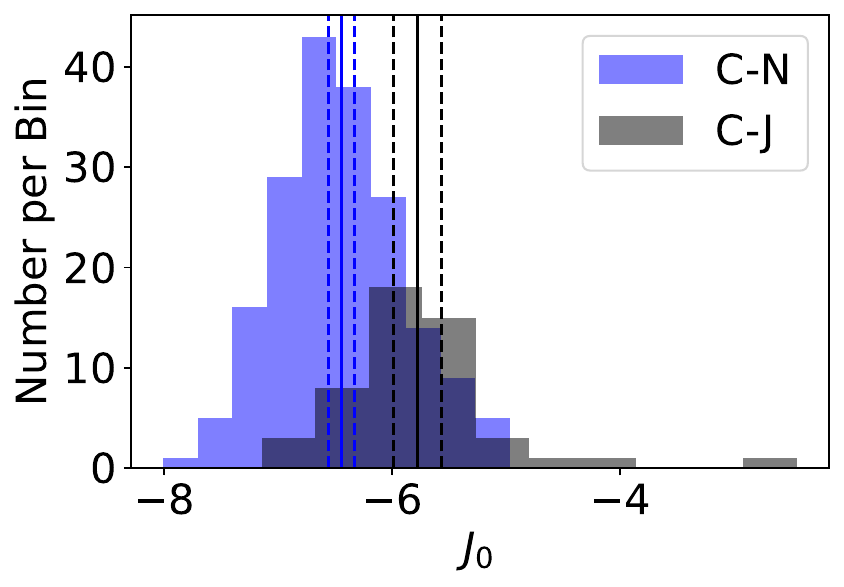}
  \end{tabular}
\caption{J-band luminosity functions for N- and J-type carbon stars retrieved from \cite{Abia_2022A&A...664A..45A}. The vertical lines correspond to the means of each luminosity function, with the dashed lines indicating the error on the means. The difference in means between the two luminosity functions is 0.5~mag.}
\label{fig:NJTypes}
\end{figure*}

We also note that the color dependence of the JAGB is not directly related to the asymmetry of some JAGB LFs, i.e., the former does not explain the latter.  One can imagine an arbitrarily  narrow color range in the center of the JAGB with a still asymmetric luminosity function.  While we do not know the cause of the asymmetry, it may be caused by a mixture of different populations/types of carbon stars with different mean luminosities. \cite{Morgan_2003MNRAS.341..534M} observed carbon stars in the LMC using the 2dF facility on the Anglo-Australian Telescope and found that the J-type carbon star sequence is offset by 0.6~mag from the K-type carbon star sequence on a K vs. J - K luminosity function. In addition, Fig. 10 in \cite{Ripoche_2020MNRAS.495.2858R} shows that both J-type and N-type carbon stars reside within the J-region used to measure the JAGB reference magnitude. Fig. 7 in \cite{Abia_2022A&A...664A..45A} shows that the bolometric luminosities of the J- and N-type carbon stars have different mean magnitudes. We retrieved the data from \cite{Abia_2022A&A...664A..45A} and find that this difference persists in the 2MASS J-band as well, with a difference in mean magnitudes between the J- and N-type carbon star luminosity functions $\sim0.5$~mag at about 2$\sigma$ (see Fig. \ref{fig:NJTypes}).
 
\begin{figure*}[ht!]
  \centering
  \begin{tabular}{cccc}
    \includegraphics[width=0.32\linewidth]{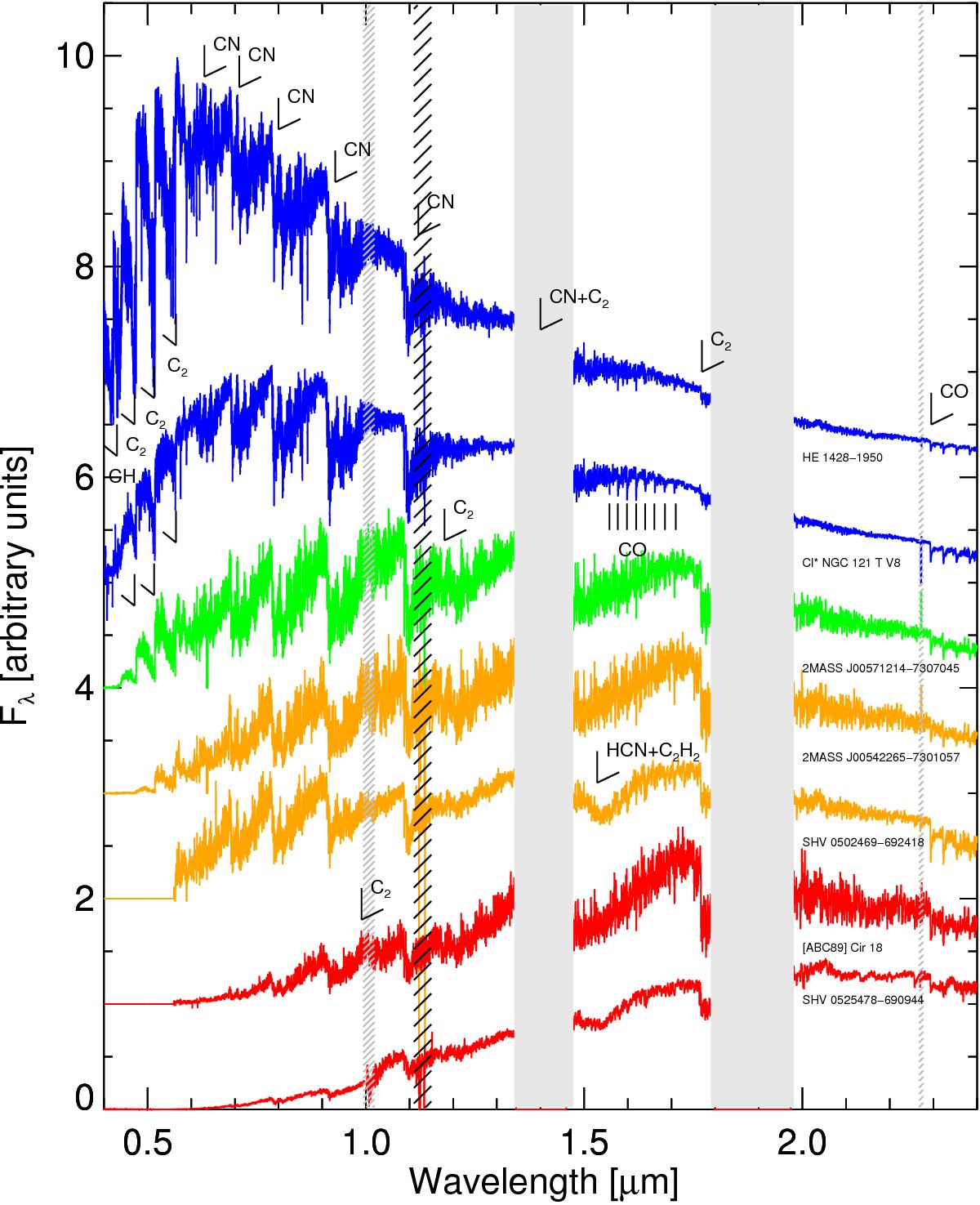}
  \end{tabular}
\caption{Fig. 5 from \cite{Gonneau_2016A&A...589A..36G}. This plot shows representative spectra from their sample.}
\label{fig:Spectra}
\end{figure*}

Finally, the spectral features distinctive to Carbon-rich stars are present in both the J, H, and $F150W$ bands suggesting that aby differences in the nature of the J-region studied above are not due to a lack of coverage of Carbon molecular bands. We refer the reader to the spectra from the X-shooter spectral library in \cite{Gonneau_2016A&A...589A..36G, Gonneau_2017AA...601A.141G}. For instance, in Fig. 5 in \cite{Gonneau_2016A&A...589A..36G} (seen in Fig. \ref{fig:Spectra}), we see that the representative carbon star in orange of group 3, which contains most of the J-region stars within the color range of 1.4 $< J - K < $ 2~mag typically used to select J-region stars. Section 4.3 in \cite{Gonneau_2016A&A...589A..36G} notes that their group 3 stars, which have colors 1.6 $< J - K < 2.2$~mag and overlap with the colors used to define the J-region, have strong $C_2$ bands in \textit{both} J- and H-bands suggesting that the spectral features of these Carbons stars do not preferentially lie in one of those two bands rather than the other. For context, the $JWST$ $F115W$ and $F150W$ bands cover between 1.013 to 1.282 $\mu$m and 1.331 to 1.668 $\mu$m, respectively.

\section{Effects of Smoothing on J-region Luminosity Functions} \label{Apdx: Effects of Smoothing on JAGB Luminosity Functions}

\begin{figure*}[ht!]
\epsscale{0.5}
\plotone{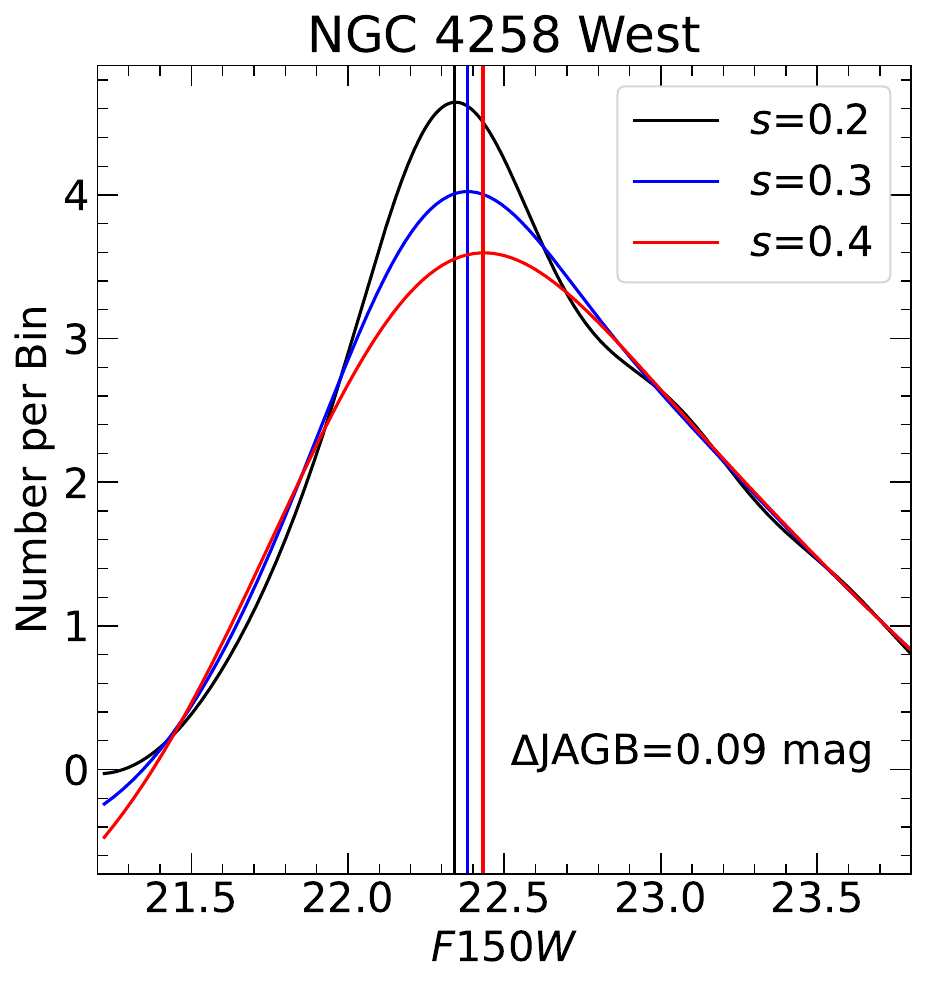}
\caption {GLOESS smoothed luminosity functions for the NGC 4258 Inner+Outer West field using smoothing paramters of $s=$ 0.2, 0.3, and 0.4~mag.}
\label{fig:JAGB_Smoothing_Test_N4258}
\end{figure*}

We find that the choice of smoothing in the mode-based method described in \cite{Lee_2022ApJ...933..201L, Lee_2023ApJ...956...15L} can bias JAGB measurements for asymmetric luminosity functions. We demonstrate this by plotting in Fig. \ref{fig:JAGB_Smoothing_Test_N4258} the smoothed luminosity function for the NGC 4258 Inner+Outer West field in \emph{F150W} using baseline color cuts of 1~mag and 1.5~mag, magnitude cuts of 21~mag and 24~mag, and three levels of smoothing ($s=$ 0.2, 0.3, and 0.4~mag). For reference, we note that smoothings of $s=$ 0.2~mag is used by \cite{Lee_2023ApJ...956...15L} and $s=$ 0.25 and 0.3~mag are used by \cite{Lee_2024ApJ...961..132L}. The difference between the brightest and faintest measured JAGBs for the range of smoothing from $s=$ 0.2 and 0.4~mag is 0.09~mag. We caution that using different smoothing parameters with this measurement method across the distance ladder can potentially introduce biases if luminosity functions are not equally asymmetric in the anchor and host galaxies. 

\section{Stability of the JAGB with Radius} \label{Apdx: Stability of the JAGB with Radius}

\cite{Lee_2022ApJ...933..201L} and \cite{Lee_2024ApJ...961..132L} show that the measured JAGB can change with radius in a galaxy and argue that the JAGB should be measured in a region where the measured JAGB converges. To find where the measured JAGB converges, \cite{Lee_2024ApJ...961..132L} construct annuli centered on their target galaxy and plot the measured JAGB as a function of the DOLPHOT sky parameter in each annulus.

We perform a similar test with our sample by constructing several annuli centered on the galaxy and measure the JAGB using the stars inside each annulus. We use the same color and magnitude cuts as described in the main body of this paper. For the purpose of this exercise, we used the mode based method and smoothing of ($s$ = 0.25~mag) \citep{Lee_2024ApJ...961..132L} to measure the JAGB.

\begin{figure*}[ht!]
  \centering
  \begin{tabular}{cccc}
    \includegraphics[width=0.45\linewidth]{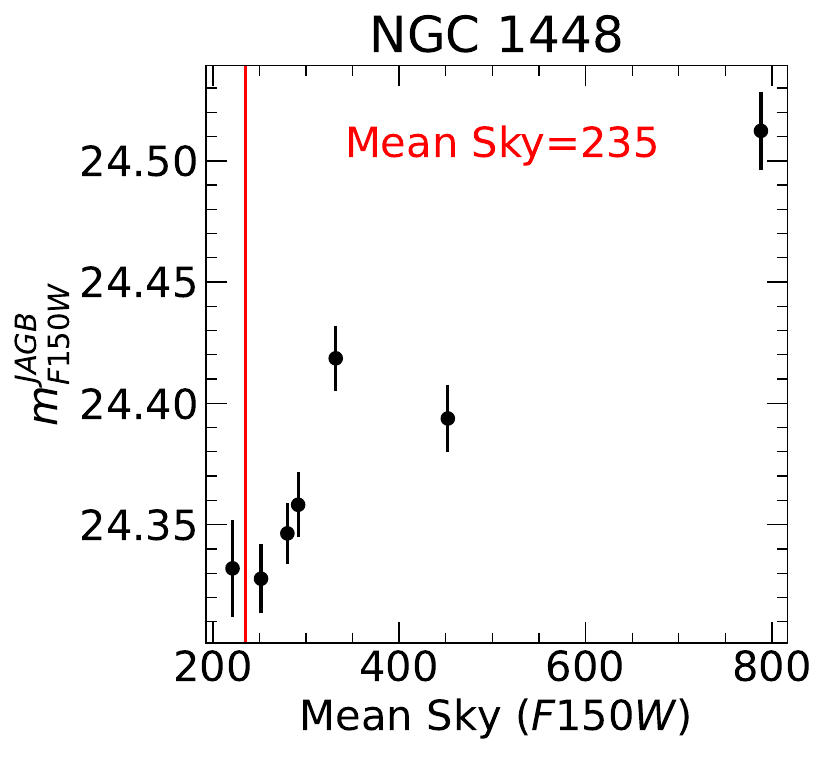} &
    \includegraphics[width=0.41\linewidth]{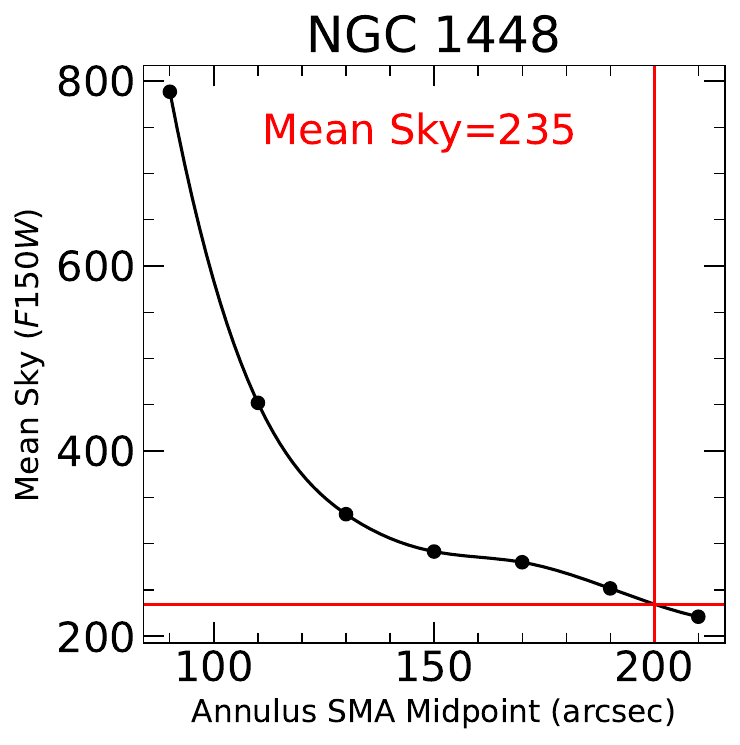} \\
    \includegraphics[width=0.45\linewidth]{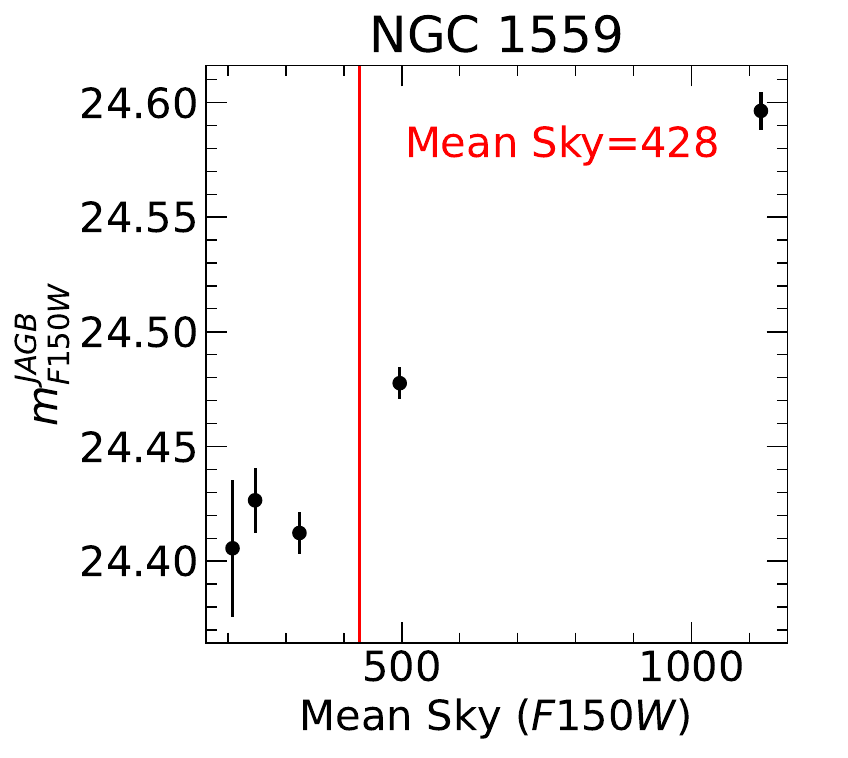} &
    \includegraphics[width=0.41\linewidth]{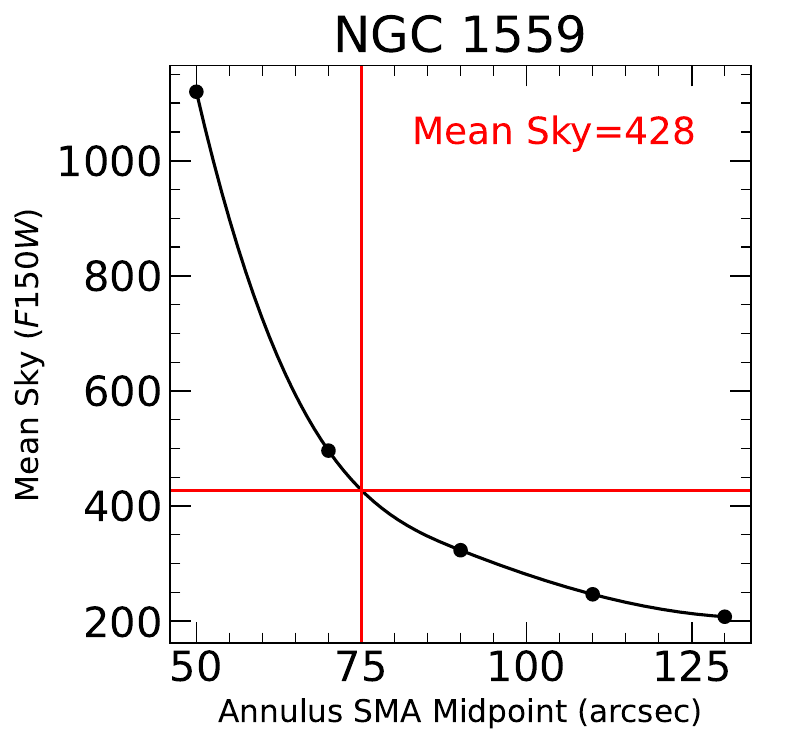} \\
    \includegraphics[width=0.45\linewidth]{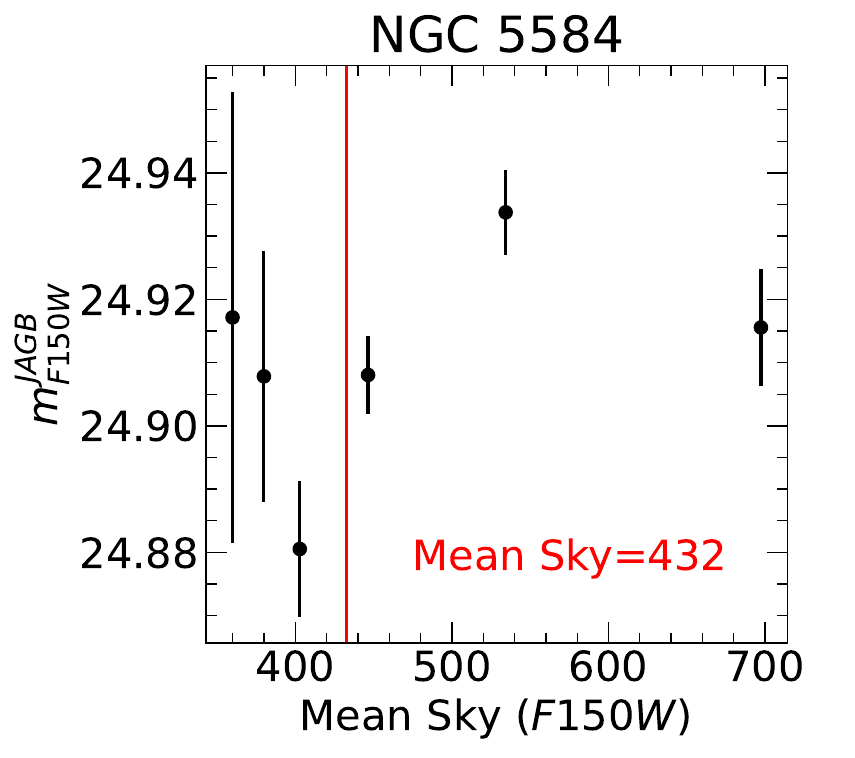} &
    \includegraphics[width=0.41\linewidth]{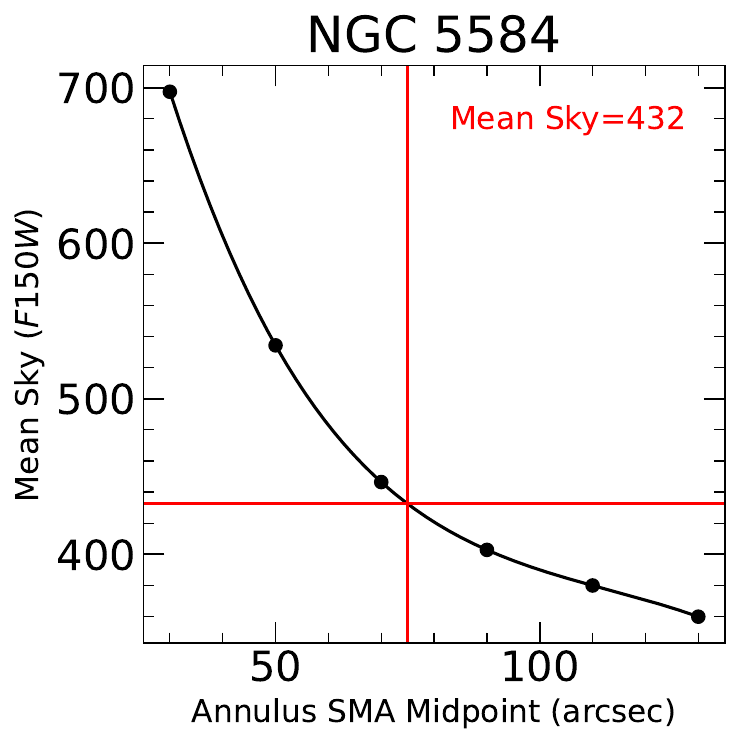} &
  \end{tabular}
\caption{Mean DOLPHOT sky value in each annuli described in Appendix \ref{Apdx: Stability of the JAGB with Radius} as a function the midpoint SMA for each annuli and the measured \emph{F150W} JAGB using the method from \cite{Lee_2024ApJ...961..132L} as a function of the mean sky in each annulus. We mark with the red lines the SMA used to construct the outer disk regions in the main body of this paper.}
\label{fig:JAGB_Sky}
\end{figure*}

For NGC 1448, we begin constructing annuli with semimajor axes (SMA) of 80$''$$<$SMA$<$100$''$ and increase each of these boundaries by 20$''$ until we reach 200$''$$<$SMA$<$220$''$. For NGC 1559 and 5584 we begin with 20$''$$<$SMA$<$40$''$ and increase each boundary in 20$''$ increments until we reach 140$''$$<$SMA$<$160$''$. We plot the mean DOLPHOT sky value as a function of the midpoint annulus SMA and the corresponding measured JAGBs as a function of the midpoint annulus SMA in Fig. \ref{fig:JAGB_Sky}. We mark with the red line the SMA used to construct the outer disk regions in the main body of this paper for reference. We find that not all galaxies show measured JAGBs that converge out to a large radius (for instance, the measured JAGBs for NGC 5584 appear to become increasingly fainter to the left of the red line without indication flattening in the same way with NGC 1559) suggesting that there may need to be additional standardization beyond a radial cut. We note that convergence does not necessarily imply that the JAGB populations across galaxies are similar and yield the same intrinsic JAGB luminosity. For instance, convergence may occur at regions of differing metaliicities which could change the JAGB luminosity (see Section \ref{sec:Intro} for a discussion on a possible metallicity trend).

\bibliography{sample631}{}

\begin{thebibliography}{}
\expandafter\ifx\csname natexlab\endcsname\relax\def\natexlab#1{#1}\fi
\providecommand{\url}[1]{\href{#1}{#1}}
\providecommand{\dodoi}[1]{doi:~\href{http://doi.org/#1}{\nolinkurl{#1}}}
\providecommand{\doeprint}[1]{\href{http://ascl.net/#1}{\nolinkurl{http://ascl.net/#1}}}
\providecommand{\doarXiv}[1]{\href{https://arxiv.org/abs/#1}{\nolinkurl{https://arxiv.org/abs/#1}}}

\bibitem[{{Abia} {et~al.}(2022){Abia}, {de Laverny}, {Romero-G{\'o}mez}, \& {Figueras}}]{Abia_2022A&A...664A..45A}
{Abia}, C., {de Laverny}, P., {Romero-G{\'o}mez}, M., \& {Figueras}, F. 2022, \aap, 664, A45, \dodoi{10.1051/0004-6361/202243595}

\bibitem[{{Anand} {et~al.}(2021){Anand}, {Rizzi}, {Tully}, {Shaya}, {Karachentsev}, {Makarov}, {Makarova}, {Wu}, {Dolphin}, \& {Kourkchi}}]{Anand_2021AJ....162...80A}
{Anand}, G.~S., {Rizzi}, L., {Tully}, R.~B., {et~al.} 2021, \aj, 162, 80, \dodoi{10.3847/1538-3881/ac0440}

\bibitem[{{Anand} {et~al.}(2024){Anand}, {Riess}, {Yuan}, {Beaton}, {Casertano}, {Li}, {Makarov}, {Makarova}, {Tully}, {Anderson}, {Breuval}, {Dolphin}, {Karachentsev}, {Macri}, \& {Scolnic}}]{Anand_2024arXiv240104776A}
{Anand}, G.~S., {Riess}, A.~G., {Yuan}, W., {et~al.} 2024, arXiv e-prints, arXiv:2401.04776, \dodoi{10.48550/arXiv.2401.04776}

\bibitem[{{Astropy Collaboration} {et~al.}(2013){Astropy Collaboration}, {Robitaille}, {Tollerud}, {Greenfield}, {Droettboom}, {Bray}, {Aldcroft}, {Davis}, {Ginsburg}, {Price-Whelan}, {Kerzendorf}, {Conley}, {Crighton}, {Barbary}, {Muna}, {Ferguson}, {Grollier}, {Parikh}, {Nair}, {Unther}, {Deil}, {Woillez}, {Conseil}, {Kramer}, {Turner}, {Singer}, {Fox}, {Weaver}, {Zabalza}, {Edwards}, {Azalee Bostroem}, {Burke}, {Casey}, {Crawford}, {Dencheva}, {Ely}, {Jenness}, {Labrie}, {Lim}, {Pierfederici}, {Pontzen}, {Ptak}, {Refsdal}, {Servillat}, \& {Streicher}}]{Astropy_2013A&A...558A..33A}
{Astropy Collaboration}, {Robitaille}, T.~P., {Tollerud}, E.~J., {et~al.} 2013, \aap, 558, A33, \dodoi{10.1051/0004-6361/201322068}

\bibitem[{{Astropy Collaboration} {et~al.}(2018){Astropy Collaboration}, {Price-Whelan}, {Sip{\H{o}}cz}, {G{\"u}nther}, {Lim}, {Crawford}, {Conseil}, {Shupe}, {Craig}, {Dencheva}, {Ginsburg}, {VanderPlas}, {Bradley}, {P{\'e}rez-Su{\'a}rez}, {de Val-Borro}, {Aldcroft}, {Cruz}, {Robitaille}, {Tollerud}, {Ardelean}, {Babej}, {Bach}, {Bachetti}, {Bakanov}, {Bamford}, {Barentsen}, {Barmby}, {Baumbach}, {Berry}, {Biscani}, {Boquien}, {Bostroem}, {Bouma}, {Brammer}, {Bray}, {Breytenbach}, {Buddelmeijer}, {Burke}, {Calderone}, {Cano Rodr{\'\i}guez}, {Cara}, {Cardoso}, {Cheedella}, {Copin}, {Corrales}, {Crichton}, {D'Avella}, {Deil}, {Depagne}, {Dietrich}, {Donath}, {Droettboom}, {Earl}, {Erben}, {Fabbro}, {Ferreira}, {Finethy}, {Fox}, {Garrison}, {Gibbons}, {Goldstein}, {Gommers}, {Greco}, {Greenfield}, {Groener}, {Grollier}, {Hagen}, {Hirst}, {Homeier}, {Horton}, {Hosseinzadeh}, {Hu}, {Hunkeler}, {Ivezi{\'c}}, {Jain}, {Jenness}, {Kanarek}, {Kendrew}, {Kern}, {Kerzendorf}, {Khvalko}, {King}, {Kirkby}, {Kulkarni},
  {Kumar}, {Lee}, {Lenz}, {Littlefair}, {Ma}, {Macleod}, {Mastropietro}, {McCully}, {Montagnac}, {Morris}, {Mueller}, {Mumford}, {Muna}, {Murphy}, {Nelson}, {Nguyen}, {Ninan}, {N{\"o}the}, {Ogaz}, {Oh}, {Parejko}, {Parley}, {Pascual}, {Patil}, {Patil}, {Plunkett}, {Prochaska}, {Rastogi}, {Reddy Janga}, {Sabater}, {Sakurikar}, {Seifert}, {Sherbert}, {Sherwood-Taylor}, {Shih}, {Sick}, {Silbiger}, {Singanamalla}, {Singer}, {Sladen}, {Sooley}, {Sornarajah}, {Streicher}, {Teuben}, {Thomas}, {Tremblay}, {Turner}, {Terr{\'o}n}, {van Kerkwijk}, {de la Vega}, {Watkins}, {Weaver}, {Whitmore}, {Woillez}, {Zabalza}, \& {Astropy Contributors}}]{Astropy_2018AJ....156..123A}
{Astropy Collaboration}, {Price-Whelan}, A.~M., {Sip{\H{o}}cz}, B.~M., {et~al.} 2018, \aj, 156, 123, \dodoi{10.3847/1538-3881/aabc4f}

\bibitem[{{Astropy Collaboration} {et~al.}(2022){Astropy Collaboration}, {Price-Whelan}, {Lim}, {Earl}, {Starkman}, {Bradley}, {Shupe}, {Patil}, {Corrales}, {Brasseur}, {N{\"o}the}, {Donath}, {Tollerud}, {Morris}, {Ginsburg}, {Vaher}, {Weaver}, {Tocknell}, {Jamieson}, {van Kerkwijk}, {Robitaille}, {Merry}, {Bachetti}, {G{\"u}nther}, {Aldcroft}, {Alvarado-Montes}, {Archibald}, {B{\'o}di}, {Bapat}, {Barentsen}, {Baz{\'a}n}, {Biswas}, {Boquien}, {Burke}, {Cara}, {Cara}, {Conroy}, {Conseil}, {Craig}, {Cross}, {Cruz}, {D'Eugenio}, {Dencheva}, {Devillepoix}, {Dietrich}, {Eigenbrot}, {Erben}, {Ferreira}, {Foreman-Mackey}, {Fox}, {Freij}, {Garg}, {Geda}, {Glattly}, {Gondhalekar}, {Gordon}, {Grant}, {Greenfield}, {Groener}, {Guest}, {Gurovich}, {Handberg}, {Hart}, {Hatfield-Dodds}, {Homeier}, {Hosseinzadeh}, {Jenness}, {Jones}, {Joseph}, {Kalmbach}, {Karamehmetoglu}, {Ka{\l}uszy{\'n}ski}, {Kelley}, {Kern}, {Kerzendorf}, {Koch}, {Kulumani}, {Lee}, {Ly}, {Ma}, {MacBride}, {Maljaars}, {Muna}, {Murphy}, {Norman},
  {O'Steen}, {Oman}, {Pacifici}, {Pascual}, {Pascual-Granado}, {Patil}, {Perren}, {Pickering}, {Rastogi}, {Roulston}, {Ryan}, {Rykoff}, {Sabater}, {Sakurikar}, {Salgado}, {Sanghi}, {Saunders}, {Savchenko}, {Schwardt}, {Seifert-Eckert}, {Shih}, {Jain}, {Shukla}, {Sick}, {Simpson}, {Singanamalla}, {Singer}, {Singhal}, {Sinha}, {Sip{\H{o}}cz}, {Spitler}, {Stansby}, {Streicher}, {{\v{S}}umak}, {Swinbank}, {Taranu}, {Tewary}, {Tremblay}, {de Val-Borro}, {Van Kooten}, {Vasovi{\'c}}, {Verma}, {de Miranda Cardoso}, {Williams}, {Wilson}, {Winkel}, {Wood-Vasey}, {Xue}, {Yoachim}, {Zhang}, {Zonca}, \& {Astropy Project Contributors}}]{Astropy_2022ApJ...935..167A}
{Astropy Collaboration}, {Price-Whelan}, A.~M., {Lim}, P.~L., {et~al.} 2022, \apj, 935, 167, \dodoi{10.3847/1538-4357/ac7c74}

\bibitem[{{Battinelli} \& {Demers}(2005)}]{Battinelli_2005AA...442..159B}
{Battinelli}, P., \& {Demers}, S. 2005, \aap, 442, 159, \dodoi{10.1051/0004-6361:20053357}

\bibitem[{{Breuval} {et~al.}(2022){Breuval}, {Riess}, {Kervella}, {Anderson}, \& {Romaniello}}]{Breuval_2022ApJ...939...89B}
{Breuval}, L., {Riess}, A.~G., {Kervella}, P., {Anderson}, R.~I., \& {Romaniello}, M. 2022, \apj, 939, 89, \dodoi{10.3847/1538-4357/ac97e2}

\bibitem[{{Dolphin}(2016)}]{Dolphin_2016ascl.soft08013D}
{Dolphin}, A. 2016, {DOLPHOT: Stellar photometry}, Astrophysics Source Code Library, record ascl:1608.013.
\newblock \doeprint{1608.013}

\bibitem[{{Dolphin}(2000)}]{Dolphin_2000PASP..112.1383D}
{Dolphin}, A.~E. 2000, \pasp, 112, 1383, \dodoi{10.1086/316630}

\bibitem[{{Freedman} \& {Madore}(2020)}]{Freedman_2020ApJ...899...67F}
{Freedman}, W.~L., \& {Madore}, B.~F. 2020, \apj, 899, 67, \dodoi{10.3847/1538-4357/aba9d8}

\bibitem[{{Freedman} {et~al.}(2021){Freedman}, {Madore}, {Hoyt}, {Jang}, {Lee}, \& {Owens}}]{Freedman_2021jwst.prop.1995F}
{Freedman}, W.~L., {Madore}, B.~F., {Hoyt}, T., {et~al.} 2021, {Answering the Most Important Problem in Cosmology Today: Is the Tension in the Hubble Constant Real?}, JWST Proposal. Cycle 1, ID. \#1995

\bibitem[{{Gonneau} {et~al.}(2016){Gonneau}, {Lan{\c{c}}on}, {Trager}, {Aringer}, {Lyubenova}, {Nowotny}, {Peletier}, {Prugniel}, {Chen}, {Dries}, {Choudhury}, {Falc{\'o}n-Barroso}, {Koleva}, {Meneses-Goytia}, {S{\'a}nchez-Bl{\'a}zquez}, \& {Vazdekis}}]{Gonneau_2016A&A...589A..36G}
{Gonneau}, A., {Lan{\c{c}}on}, A., {Trager}, S.~C., {et~al.} 2016, \aap, 589, A36, \dodoi{10.1051/0004-6361/201526292}

\bibitem[{{Gonneau} {et~al.}(2017){Gonneau}, {Lan{\c{c}}on}, {Trager}, {Aringer}, {Nowotny}, {Peletier}, {Prugniel}, {Chen}, \& {Lyubenova}}]{Gonneau_2017AA...601A.141G}
---. 2017, \aap, 601, A141, \dodoi{10.1051/0004-6361/201629750}

\bibitem[{{G{\'o}rski} {et~al.}(2020){G{\'o}rski}, {Zgirski}, {Pietrzy{\'n}ski}, {Gieren}, {Wielg{\'o}rski}, {Graczyk}, {Kudritzki}, {Pilecki}, {Narloch}, {Karczmarek}, {Suchomska}, \& {Taormina}}]{Gorski_2020ApJ...889..179G}
{G{\'o}rski}, M., {Zgirski}, B., {Pietrzy{\'n}ski}, G., {et~al.} 2020, \apj, 889, 179, \dodoi{10.3847/1538-4357/ab65ed}

\bibitem[{{Graczyk} {et~al.}(2014){Graczyk}, {Pietrzy{\'n}ski}, {Thompson}, {Gieren}, {Pilecki}, {Konorski}, {Udalski}, {Soszy{\'n}ski}, {Villanova}, {G{\'o}rski}, {Suchomska}, {Karczmarek}, {Kudritzki}, {Bresolin}, \& {Gallenne}}]{Graczyk_2014ApJ...780...59G}
{Graczyk}, D., {Pietrzy{\'n}ski}, G., {Thompson}, I.~B., {et~al.} 2014, \apj, 780, 59, \dodoi{10.1088/0004-637X/780/1/59}

\bibitem[{{Graczyk} {et~al.}(2020){Graczyk}, {Pietrzy{\'n}ski}, {Thompson}, {Gieren}, {Zgirski}, {Villanova}, {G{\'o}rski}, {Wielg{\'o}rski}, {Karczmarek}, {Narloch}, {Pilecki}, {Taormina}, {Smolec}, {Suchomska}, {Gallenne}, {Nardetto}, {Storm}, {Kudritzki}, {Ka{\l}uszy{\'n}ski}, \& {Pych}}]{Graczyk_2020ApJ...904...13G}
---. 2020, \apj, 904, 13, \dodoi{10.3847/1538-4357/abbb2b}

\bibitem[{{Green}(2018)}]{Green_2018JOSS....3..695M}
{Green}, G. 2018, The Journal of Open Source Software, 3, 695, \dodoi{10.21105/joss.00695}

\bibitem[{{Green} {et~al.}(2019){Green}, {Schlafly}, {Zucker}, {Speagle}, \& {Finkbeiner}}]{Green_2019ApJ...887...93G}
{Green}, G.~M., {Schlafly}, E., {Zucker}, C., {Speagle}, J.~S., \& {Finkbeiner}, D. 2019, \apj, 887, 93, \dodoi{10.3847/1538-4357/ab5362}

\bibitem[{{Groenewegen} \& {Marigo}(2004)}]{Groenewegen_2004agbs.book..105G}
{Groenewegen}, M. A.~T., \& {Marigo}, P. 2004, in Asymptotic Giant Branch Stars (Springer), 105--148, \dodoi{10.1007/978-1-4757-3876-6_3}

\bibitem[{{Hoffmann} {et~al.}(2016){Hoffmann}, {Macri}, {Riess}, {Yuan}, {Casertano}, {Foley}, {Filippenko}, {Tucker}, {Chornock}, {Silverman}, {Welch}, {Goobar}, \& {Amanullah}}]{Hoffmann_2016ApJ...830...10H}
{Hoffmann}, S.~L., {Macri}, L.~M., {Riess}, A.~G., {et~al.} 2016, \apj, 830, 10, \dodoi{10.3847/0004-637X/830/1/10}

\bibitem[{{Huang} {et~al.}(2020){Huang}, {Riess}, {Yuan}, {Macri}, {Zakamska}, {Casertano}, {Whitelock}, {Hoffmann}, {Filippenko}, \& {Scolnic}}]{Huang_2020ApJ...889....5H}
{Huang}, C.~D., {Riess}, A.~G., {Yuan}, W., {et~al.} 2020, \apj, 889, 5, \dodoi{10.3847/1538-4357/ab5dbd}

\bibitem[{{Joye} \& {Mandel}(2003)}]{SAO_Image_2003ASPC..295..489J}
{Joye}, W.~A., \& {Mandel}, E. 2003, in Astronomical Society of the Pacific Conference Series, Vol. 295, Astronomical Data Analysis Software and Systems XII, ed. H.~E. {Payne}, R.~I. {Jedrzejewski}, \& R.~N. {Hook}, 489

\bibitem[{{Karakas}(2014)}]{Karakas_2014MNRAS.445..347K}
{Karakas}, A.~I. 2014, \mnras, 445, 347, \dodoi{10.1093/mnras/stu1727}

\bibitem[{{Kato} {et~al.}(2007){Kato}, {Nagashima}, {Nagayama}, {Kurita}, {Koerwer}, {Kawai}, {Yamamuro}, {Zenno}, {Nishiyama}, {Baba}, {Kadowaki}, {Haba}, {Hatano}, {Shimizu}, {Nishimura}, {Nagata}, {Sato}, {Murai}, {Kawazu}, {Nakajima}, {Nakaya}, {Kandori}, {Kusakabe}, {Ishihara}, {Kaneyasu}, {Hashimoto}, {Tamura}, {Tanab{\'e}}, {Ita}, {Matsunaga}, {Nakada}, {Sugitani}, {Wakamatsu}, {Glass}, {Feast}, {Menzies}, {Whitelock}, {Fourie}, {Stoffels}, {Evans}, \& {Hasegawa}}]{Kato_2007PASJ...59..615K}
{Kato}, D., {Nagashima}, C., {Nagayama}, T., {et~al.} 2007, \pasj, 59, 615, \dodoi{10.1093/pasj/59.3.615}

\bibitem[{{Lee}(2023)}]{Lee_2023ApJ...956...15L}
{Lee}, A.~J. 2023, \apj, 956, 15, \dodoi{10.3847/1538-4357/acee69}

\bibitem[{{Lee} {et~al.}(2024){Lee}, {Freedman}, {Jang}, {Madore}, \& {Owens}}]{Lee_2024ApJ...961..132L}
{Lee}, A.~J., {Freedman}, W.~L., {Jang}, I.~S., {Madore}, B.~F., \& {Owens}, K.~A. 2024, \apj, 961, 132, \dodoi{10.3847/1538-4357/ad12c7}

\bibitem[{{Lee} {et~al.}(2021{\natexlab{a}}){Lee}, {Freedman}, {Madore}, {Owens}, {Monson}, \& {Hoyt}}]{Lee_WLM_2021ApJ...907..112L}
{Lee}, A.~J., {Freedman}, W.~L., {Madore}, B.~F., {et~al.} 2021{\natexlab{a}}, \apj, 907, 112, \dodoi{10.3847/1538-4357/abd253}

\bibitem[{{Lee} {et~al.}(2021{\natexlab{b}}){Lee}, {Freedman}, {Madore}, {Owens}, \& {Sung Jang}}]{Lee_MW_2021ApJ...923..157L}
{Lee}, A.~J., {Freedman}, W.~L., {Madore}, B.~F., {Owens}, K.~A., \& {Sung Jang}, I. 2021{\natexlab{b}}, \apj, 923, 157, \dodoi{10.3847/1538-4357/ac2f4c}

\bibitem[{{Lee} {et~al.}(2022){Lee}, {Rousseau-Nepton}, {Freedman}, {Madore}, {Cioni}, {Hoyt}, {Jang}, {Javadi}, \& {Owens}}]{Lee_2022ApJ...933..201L}
{Lee}, A.~J., {Rousseau-Nepton}, L., {Freedman}, W.~L., {et~al.} 2022, \apj, 933, 201, \dodoi{10.3847/1538-4357/ac7321}

\bibitem[{Loader(2004)}]{Loader_RePEc:zbw:caseps:200412}
Loader, C. 2004, {Smoothing: Local Regression Techniques}, Papers 2004,12, Humboldt University of Berlin, Center for Applied Statistics and Economics (CASE).
\newblock \url{https://ideas.repec.org/p/zbw/caseps/200412.html}

\bibitem[{{Macri} {et~al.}(2015){Macri}, {Ngeow}, {Kanbur}, {Mahzooni}, \& {Smitka}}]{Macri_2015AJ....149..117M}
{Macri}, L.~M., {Ngeow}, C.-C., {Kanbur}, S.~M., {Mahzooni}, S., \& {Smitka}, M.~T. 2015, \aj, 149, 117, \dodoi{10.1088/0004-6256/149/4/117}

\bibitem[{{Madore} \& {Freedman}(2020)}]{Madore_2020ApJ...899...66M}
{Madore}, B.~F., \& {Freedman}, W.~L. 2020, \apj, 899, 66, \dodoi{10.3847/1538-4357/aba045}

\bibitem[{{Madore} {et~al.}(2022){Madore}, {Freedman}, {Lee}, \& {Owens}}]{Madore_2022ApJ...938..125M}
{Madore}, B.~F., {Freedman}, W.~L., {Lee}, A.~J., \& {Owens}, K. 2022, \apj, 938, 125, \dodoi{10.3847/1538-4357/ac92fd}

\bibitem[{{Marigo} {et~al.}(2008){Marigo}, {Girardi}, {Bressan}, {Groenewegen}, {Silva}, \& {Granato}}]{Marigo_2008A&A...482..883M}
{Marigo}, P., {Girardi}, L., {Bressan}, A., {et~al.} 2008, \aap, 482, 883, \dodoi{10.1051/0004-6361:20078467}

\bibitem[{{M{\'e}nard} {et~al.}(2010){M{\'e}nard}, {Scranton}, {Fukugita}, \& {Richards}}]{Menard_2010MNRAS.405.1025M}
{M{\'e}nard}, B., {Scranton}, R., {Fukugita}, M., \& {Richards}, G. 2010, \mnras, 405, 1025, \dodoi{10.1111/j.1365-2966.2010.16486.x}

\bibitem[{{Morgan} {et~al.}(2003){Morgan}, {Cannon}, {Hatzidimitriou}, \& {Croke}}]{Morgan_2003MNRAS.341..534M}
{Morgan}, D.~H., {Cannon}, R.~D., {Hatzidimitriou}, D., \& {Croke}, B.~F.~W. 2003, \mnras, 341, 534, \dodoi{10.1046/j.1365-8711.2003.06424.x}

\bibitem[{{Parada} {et~al.}(2021){Parada}, {Heyl}, {Richer}, {Ripoche}, \& {Rousseau-Nepton}}]{Parada_2021MNRAS.501..933P}
{Parada}, J., {Heyl}, J., {Richer}, H., {Ripoche}, P., \& {Rousseau-Nepton}, L. 2021, \mnras, 501, 933, \dodoi{10.1093/mnras/staa3750}

\bibitem[{{Parada} {et~al.}(2023){Parada}, {Heyl}, {Richer}, {Ripoche}, \& {Rousseau-Nepton}}]{Parada_2023MNRAS.522..195P}
---. 2023, \mnras, 522, 195, \dodoi{10.1093/mnras/stad965}

\bibitem[{{Persson} {et~al.}(2004){Persson}, {Madore}, {Krzemi{\'n}ski}, {Freedman}, {Roth}, \& {Murphy}}]{Persson_2004AJ....128.2239P}
{Persson}, S.~E., {Madore}, B.~F., {Krzemi{\'n}ski}, W., {et~al.} 2004, \aj, 128, 2239, \dodoi{10.1086/424934}

\bibitem[{{Pietrzy{\'n}ski} {et~al.}(2019){Pietrzy{\'n}ski}, {Graczyk}, {Gallenne}, {Gieren}, {Thompson}, {Pilecki}, {Karczmarek}, {G{\'o}rski}, {Suchomska}, {Taormina}, {Zgirski}, {Wielg{\'o}rski}, {Ko{\l}aczkowski}, {Konorski}, {Villanova}, {Nardetto}, {Kervella}, {Bresolin}, {Kudritzki}, {Storm}, {Smolec}, \& {Narloch}}]{Pietrzynski_2019Natur.567..200P}
{Pietrzy{\'n}ski}, G., {Graczyk}, D., {Gallenne}, A., {et~al.} 2019, \nat, 567, 200, \dodoi{10.1038/s41586-019-0999-4}

\bibitem[{{Reid} {et~al.}(2019){Reid}, {Pesce}, \& {Riess}}]{Reid_2019ApJ...886L..27R}
{Reid}, M.~J., {Pesce}, D.~W., \& {Riess}, A.~G. 2019, \apjl, 886, L27, \dodoi{10.3847/2041-8213/ab552d}

\bibitem[{{Riess} {et~al.}(2021){Riess}, {Anderson}, {Breuval}, {Casertano}, {Macri}, {Scolnic}, \& {Yuan}}]{Riess_2021jwst.prop.1685R}
{Riess}, A., {Anderson}, R.~I., {Breuval}, L., {et~al.} 2021, {Uncrowding the Cepheids for an Improved Determination of the Hubble Constant}, JWST Proposal. Cycle 1, ID. \#1685

\bibitem[{{Riess} {et~al.}(2022){Riess}, {Yuan}, {Macri}, {Scolnic}, {Brout}, {Casertano}, {Jones}, {Murakami}, {Anand}, {Breuval}, {Brink}, {Filippenko}, {Hoffmann}, {Jha}, {D'arcy Kenworthy}, {Mackenty}, {Stahl}, \& {Zheng}}]{Riess_2022ApJ...934L...7R}
{Riess}, A.~G., {Yuan}, W., {Macri}, L.~M., {et~al.} 2022, \apjl, 934, L7, \dodoi{10.3847/2041-8213/ac5c5b}

\bibitem[{{Riess} {et~al.}(2023){Riess}, {Anand}, {Yuan}, {Casertano}, {Dolphin}, {Macri}, {Breuval}, {Scolnic}, {Perrin}, \& {Anderson}}]{Riess_2023ApJ...956L..18R}
{Riess}, A.~G., {Anand}, G.~S., {Yuan}, W., {et~al.} 2023, \apjl, 956, L18, \dodoi{10.3847/2041-8213/acf769}

\bibitem[{{Ripoche} {et~al.}(2020){Ripoche}, {Heyl}, {Parada}, \& {Richer}}]{Ripoche_2020MNRAS.495.2858R}
{Ripoche}, P., {Heyl}, J., {Parada}, J., \& {Richer}, H. 2020, \mnras, 495, 2858, \dodoi{10.1093/mnras/staa1346}

\bibitem[{{Scolnic} {et~al.}(2023){Scolnic}, {Riess}, {Wu}, {Li}, {Anand}, {Beaton}, {Casertano}, {Anderson}, {Dhawan}, \& {Ke}}]{Scolnic_2023ApJ...954L..31S}
{Scolnic}, D., {Riess}, A.~G., {Wu}, J., {et~al.} 2023, \apjl, 954, L31, \dodoi{10.3847/2041-8213/ace978}

\bibitem[{{Scowcroft} {et~al.}(2016){Scowcroft}, {Freedman}, {Madore}, {Monson}, {Persson}, {Rich}, {Seibert}, \& {Rigby}}]{Skowcroft_2016ApJ...816...49S}
{Scowcroft}, V., {Freedman}, W.~L., {Madore}, B.~F., {et~al.} 2016, \apj, 816, 49, \dodoi{10.3847/0004-637X/816/2/49}

\bibitem[{{Skowron} {et~al.}(2021){Skowron}, {Skowron}, {Udalski}, {Szyma{\'n}ski}, {Soszy{\'n}ski}, {Wyrzykowski}, {Ulaczyk}, {Poleski}, {Koz{\l}owski}, {Pietrukowicz}, {Mr{\'o}z}, {Rybicki}, {Iwanek}, {Wrona}, \& {Gromadzki}}]{Skowron_2021ApJS..252...23S}
{Skowron}, D.~M., {Skowron}, J., {Udalski}, A., {et~al.} 2021, \apjs, 252, 23, \dodoi{10.3847/1538-4365/abcb81}

\bibitem[{{Skrutskie} {et~al.}(2006){Skrutskie}, {Cutri}, {Stiening}, {Weinberg}, {Schneider}, {Carpenter}, {Beichman}, {Capps}, {Chester}, {Elias}, {Huchra}, {Liebert}, {Lonsdale}, {Monet}, {Price}, {Seitzer}, {Jarrett}, {Kirkpatrick}, {Gizis}, {Howard}, {Evans}, {Fowler}, {Fullmer}, {Hurt}, {Light}, {Kopan}, {Marsh}, {McCallon}, {Tam}, {Van Dyk}, \& {Wheelock}}]{Skrutskie_2006AJ....131.1163S}
{Skrutskie}, M.~F., {Cutri}, R.~M., {Stiening}, R., {et~al.} 2006, \aj, 131, 1163, \dodoi{10.1086/498708}

\bibitem[{Tange(2011)}]{Tange_2011a}
Tange, O. 2011, ;login: The USENIX Magazine, 36, 42.
\newblock \url{http://www.gnu.org/s/parallel}

\bibitem[{{Trentin} {et~al.}(2023){Trentin}, {Ripepi}, {Molinaro}, {Catanzaro}, {Storm}, {De Somma}, {Marconi}, {Bhardwaj}, {Gatto}, {Testa}, {Musella}, {Clementini}, \& {Leccia}}]{Trentin_2023arXiv231003603T}
{Trentin}, E., {Ripepi}, V., {Molinaro}, R., {et~al.} 2023, arXiv e-prints, arXiv:2310.03603, \dodoi{10.48550/arXiv.2310.03603}

\bibitem[{{Tully} {et~al.}(2009){Tully}, {Rizzi}, {Shaya}, {Courtois}, {Makarov}, \& {Jacobs}}]{Tully_2009AJ....138..323T}
{Tully}, R.~B., {Rizzi}, L., {Shaya}, E.~J., {et~al.} 2009, \aj, 138, 323, \dodoi{10.1088/0004-6256/138/2/323}

\bibitem[{{Warfield} {et~al.}(2023){Warfield}, {Richstein}, {Kallivayalil}, {Cohen}, {Savino}, {Boyer}, {Garling}, {Gennaro}, {McQuinn}, {Newman}, {Anderson}, {Cole}, {Correnti}, {Dolphin}, {Geha}, {Sandstrom}, {Weisz}, \& {Williams}}]{Warfield_2023RNAAS...7...23W}
{Warfield}, J.~T., {Richstein}, H., {Kallivayalil}, N., {et~al.} 2023, Research Notes of the American Astronomical Society, 7, 23, \dodoi{10.3847/2515-5172/acbb72}

\bibitem[{{Weinberg} \& {Nikolaev}(2001)}]{Weinberg_2001ApJ...548..712W}
{Weinberg}, M.~D., \& {Nikolaev}, S. 2001, \apj, 548, 712, \dodoi{10.1086/319001}

\bibitem[{{Zgirski} {et~al.}(2021){Zgirski}, {Pietrzy{\'n}ski}, {Gieren}, {G{\'o}rski}, {Wielg{\'o}rski}, {Karczmarek}, {Bresolin}, {Kervella}, {Kudritzki}, {Storm}, {Graczyk}, {Hajdu}, {Narloch}, {Pilecki}, {Suchomska}, \& {Taormina}}]{Zgirski_2021ApJ...916...19Z}
{Zgirski}, B., {Pietrzy{\'n}ski}, G., {Gieren}, W., {et~al.} 2021, \apj, 916, 19, \dodoi{10.3847/1538-4357/ac04b2}

\end{thebibliography}
\bibliographystyle{aasjournal}

\end{document}